\documentclass[11pt]{article}
\usepackage{amsmath,amssymb,graphicx,subfig}
\usepackage[english]{babel}
\usepackage{longtable}
\usepackage{lscape}
\usepackage{textcomp}

\usepackage{mathtools}
\DeclarePairedDelimiter{\floor}{\lfloor}{\rfloor}
\DeclarePairedDelimiter{\ceil}{\lceil}{\rceil}

\usepackage{hyperref}
\hypersetup{
    colorlinks=true,
    linkcolor=blue,
    filecolor=magenta,      
    urlcolor=cyan,
}

\newcommand{\Covid}{\mbox{COVID-19}}

\newcommand{\IID}{IID}
\newcommand{\trp}{^{\tt T}}
\DeclareMathOperator{\tr}{tr}

\title{Prediction Regions for Poisson and Over-Dispersed Poisson Regression Models with Applications to Forecasting Number of Deaths during the \Covid\ Pandemic}
\author{T.\ Kim\footnote{Post-doctoral Researcher, Department of Statistics, University of Haifa, Haifa, Israel 3498838. {\em Email:} ktaeho@campus.haifa.ac.il.} \and B.\ Lieberman\footnote{Graduate Student, Department of Statistics, University of South Carolina, Columbia, SC 29208. {\em Email:} LIEBERB@email.sc.edu.} \and G.\ Luta\footnote{Professor, Department of Biostatistics, Bioinformatics \& Biomathematics, Georgetown University, Washington DC. {\em Email:} George.Luta@georgetown.edu.} \and E.\ Pe\~na\footnote{{\em Corresponding Author.} Professor, Department of Statistics, University of South Carolina, Columbia, SC 29208. {\em Email:} pena@stat.sc.edu.}}

\begin{document}
\maketitle

\begin{abstract}
Motivated by the current Coronavirus Disease (\Covid) pandemic, which is due to the SARS-CoV-2 virus, and the important problem of forecasting daily deaths and cumulative deaths, this paper examines the construction of prediction regions or intervals under the Poisson regression model and for an over-dispersed Poisson regression model. For the Poisson regression model, several prediction regions are developed and their performance are compared through simulation studies. The methods are applied to the problem of forecasting daily and cumulative deaths in the United States (US) due to \Covid.  To examine their performance relative to what actually happened, daily deaths data until May 15th were used to forecast cumulative deaths by June 1st. It was observed that there is over-dispersion in the observed data relative to the Poisson regression model. An over-dispersed Poisson regression model is therefore proposed. This new model builds on frailty ideas in Survival Analysis and over-dispersion is quantified through an additional parameter. The Poisson regression model is a hidden model in this over-dispersed Poisson regression model and obtains as a limiting case when the over-dispersion parameter increases to infinity.  A prediction region for the cumulative number of US deaths due to \Covid\ by July 16th, given the data until July 2nd, is presented. Finally, the paper discusses limitations of proposed procedures and mentions open research problems, as well as the dangers and pitfalls when forecasting on a long horizon, with focus on this pandemic where events, both foreseen and unforeseen, could have huge impacts on point predictions and prediction regions.

\medskip


\medskip

\end{abstract}

\section{Introduction}
\label{sec-Introduction}

The current Coronavirus Disease (\Covid) pandemic  \cite{AboutCovid19}, caused by the SARS-CoV-2 virus,  is providing statisticians, data scientists, machine learners, and other modelers a real-time laboratory to test and demonstrate their forecasting skills and abilities, with the quality of their forecasts assessable in a matter of days, weeks, or months.  See, for instance, \url{https://covid19-projections.com} from the Masachussetts Institute of Technology (MIT) and the Institute of Health Metrics (IHME)'s \url{https://covid19.healthdata.org/united-states-of-america} based at the University of Washington in Seattle, as well as \cite{Kra20} discussing the complexities of modeling pandemics. Of particular interests are the forecasting of the numbers of daily cases\footnote{\Covid\ cases include confirmed and probable cases (infected people) and deaths according to a statement by the Council of State and Territorial Epidemiologists (CSTE) issued on April 5, 2020. See \cite{CSTE}.}, deaths, and hospitalizations,  or the cumulative cases, deaths, and hospitalizations attributable to \Covid\ at a future date in a specified country or a locality (e.g., a county, state, or province) on the basis of currently observed cases, deaths, and hospitalizations data. Such forecasts are of critical importance since they are major components in the decision-making process by government officials, business leaders, and educational and university administrators regarding the termination of lockdowns, lessening of social distancing and other mitigation regulations, opening of businesses, or continuing with online class formats in K-12 schools, colleges, and universities. 

\begin{figure}[h]
\caption{Scatterplot of the mumber of reported daily deaths and the cumulative deaths due to \Covid\ in the United States with respect to the number of days starting from December 31, 2019 (Day 62) until May 15, 2020 (Day 137), as reported by the European Center for Disease Control.}
\label{deaths and cumulative deaths data USA}
\begin{tabular}{cc}
\includegraphics[width=3in]{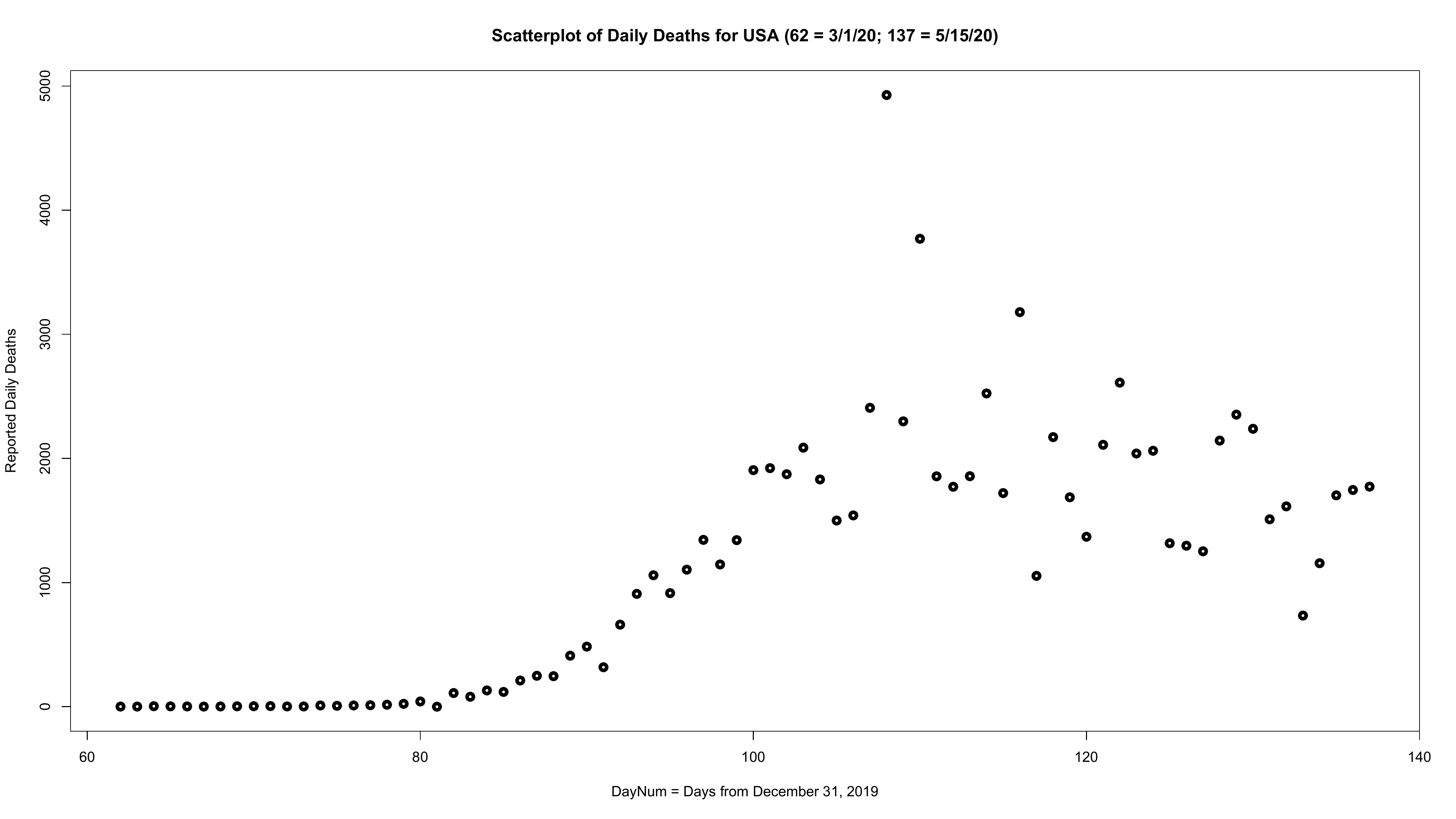} &
\includegraphics[width=3in]{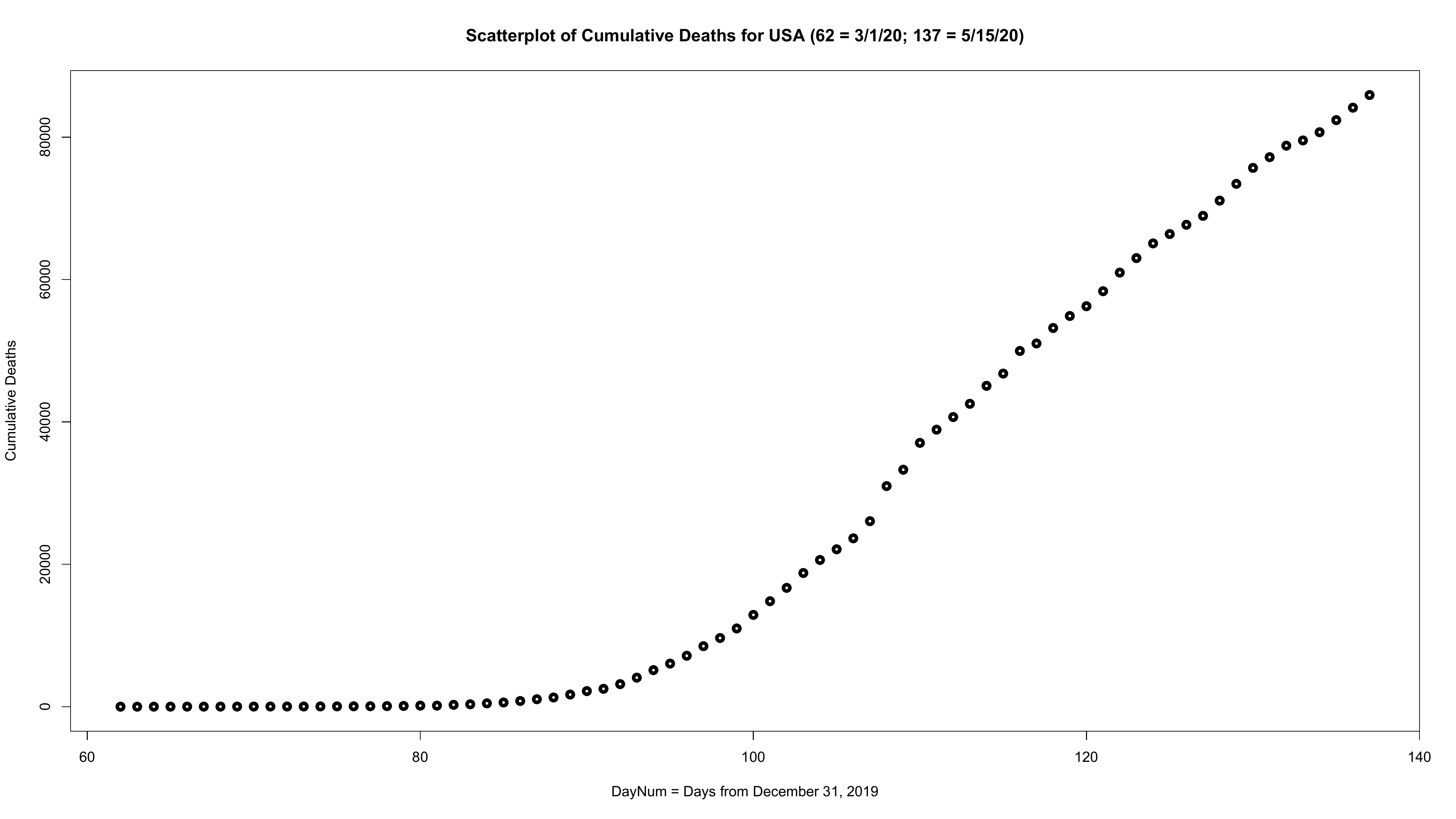}
\end{tabular}
\end{figure}

The left panel of Figure \ref{deaths and cumulative deaths data USA} provides the daily number of reported deaths due to \Covid\ for the United States (US) with respect to the number of days since December 31, 2019 until May 15, 2020, which is Day 137 in the figures, 
as reported by the \href{https://www.ecdc.europa.eu/en/covid-19-pandemic}{European Center for Disease Control (ECDC)} \cite{ECDCData} [see Section \ref{supp-ECDC}].  For a given date/day, including weekends, in the data set, the numbers reported are from the preceding day, which is due to a processing lag in reporting. The right panel of Figure \ref{deaths and cumulative deaths data USA} is a scatterplot of the cumulative number of deaths in the US due to \Covid. 
Given these daily and cumulative deaths data set, it is of interest to forecast the number of cumulative deaths in the US by, say, May 25, 2020 (corresponding to Day 147), which is Memorial Day, and to ask whether by that day the cumulative number of deaths in the US due to \Covid\ will have surpassed the ominously depressing and grim milestone of 100,000 cumulative deaths. Later, for our illustration, we will consider the problem of forecasting the cumulative number of deaths in the US due to \Covid\ at the end of May 2020, and compare our forecast with what eventually occurred. Finally, we attempt to forecast the cumulative number of deaths by July 16th based on the data on July 2nd.
%

Such forecasting problems are clearly non-trivial since there is the distinct possibility that whatever model we had fitted in the observed time-frame may not apply to the time period under forecast, the ever-present danger and risk of extrapolation. Aside from the fitted model most likely not being the true data generating model -- recalling the aphorism attributed to George E.\ P.\ Box \cite{Box76} that {\em all models are wrong, but some are useful} --  there are other factors, some beyond our control, that could impact the number of reported deaths at a future time, such as premature easing of social distancing and re-opening of business establishments, virus mutations, better diagnostic tools, changing hotspots, overburdened health care facilities, introduction of effective treatments, beneficial or detrimental actions by local, state, and/or federal entities, changing definition deaths due to \Covid, under- or over-reporting of deaths, timely development of a vaccine, protests and riots arising from social unrests, and others. But high-level decision-makers such as government officials, business leaders, educational administrators, and society itself, demand some beacon, however dim such beacons may be, to guide them in their decision-making. Statisticians, data scientists, machine learners, and other modelers are always ready and willing to provide such beacons. 

This paper is in this spirit. We will examine existing methods and develop new methods for constructing prediction regions for random variables that pertain to the number of occurrences of an event of interest. A prediction region contains more information compared to just a point prediction since it provides information about the uncertainty inherent in the prediction. Note that with a prediction region we are interested in the would-be realized value of a random variable, not the value of a parameter, hence instead of referring to it as a {\em confidence} region, it is instead called a {\em prediction} region. The events of particular interest are those that are `rare' in the sense that, informally, the probability of an event occurring in an infinitesimal interval is also infinitesimal. Consequently, our starting point will be the Poisson distribution which is a model for the number of occurrences of a rare event, and transition to the more general Poisson regression model, and eventually to an over-dispersed Poisson regression model which turns out to be a better model in the \Covid\ application. The real-life and practical application for which our methods will be applied is the construction of prediction regions for the daily and cumulative number of deaths due to \Covid\ in the US for a future date given {\em only} the daily deaths data until a current date.  Note that such predictions or forecasts could probably be improved by utilizing other information (such as the capacities of health care facilities; movements of people in a region; information about sensitivity and specificity of diagnostic tests; transmission rates ($R_0$) of the virus; and others), or via stratification by states, cities, or counties and then combining the results from these strata to obtain a point prediction and a prediction region for the whole US. However, we approach the construction of the prediction regions for the daily and cumulative deaths at a future date by {\em just} utilizing the observed reported daily deaths data for the whole US, which in a sense is the most reliable {\em available} data regarding this \Covid\ pandemic. It might be possible to utilize information about the number of cases or infected individuals, which is also reported daily, but we feel that this is not a reliable information since it is highly dependent on the number of tests that are performed and on the sensitivity and specificity of the diagnostic tests used. In addition, if such information is to be used in the prediction model, then we may not have their realized values at the future date on which the prediction region is desired. We point out that even though we are employing probabilistic models in the form of the Poisson or an over-dispersed Poisson model, which are derivable from intuitive conditions when dealing with rare events (cf., \cite{Res92,LiuPen16}), our prediction method is still purely {\em data-driven} being {\em only} reliant on the observed data. 

\section{Poisson Model}
\label{sec-Poisson Model}

The occurrence of a death due to \Covid\ could still be considered as a rare event when viewed in the context of the whole population, though even if it is rare, deaths are still significant and dire events. This is because to die of \Covid, generally one first needs to get infected, which at this point is still a rare event, and then having been infected, to die from it. The rate of dying when infected with \Covid, if not age-adjusted, is still rather low, less than 2\% (see, for instance, \href{https://www.worldometers.info/coronavirus/coronavirus-death-rate/}{Coronavirus (COVID-19) Mortality Rate}). Because of its rarity, a plausible probability model for the number of deaths due to \Covid\ is therefore the Poisson model whose probability mass function (PMF) is given by
\begin{equation}
\label{Poisson PMF}
p(k|\lambda) = \frac{\exp(-\lambda) \lambda^k}{k!}I\{k \in \mathbb{Z}_{0,+}\}
\end{equation}
with $\mathbb{Z}_{0,+} = \{0,1,2,\ldots\}$, and $\lambda > 0$ is the rate parameter, which is also the mean and variance of the distribution, and $I\{\cdot\}$ is the indicator function. For a variable $Y$ with this Poisson distribution, we write $Y \sim POI(\lambda)$. The cumulative distribution function of a $POI(\lambda)$ is 
\begin{equation}
\label{Poisson CDF}
P(w|\lambda) = \Pr\{Y \le w | \lambda\} = \sum_{\{k \in \mathbb{Z}_{0,+}:\ k \le w\}} p(k|\lambda) I\{w \ge 0 \}.
\end{equation}
We start our investigations with this no-covariate Poisson model, equivalently, a model with intercept only, since results for the Poisson regression model build on this no-covariate model.

\subsection{Prediction Regions when Rate $\lambda$ is Known}

Suppose now that $Y_0 \sim POI(\lambda)$, where for the moment we assume that we know the rate parameter $\lambda$. A $100(1-\alpha)\%$ prediction region for $Y_0$ is a subset $\Gamma(\lambda,\alpha) \subset \mathbb{Z}_{0,+}$ such that
\begin{equation}
\label{prediction region1}
\Pr\{Y_0 \in \Gamma(\lambda,\alpha) | \lambda\} = \sum_{k \in \Gamma(\lambda,\alpha)} p(k|\lambda) \ge 1 - \alpha.
\end{equation}
Note that this region will not be an {\em interval} being a subset of $\mathbb{Z}_{0,+}$, though if this region is formed as the intersection between $\mathbb{Z}_{0,+}$ and an interval in $\Re$, then we may call it imprecisely as an interval.
Subject to this condition, a desirable property of such a region is that its cardinality is as small as possible. If we allow for $\Gamma(\lambda,\alpha)$ to depend on a randomizer $U$, a standard uniform random variable independent of $Y_0$, the smallest cardinality $100(1-\alpha)\%$ prediction region is, using a Neyman-Pearson Lemma type argument, given by
\begin{equation}
\label{smallest cardinality prediction interval0}
\Gamma_0(U;\lambda,\alpha) = A(\lambda;c(\alpha)) \bigcup \left\{\{U \le \gamma(\alpha) \} \bigcap a(\lambda;c(\alpha)) \right\},
\end{equation}
where, for $d \in \Re$, we define the subsets of $\mathbb{Z}_{0,+}$ given by
\begin{eqnarray*}
A(\lambda;d) & = & \left\{k \in \mathbb{Z}_{0,+}:\ p(k|\lambda) > d\right\}; \\
a(\lambda;d) & = & \left\{k \in \mathbb{Z}_{0,+}:\ p(k|\lambda) = d\right\},
\end{eqnarray*}
and $c(\alpha)$ and $\gamma(\alpha)$ determined via
\begin{eqnarray*}
c(\alpha) & = & \inf\left\{d \ge 0:\ \Pr\{Y_0 \in A(\lambda;d) | \lambda\} \le 1 - \alpha\right\}; \\
\gamma(\alpha) & = & \frac{(1-\alpha) - \Pr\{Y_0 \in A(\lambda;c(\alpha)) |\lambda\}}{\Pr\{Y_0 \in a(\lambda;c(\alpha)) | \lambda\}},
\end{eqnarray*}
with $0/0 = 0$. Observe that by allowing randomized prediction regions, we have
\begin{displaymath}
\Pr\{Y_0 \in \Gamma_0(U;\lambda,\alpha) | \lambda\} =  1 - \alpha.
\end{displaymath}
If we do not admit randomized prediction regions, which is achieved by always taking $U = 0$ in $\Gamma_0(U;\lambda,\alpha)$, then unless $1-\alpha$ is a `natural' prediction coefficient, we will not achieve equality in the preceding probability statement. The use of the adjective `natural' is analogous to its use in constructing nonparametric confidence intervals, cf., \cite{RanWol79}. See \cite{PenKim19} on the application of Neyman-Pearson-type arguments to construct optimal confidence regions, which could be adapted to the construction of prediction regions.

There are two other ways of constructing prediction intervals for $Y_0$ when $\lambda$ is large using normal approximations. To obtain the prediction regions, these intervals are then intersected with $\mathbb{Z}_{0,+}$. Letting $N(\mu,\sigma^2)$ denote a normal distribution with mean $\mu$ and variance $\sigma^2$, we recall that when $\lambda$ is large owing to the Central Limit Theorem and the Delta Method (cf., \cite{CasBer90}), then we have the normal approximations
\begin{displaymath}
Y_0 \stackrel{\bullet}{\sim} N(\lambda,\lambda) \quad \mbox{and} \quad \sqrt{Y_0} \stackrel{\bullet}{\sim} N\left(\sqrt{\lambda}, 1/4\right).
\end{displaymath}
Let $\phi(\cdot)$ and $\Phi(\cdot)$ be the probability density and cumulative distribution functions of a standard normal random variable so that
\begin{displaymath}
\phi(z) = \frac{1}{\sqrt{2\pi}} \exp\left\{-\frac{1}{2}z^2\right\} \quad \mbox{and} \quad
\Phi(z) = \int_{-\infty}^z \phi(w) dw.
\end{displaymath}
Let $z_\alpha = \Phi^{-1}(1-\alpha)$ be its $(1-\alpha)$th quantile. Two approximate prediction regions for $Y_0$ when $\lambda$ is large, which are based on the above normal approximations, are given by
\begin{eqnarray}
\Gamma_1(\lambda,\alpha) & = & \left[\max\{0,\lambda - z_{\alpha/2} \sqrt{\lambda}\},\lambda + z_{\alpha/2} \sqrt{\lambda}\right] \bigcap \mathbb{Z}_{0,+}; \label{first approximate pred int} \\ 
\Gamma_2(\lambda,\alpha) & = & \left[\left(\max\{0,\sqrt{\lambda} - (z_{\alpha/2})/2\}\right)^2,\left(\sqrt{\lambda} + (z_{\alpha/2})/2\right)^2\right]  \bigcap \mathbb{Z}_{0,+}. \label{second approximate pred int}
\end{eqnarray}
When $\lambda$ is large, as noted in the construction of $\Gamma_1$, we may approximate the Poisson probabilities by normal probabilities, via
\begin{displaymath}
p(k|\lambda) \approx \frac{1}{\sqrt{\lambda}} \phi\left(\frac{k-\lambda}{\sqrt{\lambda}}\right).
\end{displaymath}
As such, we obtain the approximation
\begin{displaymath}
A(\lambda;c(\alpha)) \approx \left[\lambda - z_{\alpha/2} \sqrt{\lambda}, \lambda + z_{\alpha/2} \sqrt{\lambda}\right] \bigcap \mathbb{Z}_{0,+}.
\end{displaymath}
Consequently, when $\lambda$ is large, the regions $\Gamma_0(U;\lambda,\alpha)$ and $\Gamma_1(\lambda,\alpha)$ should be close to each other.

\begin{figure}[h]
\caption{Exact properties of the four prediction regions $\Gamma_{0R}$, $\Gamma_{0N}$, $\Gamma_1$, and $\Gamma_2$ for $Y_0 \sim POI(\lambda)$ under the situation when the Poisson rate $\lambda$ is known.}
\label{fig-PlotExactProperties}
\includegraphics[width=\textwidth,height=2in]{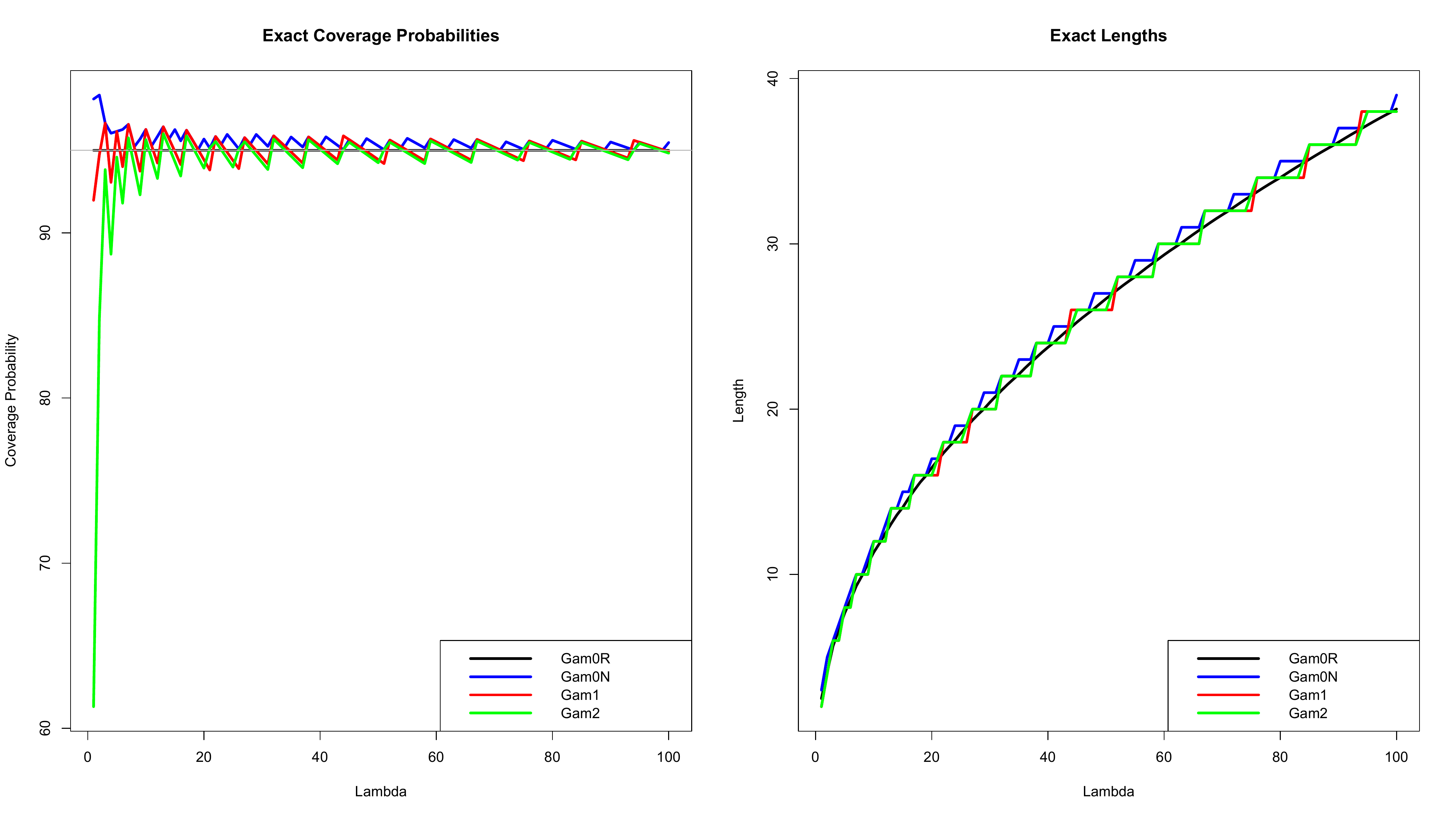}
\end{figure}

For these prediction regions $\Gamma_{0R}$ (randomized), $\Gamma_{0N}$ (nonrandomized), $\Gamma_1$, and $\Gamma_2$, the exact coverage probabilities and their exact lengths (mean length for $\Gamma_{0R}$) could be computed under the $POI(\lambda)$ distribution, since $\lambda$ is known. Note that the lengths, which are the differences between the upper and lower integer limits of the prediction regions, are equivalent surrogates of the cardinalities of the regions. Figure \ref{fig-PlotExactProperties} depicts the exact coverage probabilities (CP), expressed in percentages, and their lengths (expected length for $\Gamma_{0R}$), for different values of $\lambda$.
Except when $\lambda$ takes small values where the coverage probabilities of $\Gamma_1$ and $\Gamma_2$ are degraded, especially for the latter, the performance of these prediction regions are quite similar. The coverage probability of $\Gamma_{0R}$ is exactly equal to $1 - \alpha$, whereas that for $\Gamma_{0N}$ is always at least equal to $1 - \alpha$. Both $\Gamma_1$ and $\Gamma_2$ could have coverage probabilities that could be below the nominal coverage level, though as $\lambda$ increases, these differences become negligible. By construction, $\Gamma_{0R}$ has a shorter interval than $\Gamma_{0N}$; for some values of $\lambda$, the length of $\Gamma_{0R}$ exceeds that of $\Gamma_1$ and $\Gamma_2$, but this is because the coverage probabilities of $\Gamma_1$ and $\Gamma_2$ are lower than the nominal coverage level.

\subsection{Prediction Regions when Rate $\lambda$ is Unknown}

But, in the preceding developments, we have assumed that the rate parameter $\lambda$ is known, an unrealistic assumption. How do we deal with the situation when $\lambda$ is unknown? Suppose that we had observed a realization $\mathbf{y} = (y_1, y_2, \ldots, y_n)$ of a random sample $\mathbf{Y} = (Y_1, Y_2, \ldots, Y_n)$ from $POI(\lambda)$, so the components of $\mathbf{Y}$ are independent and identically distributed (\IID) from $POI(\lambda)$. Our goal is to utilize $\mathbf{y}$ to construct a $100(1-\alpha)\%$ prediction region for an unobserved $Y_0$, which is independent of $\mathbf{Y}$ and whose distribution is also $POI(\lambda)$. How will we achieve our goal? Note that through the Sufficiency Principle, we may reduce the problem by simply assuming that we had observed $t = \sum_{i=1}^n y_i$, the realization of the sufficient statistic for $\lambda$ given by $T = \sum_{i=1}^n Y_i$, which has a $POI(n\lambda)$ distribution. The reduced problem therefore is that we have $(T,Y_0)$ which are independent random variables with $T \sim POI(n\lambda)$ and $Y_0 \sim POI(\lambda)$ and our goal is to construct a $100(1-\alpha)\%$ prediction region $\tilde{\Gamma}(T,U;\alpha)$ for $Y_0$, which utilizes $T$, and possibly a randomizer $U$ which is independent of $(T,Y_0)$.

Given $T = t$, the maximum likelihood estimate (MLE) of $\lambda$ is $\hat{\lambda}(t) = t/n$. By virtue of the consistency of $\hat{\lambda}(T)$ for $\lambda$ as $n \rightarrow \infty$, a seemingly straight-forward approach to constructing a prediction region for $Y_0$ is to replace $\lambda$ in $\Gamma_0(U;\lambda,\alpha)$, $\Gamma_1(\lambda,\alpha)$, and $\Gamma_2(\lambda,\alpha)$ in (\ref{smallest cardinality prediction interval0}), (\ref{first approximate pred int}), and (\ref{second approximate pred int}), respectively, by $\hat{\lambda}(t)$ to obtain
\begin{eqnarray}
\tilde{\Gamma}_0(T,U;\alpha) & = & \Gamma_0(U;\hat{\lambda}(T),\alpha); \label{tilde Gamma0}\\
\tilde{\Gamma}_1(T;\alpha) & = & \Gamma_1(\hat{\lambda}(T),\alpha); \label{tilde Gamma1}
\\
\tilde{\Gamma}_2(T;\alpha) & = & \Gamma_2(\hat{\lambda}(T),\alpha) \label{tilde Gamma2}.
\end{eqnarray}
How do these prediction regions compare with each other in terms of performance, both in the context of their coverage probabilities and also their cardinalities, whose surrogate are lengths?
%
In particular, by substituting $\hat{\lambda}(T) = T/n$ for $\lambda$, how does this impact the coverage probabilities of these prediction regions and are they still valid, even in an asymptotic sense?

It is not clear how the substitution of $\lambda$ by $\hat{\lambda}(T) = T/n$ will impact the {\em exact} performance of the first prediction region $\tilde{\Gamma}_0$. However, for the second and third prediction regions $\tilde{\Gamma}_1$ and $\tilde{\Gamma}_2$, we could alter them to take into account the substitutions, provided that $\lambda$ is large. As noted earlier, when $\lambda$ is large, $\tilde{\Gamma}_0 \approx \tilde{\Gamma}_1$, so the alteration of $\tilde{\Gamma}_1$ should also apply, approximately, to $\tilde{\Gamma}_0$. The change in distributions of the pivotal quantities arising from these substitutions are reflected below, a consequence of the Delta-Method.
\begin{eqnarray*}
& \frac{Y_0 - \hat{\lambda}(T)}{\sqrt{\hat{\lambda}(T)}}  \stackrel{\bullet}{\sim}  N\left(0,1 + \frac{1}{n}\right) \quad \mbox{and} \quad
\frac{\sqrt{Y_0} - \sqrt{\hat{\lambda}(T)}}{\sqrt{1/4}}  \stackrel{\bullet}{\sim}  N\left(0,1 + \frac{1}{n}\right). &
\end{eqnarray*}
From these normal approximations, we could improve the prediction intervals $\tilde{\Gamma}_1$ and $\tilde{\Gamma}_2$ into the following prediction intervals, which take into account the impact of these substitutions where, for notational economy, we write $\hat{\lambda}$ for $\hat{\lambda}(T)$ and `$\vee$' for $\max$; $\lceil\rceil$ for the ceiling function; and $\lfloor\rfloor$ for the floor function:
{\small
\begin{eqnarray}
& \check{\Gamma}_1(n,T;\alpha)  =  \left[\ceil*{0 \vee \left\{\hat{\lambda} - z_{\alpha/2} \sqrt{\hat{\lambda}(1+1/n)}\right\}},\; 
\floor*{\hat{\lambda} + z_{\alpha/2} \sqrt{\hat{\lambda}(1+1/n)}}\right] \bigcap \mathbb{Z}_{0,+};  & \label{check Gamma1} \\ 
& \check{\Gamma}_2(n,T;\alpha)  =  \left[\ceil*{\left(0 \vee \left\{\sqrt{\hat{\lambda}} - (z_{\alpha/2})\sqrt{\frac{1}{4}\left(1 + \frac{1}{n}\right)}\right\}\right)^2},\; 
\floor*{\left(\sqrt{\hat{\lambda}} + (z_{\alpha/2})\sqrt{\frac{1}{4}\left(1 + \frac{1}{n}\right)}\right)^2}\right] \bigcap \mathbb{Z}_{0,+}. & \label{check Gamma2}
\end{eqnarray}
}
Note that by intersecting the intervals with $\mathbb{Z}_{0,+}$, the floor and ceiling functions are actually not needed, but we retain them in the formula since when we consider the `length', this pertains to the length of the interval. Observe that if the lower limits of these intervals are not zeros, which will usually be the case for large $\lambda$, then it is a simple exercise to show that these two prediction intervals have the same lengths, but they are not identical regions.

In trying to adapt the prediction region $\Gamma_0(U;\lambda,\alpha)$ in (\ref{smallest cardinality prediction interval0}) to the situation where $\lambda$ is unknown, the main idea is to replace $\lambda$ by an estimate obtained from the observed data. Doing so leads to estimates of $p(k|\lambda), k \in \mathbb{Z}_{0,+}$ which are then used in determining $c(\alpha)$ and $\gamma(\alpha)$ in (\ref{smallest cardinality prediction interval0}). Thus, $\tilde{\Gamma}_0$ in (\ref{tilde Gamma0}) is obtained by using the ML estimates of $\{p(k|\lambda), k=0,1,\ldots\}$ given by $\{p(k|\hat{\lambda}(n,t)),k=0,1,\ldots\}$ with $\hat{\lambda}(n,t) = t/n$. This begs the question on whether other possible estimates of $\{p(k|\lambda), k=0,1,\ldots\}$ could be utilized which may have better performances than the use of the ML estimates. An approach based on a second-order Taylor expansion adjusts $p(k|\hat{\lambda}(n,T))$ and leads to the approximation
\begin{equation}
\label{Taylor approx}
p(k|\lambda) \approx \hat{p}_3(k;(n,t)) \equiv \frac{p(k|\hat{\lambda}(n,t))}
{1 + \frac{1}{2} \left[\left(1 - \frac{k}{\hat{\lambda}}\right)^2 - \frac{k}{\hat{\lambda}^2}\right] \frac{\hat{\lambda}}{n}}, k \in \mathbb{Z}_{0,+}.
\end{equation}
By using $\hat{p}_3(k;(n,t))$ in place of $p(k|\lambda)$ in (\ref{smallest cardinality prediction interval0}) results in the prediction region denoted by $\check{\Gamma}_3(n,t;\alpha)$. Another intriguing possibility is to utilize the uniformly minimum variance unbiased estimator (UMVUE) (see \cite{CasBer90}) of $p(k|\lambda)$, given the data $(n,T)$ with $T \sim POI(n\lambda)$, or equivalently, $Y_1, Y_2, \ldots, Y_n$ which are \IID\ $POI(\lambda)$. The UMVUE of $p(k|\lambda)$, usually obtained through the Rao-Blackwell Theorem and Lehmann-Scheffe Theorem \cite{CasBer90}, is given by
\begin{equation}
\label{UMVUE of p(k)}
\hat{p}_4(k;(n,t)) = {t \choose k} \left(\frac{1}{n}\right)^k \left(1 - \frac{1}{n}\right)^{t-k} I\{k \in \{0,1,\ldots,t\}\},
\end{equation}
the binomial probability at $k$ with parameters $(t,1/n)$. Observe, however, that this approximation will lead to zero probabilities for $k$ outside of the set $\{0,1,\ldots,t\}$. Using $\hat{p}_4(k;(n,t))$ in lieu of $p(k|\lambda)$ in (\ref{smallest cardinality prediction interval0}) leads to the prediction region denoted by $\check{\Gamma}_4(n,t;\alpha)$.

As yet another idea is to develop a procedure by borrowing from the Bayesian playbook \cite{CasBer90}. We suppose that our prior knowledge of the value of the Poisson rate $\lambda$ is represented by a distribution function $G$. Having observed $\mathbf{Y} = \mathbf{y} = (y_1, y_2, \ldots, y_n)$, the posterior distribution of $\lambda$ is given by
\begin{equation}
\label{posterior of lambda}
G(\lambda|\mathbf{y}) = \frac{\int_0^\lambda \exp\{-nw\} w^t G(dw)}{\int_0^\infty \exp\{-nw\} w^t G(dw)}
\end{equation}
with $t = t(\mathbf{y}) = \sum_{i=1}^n y_i$. The conditional probability mass function of $Y_0$, given $\mathbf{Y} = \mathbf{y}$, also called the posterior predictive PMF, is
\begin{equation}
\label{predictive pmf of Y0}
p(y_0|\mathbf{y};G) = \frac{1}{y_0!} \frac{\int_0^\infty \exp\{-(n+1)w\} w^{t+y_0} G(dw)}{\int_0^\infty \exp\{-nw\} w^t G(dw)}.
\end{equation}
If we are pure Bayesians, then we will completely know, or trust, our $G$, so we could use the predictive PMF $p(\cdot|\mathbf{y};G)$ in lieu of the Poisson PMF in (\ref{smallest cardinality prediction interval0}) to form a Bayesian prediction region for $Y_0$.  Usually, however, we may try to estimate $G$ by a $\hat{G}(\cdot;\mathbf{y})$ based on $\mathbf{y}$. This brings us to the realm of the Empirical Bayes (EB) approach, pioneered by Herbert Robbins; see \cite{Rob56,Rob80,Rob83}. An extreme case is to `estimate' $G$ by a {\em degenerate} distribution at the ML estimate $\hat{\lambda} = t/n$, which leads to just substituting $\hat{\lambda}$ in the Poisson PMF, hence results in the prediction region $\tilde{\Gamma}_0$ in (\ref{tilde Gamma0}). Another possibility is to try to estimate $G$ non-parametrically. However, here we implement this Bayesian and EB approaches using a family of conjugate priors, so we assume $G$ is a gamma distribution with mean $\kappa/\beta$ and variance $\kappa/\beta^2$, denoted by $\mathfrak{G}_{\kappa,\beta}$, whose density function is 
\begin{displaymath}
g(\lambda|\kappa,\beta) = \frac{\beta^\kappa}{\Gamma(\kappa)} \lambda^{\kappa-1} \exp\{-\beta\lambda\} I\{\lambda > 0\}
\end{displaymath}
where $\kappa > 0$ and $\beta > 0$. Note that $\kappa$ is the shape parameter and $\beta$ is the scale parameter. Under $G = \mathfrak{G}_{\kappa,\beta}$, simplifying (\ref{predictive pmf of Y0}) we obtain, for $y_0 \in \mathbb{Z}_{0,+}$,
\begin{equation}
\label{predictive pmf of Y0 under gamma}
p(y_0|\mathbf{y};\mathfrak{G}_{\kappa,\beta}) = \frac{\Gamma(\kappa + t + y_0)}{\Gamma(y_0+1) \Gamma(\kappa + t)} \left(\frac{\beta+n}{\beta+n+1}\right)^{t+\kappa} \left(\frac{1}{\beta+n+1}\right)^{y_0}.
\end{equation}
When $\kappa$ is a positive integer, the PMF in (\ref{predictive pmf of Y0 under gamma}) corresponds to a negative binomial distribution with parameters $\kappa+t$ and $(\beta+n)/(\beta+n+1)$. The PMF in (\ref{predictive pmf of Y0 under gamma}) could be used in place of the Poisson PMF in (\ref{smallest cardinality prediction interval0}) to form a Bayesian prediction region for $Y_0$, given $(\kappa,\beta)$, denoted by $\check{\Gamma}_5(U,\mathbf{Y};(\kappa,\beta))$. An approach to specifying $(\kappa,\beta)$ is to specify a prior mean and prior standard deviation for $\lambda$, say $M$ and $S$, respectively, which yield 
%
$\kappa = M^2/S^2$ and $\beta = M/S^2$.
%
The EB approach estimates $\kappa$ and $\beta$ from the data $\mathbf{y} = (y_1, y_2, \ldots, y_n)$. Such estimation could be done via maximum likelihood using the likelihood function obtained from the joint marginal distribution of $(Y_1, Y_2, \ldots, Y_n)$ based on the model $Y_i | \lambda_i \sim POI(\lambda_i)$ and $\lambda_i \sim \mathfrak{G}_{\kappa,\beta}$. This likelihood function is given by
\begin{equation}
\label{marginal likelihood of ys}
L(\kappa,\beta|\mathbf{y}) = \frac{\prod_{i=1}^n \Gamma(\kappa + y_i)}{[\Gamma(\kappa)]^n \prod_{i=1}^n \Gamma(y_i+1)} \left(\frac{\beta}{\beta+1}\right)^{n\kappa} \left(\frac{1}{\beta+1}\right)^{\sum_{i=1}^n y_i}.
\end{equation}
A method-of-moments approach to estimating $(\kappa,\beta)$ based on $\mathbf{y}$ fails, however, because  negative estimates of $\kappa$ and $\beta$ are obtained when the sample variance of $\mathbf{y}$ is smaller than its sample mean.

At this point we mention previous works dealing with prediction intervals under the Poisson model. Prediction interval methods for the Poisson model have been incorporated in the {\tt R} package {\em envStats} 
\cite{Millard13}. An object function in this package is {\tt predIntPois} dealing with construction of prediction intervals under the Poisson model. It provides four options for the type of prediction interval to construct. The methods are based on procedures presented in \cite{CoxHin74,GibBhaAry09,Nelson82}. The option {\tt normal.approx} in {\tt predIntPois} coincides with the prediction region $\tilde{\Gamma}_1$ based on the normal approximation. In these earlier procedures, randomization was not utilized, hence generally conservative prediction intervals are obtained. Other approaches for prediction interval construction under the Poisson model, including Poisson regression models, are based on bootstrapping and simulation techniques, hence are computationally-intensive \cite{RblogPoiPI}.

\begin{figure}
\caption{Simulated coverage probabilities and mean lengths of the prediction intervals $\tilde{\Gamma}_0$ and $\check{\Gamma}_j, j=1,2,3,4,5$, plotted with respect to $n$ for values of  $\lambda \in \{1, 5, 30, 100\}$. The number of replications for each combination of $(n,\lambda)$ was 10000.}
\label{figs-simulation 1} 
\includegraphics[width=\textwidth,height=1.75in]{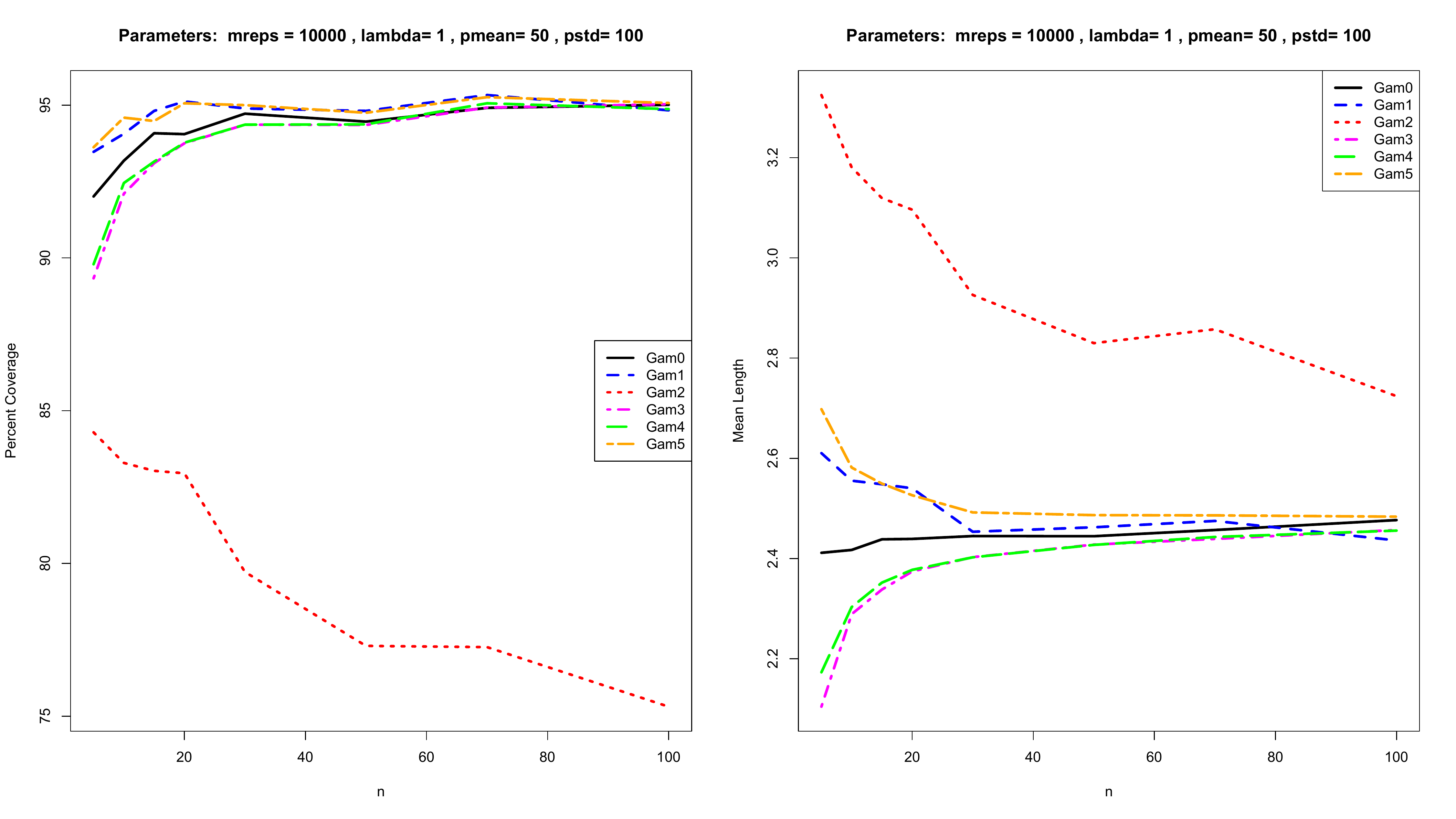} \\
\includegraphics[width=\textwidth,height=1.75in]{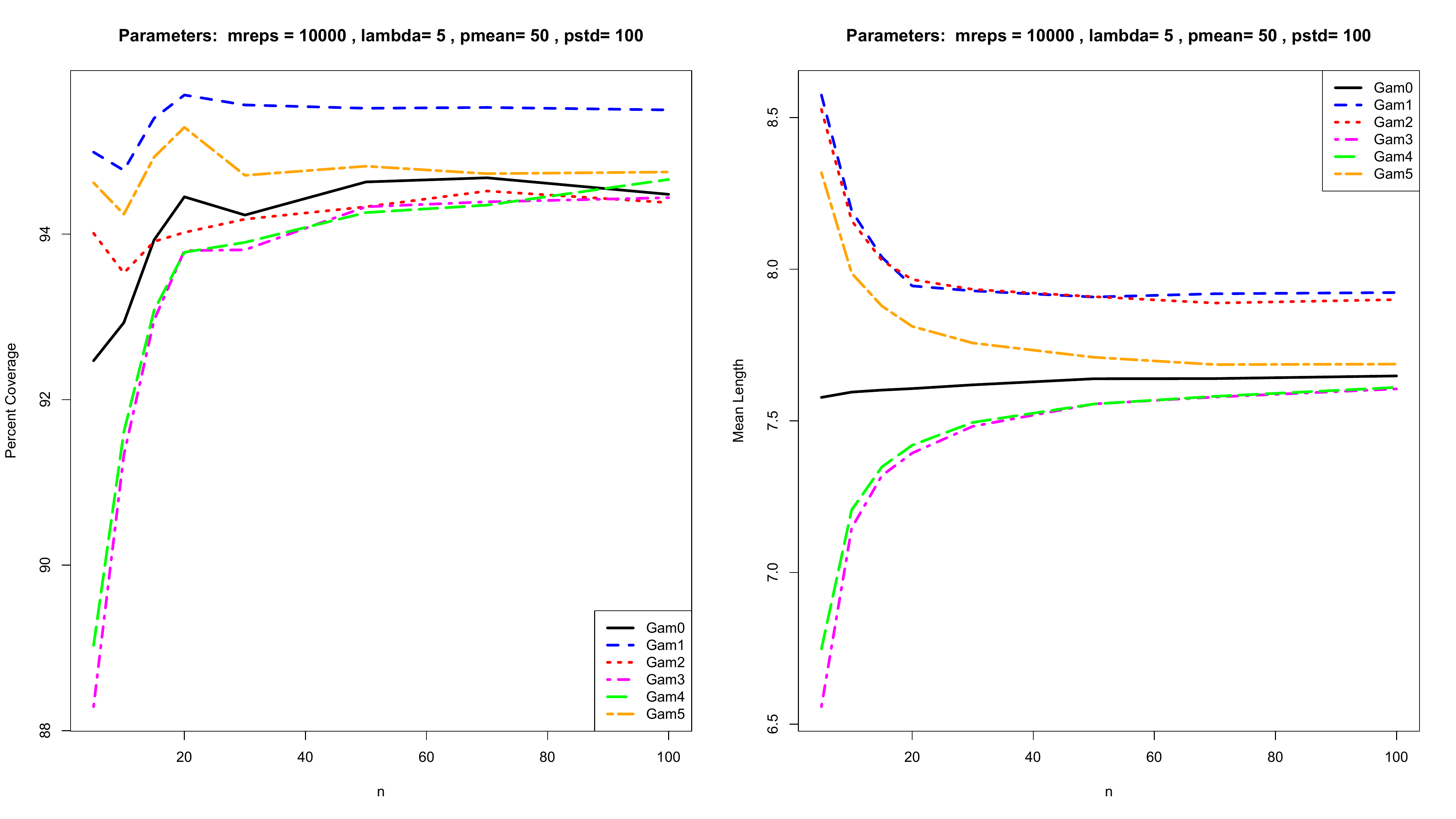} \\
%
\includegraphics[width=\textwidth,height=1.75in]{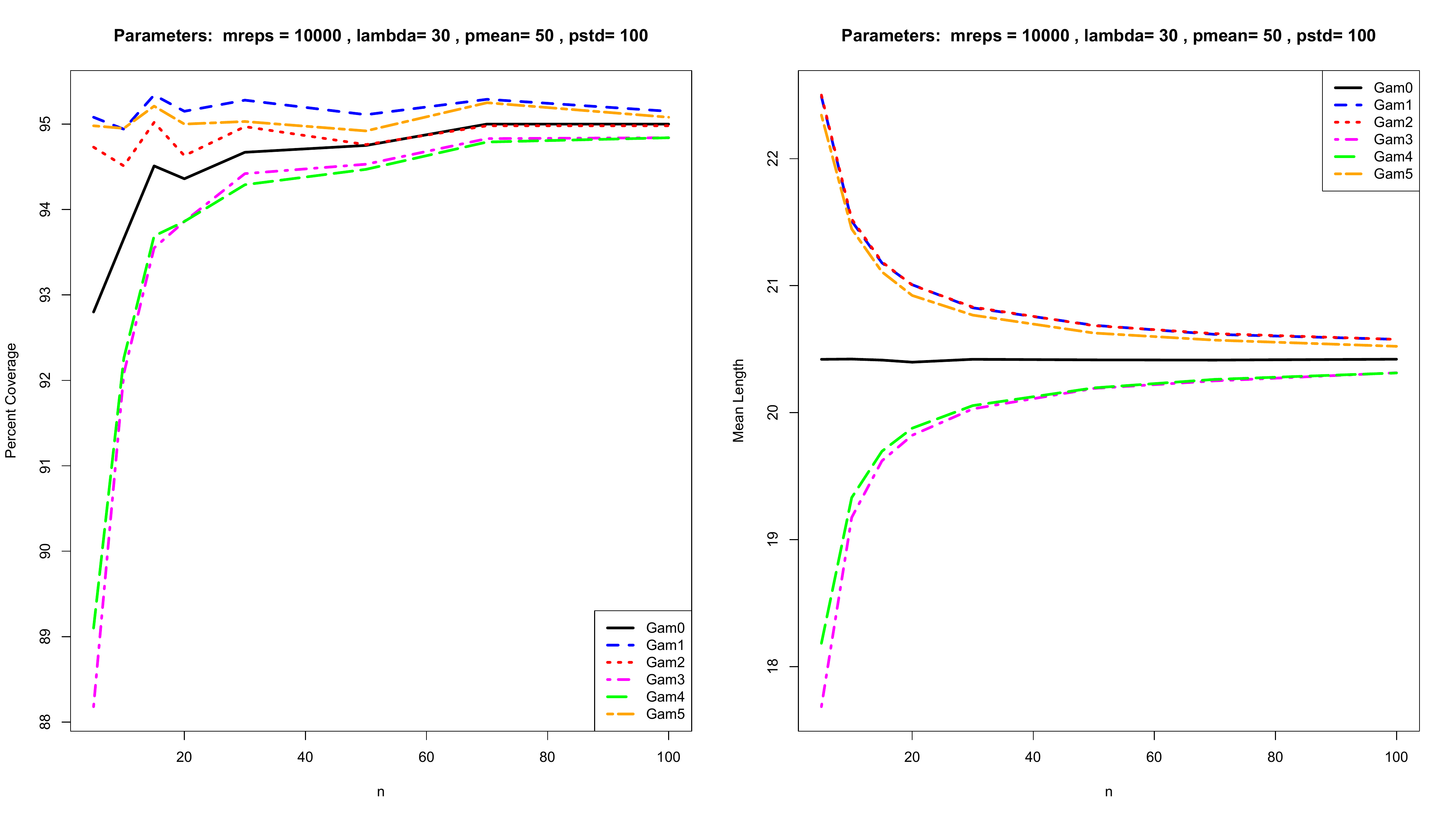} \\
\includegraphics[width=\textwidth,height=1.75in]{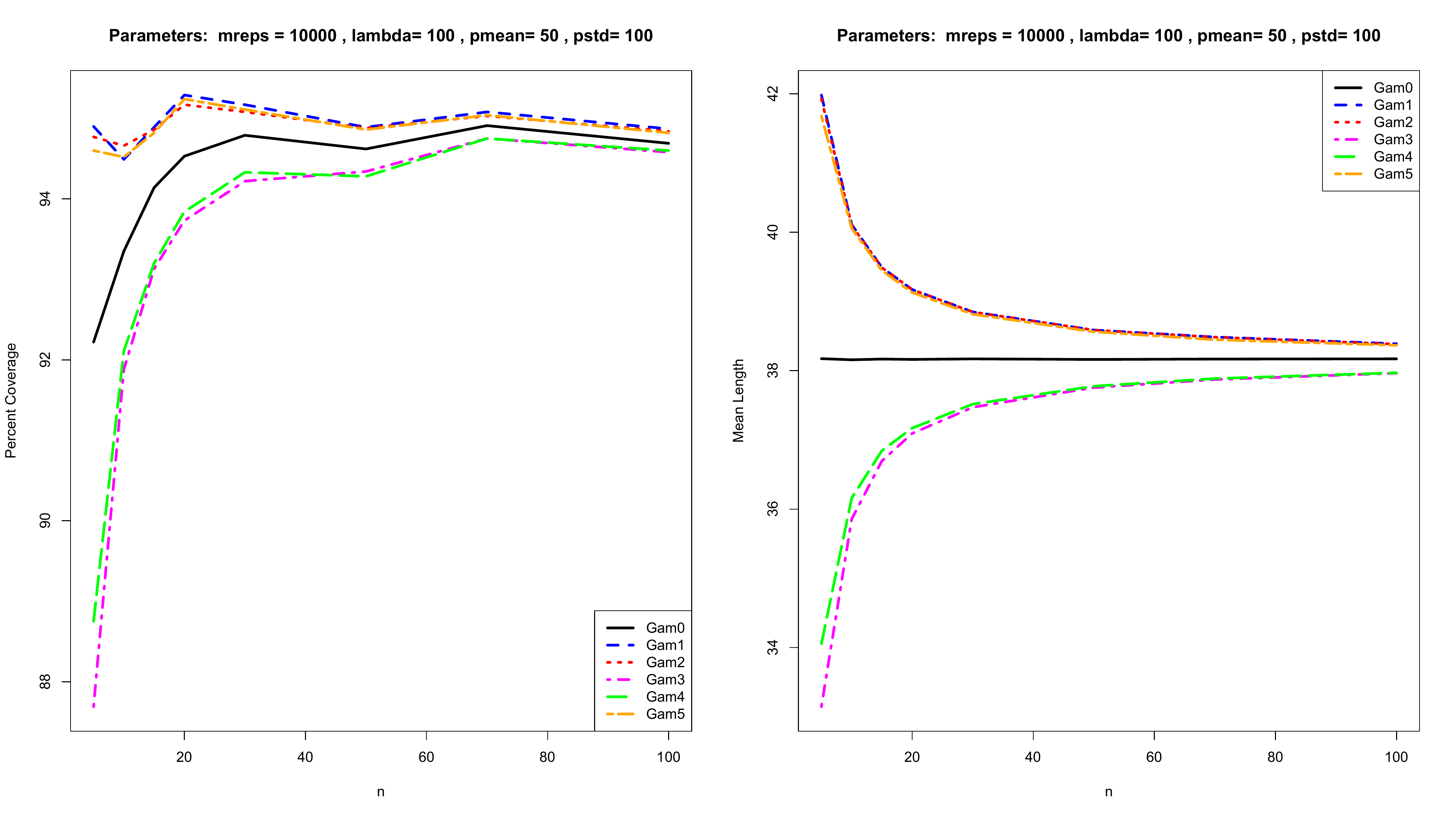}
\end{figure}

To compare performance of the prediction regions $\tilde{\Gamma}_0$ (randomized version), $\check{\Gamma}_1$, $\check{\Gamma}_2$, $\check{\Gamma}_3$, $\check{\Gamma}_4$, and $\check{\Gamma}_5$ with $M = 50, S = 100$, and for $n \in \{5,10,15,20,30,50,70,100\}$ and $\lambda \in \{1,5, 15, 30, 50,100, 200\}$, we performed simulation studies, with program codes in the {\tt R} \cite{R} environment, to determine the coverage probabilities and the lengths of the regions (recall that length is an equivalent surrogate for the cardinality of the regions since we took the ceiling and the floor of the lower and upper limits, respectively, for the intervals that leads to $\check{\Gamma}_1$ and $\check{\Gamma}_2$). For each combination of $n$ and $\lambda$, 10000 simulation replications of the basic simulation experiment were performed. Table \ref{table-simulation 1 results} in the Appendix presents the results on the coverage percentages, mean lengths of the prediction intervals, and standard deviations of the lengths for different values of $\lambda$. The basic simulation experiment is, for a fixed $n$ and $\lambda$, to generate $T \sim POI(n\lambda)$ and $Y_0 \sim POI(\lambda)$. The $T$ variable could be viewed as $T = \sum_{i=1}^n Y_i$ where $Y_1, Y_2, \ldots, Y_n$ are \IID\ from a $POI(\lambda)$, The prediction regions are then constructed based on the observed $(n,T)$, with prediction coefficient of 95\%. Note that since $\check{\Gamma}_5$ is the Bayes prediction region instead of the EB, we only needed the value of $T = \sum_{i=1}^n Y_i$, but if we also used the EB approach, then we would have needed the values of $(y_1,y_2,\ldots,y_n)$ to estimate $(\kappa,\beta)$. After constructing the prediction regions, it is then determined if $Y_0$ is contained in these regions. Coverage percentage is the percentage out of the 10000 prediction regions that contain the $Y_0$; mean (standard deviation) length is the average (standard deviation) of the lengths of the 10000 prediction intervals. Figure \ref{figs-simulation 1} presents plots with respect to $n$ of the coverage probabilities (CP) and mean lengths (ML) for $\lambda \in \{1, 5, 30, 100\}$.

Examining Table \ref{table-simulation 1 results} and Figure \ref{figs-simulation 1} we observe that when $\lambda = 1$, the CP of $\check{\Gamma}_2$ is very poor and {\em even deteriorates} as $n$ increases. The reason for this is that the realized $Y_0$ tends to equal 0, but the square-root transformation has a tendency to shift to the right the prediction interval, hence the interval tends to miss $Y_0$. This result for $\check{\Gamma}_2$ is consistent with the result when the rate $\lambda$ is known. When $n = 5$, $\check{\Gamma}_3$ and $\check{\Gamma}_4$ have unacceptably lower CPs compared to the nominal level, while $\tilde{\Gamma}_0$ also has CPs which are below the nominal level, as well as $\check{\Gamma}_1$ and $\check{\Gamma}_5$, though the last two regions have CPs closer to the desired level. The length of $\tilde{\Gamma}_0$ tends to be shorter than $\check{\Gamma}_1$ and $\check{\Gamma}_5$. As $n$ increases, the CPs of $\check{\Gamma}_3$ and $\check{\Gamma}_4$ get closer to the desired level, and their lengths tend to be a bit shorter than $\tilde{\Gamma}_0$ and $\check{\Gamma}_1$. When $\lambda = 5$, the CPs of $\tilde{\Gamma}_0$, $\check{\Gamma}_2$, $\check{\Gamma}_3$, and $\check{\Gamma}_4$ are all below the nominal level, whereas for $\check{\Gamma}_1$ and $\check{\Gamma}_5$, their CPs exceed or are quite close to the nominal level, except when $n = 5$. As a consequence, they ended up having longer mean lengths. These behaviors continue to hold as $\lambda$ was increased, but with the CPs getting closer to the nominal level, especially as $n$ increases. When $n$ is small, the CPs of $\check{\Gamma}_3$ and $\check{\Gamma}_4$ are still appreciably lower than the nominal level. When $\lambda$ is large, $\check{\Gamma}_1$, $\check{\Gamma}_2$, and $\check{\Gamma}_5$ almost have the same performance. Summing up our observations from these simulation studies for this no-covariate or  intercept only Poisson model, in terms of adapting to the estimation of the unknown rate $\lambda$, $\check{\Gamma}_1$ and $\check{\Gamma}_5$ possess the best performance among these six prediction regions in terms of achieving the nominal level, but they also tend to be longer than the others.

\section{Prediction Regions under Poisson Regression}
\label{sec-Poisson Regression}

\subsection{With Known Parameter Vector}

Next, we consider the situation where there is a $1 \times (p+1)$ covariate vector $\mathbf{x} = (x_0,x_1,\ldots,x_{p})$ that could affect the rate parameter, where in our implementation we take $x_0 = 1$, so the model has an intercept component. Thus, we suppose that the rate parameter is $\lambda(\mathbf{x};\mathbf{\theta})$, where $\mathbf{\theta} = (\theta_0,\theta_1,\ldots,\theta_{p})\trp$ is a $(p+1) \times 1$ vector of parameters. We further assume that there exists a non-negative continuously-differentiable function $\rho(\cdot)$, called the inverse link function, such that
\begin{equation}
\label{assumption on rho}
\lambda(\mathbf{x};\mathbf{\theta}) = \rho(\mathbf{x}\mathbf{\theta}) = \rho\left(\sum_{j=0}^p x_j \theta_j\right).
\end{equation}
This is the so-called Poisson regression model and belongs to the class of generalized linear models or the class of non-linear models \cite{RenSch08,CasBer90}.  
If $Y_0$, given $\mathbf{x} = \mathbf{x}_0$, has a Poisson distribution with rate $\lambda(\mathbf{x}_0;\theta)$, and $\theta$ is {\em known}, then we could construct prediction regions for $Y_0$ according to the methods described in the first part of Section \ref{sec-Poisson Model} when $\lambda$ was assumed known.

\subsection{With Unknown Parameter Vector}

When $\theta$ is not known, then there is a need to estimate it. Let us therefore assume that we are able to observe the sample $\{(Y_i,\mathbf{x}_i), i=1,2,\ldots,n\}$ with $Y_i | \mathbf{x}_i \sim POI(\lambda(\mathbf{x}_i;\theta))$ and with the $Y_i$s independent and the $\mathbf{x}_i$s fixed. We seek to construct a prediction region for $Y_0$ associated with the covariate vector $\mathbf{x}_0$. First, we introduce the following functions:
\begin{eqnarray*}
& \stackrel{\bullet}{\rho}(w) = \frac{d}{dw} \rho(w)\quad \mbox{and} \quad \stackrel{\bullet\bullet}{\rho}(w) = \frac{d^2}{dw^2} \rho(w); & \\ & \psi(w) = \frac{d}{dw} \log \rho(w) = \frac{\stackrel{\bullet}{\rho}(w)}{\rho(w)} \quad \mbox{and} \quad \Psi(w) = \frac{d^2}{dw^2} \log \rho(w) = \frac{\stackrel{\bullet\bullet}{\rho}(w)}{\rho(w)} - \psi(w)^2. &
\end{eqnarray*}
The log-likelihood function for $\mathbf{\theta}$, given $\{(y_i,\mathbf{x}_i), i=1,2,\ldots,n\}$, is given by
\begin{displaymath}
l(\mathbf{\theta}) = -\sum_{i=1}^n \rho(\mathbf{x}_i\mathbf{\theta}) + \sum_{i=1}^n y_i \log \rho(\mathbf{x}_i\mathbf{\theta}).
\end{displaymath}
The associated score vector function is
\begin{equation}
\label{score function}
\mathbf{U}(\mathbf{\theta}) = \nabla_\mathbf{\theta} l(\mathbf{\theta}) = \sum_{i=1}^n [y_i - \rho(\mathbf{x}_i\mathbf{\theta})] \mathbf{x}_i\trp \psi(\mathbf{x}_i\mathbf{\theta});
\end{equation}
whereas, the {\em observed} Fisher information matrix is, with $\mathbf{x}^{\otimes 2} = \mathbf{x}\trp\mathbf{x}$,
\begin{equation}
\label{observed Fisher information}
\mathbf{I}(\mathbf{\theta}) = -\nabla_{\mathbf{\theta}\trp} \nabla_\mathbf{\theta} l(\mathbf{\theta}) = \sum_{i=1}^n \mathbf{x}_i^{\otimes 2} \stackrel{\bullet\bullet}{\rho}(\mathbf{x}_i\mathbf{\theta}) - \sum_{i=1}^n y_i \mathbf{x}_i^{\otimes 2} \Psi(\mathbf{x}_i\mathbf{\theta}).
\end{equation}
Thus, the {\em expected} Fisher information matrix is
\begin{equation}
\label{expected Fisher information}
\mathfrak{I}(\mathbf{\theta}) = \sum_{i=1}^n \mathbf{x}_i^{\otimes 2} [\psi(\mathbf{x}_i\mathbf{\theta})]^2 \rho(\mathbf{x}_i\mathbf{\theta}).
\end{equation}
The MLE of $\mathbf{\theta}$ based on $\{(Y_i,\mathbf{x}_i), i=1,2,\ldots,n\}$, denoted by $\hat{\mathbf{\theta}}$, solves the equation
$$\mathbf{U}(\mathbf{\theta}) = \sum_{i=1}^n [y_i - \rho(\mathbf{x}_i\mathbf{\theta})] \mathbf{x}_i\trp \psi(\mathbf{x}_i\mathbf{\theta}) = \mathbf{0}.$$
This will usually be obtained through iterative procedures, such as the iterative Newton-Raphson method, with the iteration given by
\begin{equation}
\label{NR Iteration}
\mathbf{\theta} \leftarrow \mathbf{\theta} + [I(\mathbf{\theta})]^{-1} U(\mathbf{\theta}).
\end{equation}
By the large-sample theory of ML estimation (cf., \cite{CasBer90,vanderVaart98}), as $n \rightarrow \infty$ and under regularity conditions on the sequence of covariate vectors $\mathbf{x}_i, i=1,2,\ldots,n$, we have that
\begin{displaymath}
\hat{\mathbf{\theta}} \sim AN\left[\mathbf{\theta}, \mathfrak{I}(\mathbf{\theta})^{-1}\right].
\end{displaymath}
A consistent estimator of $\mathfrak{I}(\mathbf{\theta})$ is $\mathbf{I}(\hat{\mathbf{\theta}})$. By the Delta-Method, it then follows that the ML estimator of $\lambda(\mathbf{x}_0;\mathbf{\theta})$ satisfies, as $n \rightarrow \infty$,
\begin{displaymath}
\widehat{\lambda(\mathbf{x}_0)} = \lambda(\mathbf{x}_0;\hat{\mathbf{\theta}}) = \rho(\mathbf{x}_0\hat{\mathbf{\theta}}) \sim
AN\left(\rho(\mathbf{x}_0\mathbf{\theta}), [\psi(\mathbf{x}_0\mathbf{\theta})]^2 [\rho(\mathbf{x}_0\mathbf{\theta})]^2 \tr\left[\mathfrak{I}(\mathbf{\theta})^{-1} \mathbf{x}_0^{\otimes 2}\right]\right),
\end{displaymath}
where `$\tr$' means trace of a matrix.

Using this result and when $\lambda_0 = \rho(\mathbf{x}_0\mathbf{\theta})$ is large, we obtain the approximate distributions of relevant pivotal quantities for constructing prediction regions. We write $\hat{\lambda}_0$ for $\lambda(\mathbf{x}_0;\hat{\mathbf{\theta}})$ and $\hat{\psi}_0$ for $\psi(\mathbf{x}_0\hat{\mathbf{\theta}})$. These pivotal quantities are:
\begin{eqnarray*}
\frac{Y_0 - \hat{\lambda}_0}{\sqrt{\hat{\lambda}_0}} & \stackrel{\bullet}{\sim} & N\left[0,\hat{V} \equiv 1 + \hat{\psi}_0^2 \hat{\lambda}_0 \tr\left([\mathbf{I}(\hat{\theta})]^{-1} \mathbf{x}_0^{\otimes 2}\right)\right]; \\
\frac{\sqrt{Y_0} - \sqrt{\hat{\lambda}_0}}{\sqrt{1/4}} & \stackrel{\bullet}{\sim} & N\left[0,\hat{V} \equiv 1 + \hat{\psi}_0^2 \hat{\lambda}_0 \tr\left([\mathbf{I}(\hat{\theta})]^{-1} \mathbf{x}_0^{\otimes 2}\right)\right].
\end{eqnarray*}
Writing $(\mathbf{Y},\mathbf{X}) = \{(Y_i,\mathbf{x}_i), i=1,2,\ldots,n\}$, from these pivotal quantities, we are then able to obtain approximate prediction regions for $Y_0$ given by:
\begin{equation}
\label{check Gamma1 cov}
\check{\Gamma}_1(\mathbf{x}_0,(\mathbf{Y},\mathbf{X});\alpha) = \left[
0 \vee \left(\hat{\lambda}_0 -  z_{\alpha/2} \sqrt{\hat{\lambda}_0} \sqrt{\hat{V}}\right),
\hat{\lambda}_0 + z_{\alpha/2} \sqrt{\hat{\lambda}_0} \sqrt{\hat{V}}\right] \bigcap \mathbb{Z}_{0,+};
\end{equation}
and
{\small
\begin{eqnarray}
\check{\Gamma}_2(\mathbf{x}_0,(\mathbf{Y},\mathbf{X});\alpha) & = & \left[
\left\{0 \vee \left(\sqrt{\hat{\lambda}_0} -  z_{\alpha/2} \left(\frac{1}{2}\right)\sqrt{\hat{V}}\right)\right\}^2,
\left\{\sqrt{\hat{\lambda}_0} +  z_{\alpha/2} \left(\frac{1}{2}\right) \sqrt{\hat{V}}\right\}^2\right]  \bigcap \mathbb{Z}_{0,+}.
\label{check Gamma2 cov}
\end{eqnarray}
}
We could also have the prediction region based on $\Gamma_0$ from Section \ref{sec-Poisson Model} given by
\begin{equation}
\label{tilde Gamma0 cov}
\tilde{\Gamma}_0(U,\mathbf{x}_0,(\mathbf{Y},\mathbf{X});\alpha) = \Gamma_0(U;\hat{\lambda}_0,\alpha),
\end{equation}
where we note that the dependence on $\mathbf{x}_0$ and $(\mathbf{Y},\mathbf{X})$ is through $\hat{\lambda}_0$. Note, however, that we are simply plugging in the estimate of $\lambda(\mathbf{x}_0;\mathbf{\theta})$, but {\em without} taking into consideration the variability inherent in the estimator $\lambda(\mathbf{x}_0;\hat{\mathbf{\theta}})$.

A specific inverse link function $\rho(\cdot)$, which we will consider in the application to forecasting deaths in the US due to \Covid, is the exponential function
%
$\rho(w) = \exp(w)$,
%
so that 
\begin{displaymath}
\rho(w) =\ \stackrel{\bullet}{\rho}(w) =\ \stackrel{\bullet\bullet}{\rho}(w) = \exp(w);\quad \psi(w) = 1; \quad \mbox{and}\quad \Psi(w) = 0.
\end{displaymath}
For this special inverse link function, we obtain the simplifications for the score vector and information matrices functions given by
\begin{equation}
\label{score and info exp link}
\mathbf{U}(\mathbf{\theta}) = \sum_{i=1}^n \mathbf{x}_i\trp [y_i - \rho(\mathbf{x}_i\mathbf{\theta})] \quad \mbox{and} \quad \mathbf{I}(\mathbf{\theta}) = \mathfrak{I}(\mathbf{\theta}) = \sum_{i=1}^n \mathbf{x}_i^{\otimes 2} \rho(\mathbf{x}_i\mathbf{\theta}).
\end{equation}

We also mention the extension of the Bayesian/EB approaches to constructing prediction regions in the regression setting. We suppose that the parameter $\mathbf{\theta}$ in $\lambda(\mathbf{x};\mathbf{\theta}) = \rho(\mathbf{x}\mathbf{\theta})$ takes values in a parameter space $\Theta$. The approach then proceeds by starting with a prior distribution $\Pi(\cdot)$ on $\Theta$ which quantifies our prior knowledge about $\mathbf{\theta}$. The posterior predictive distribution of $Y_0$, the response at $\mathbf{x}_0$, given the data $(\mathbf{Y},\mathbf{X}) = \{(y_i,\mathbf{x}_i), i=1,2,\ldots,n\}$, is given by
\begin{equation}
\label{posterior predictive in regression}
p(y_0|\mathbf{x}_0, (\mathbf{Y},\mathbf{X})) = \frac{1}{y_0!} \frac{H[(\mathbf{Y}_0,\mathbf{X}_0)]}{H[(\mathbf{Y},\mathbf{X})]}
\end{equation}
where $(\mathbf{Y}_0,\mathbf{X}_0) = \{(Y_i,\mathbf{x}_i), i = 0,1,2,\ldots,n\}$ and
\begin{displaymath}
H[(\mathbf{Y},\mathbf{X})] = \int_\Theta \exp\left\{-\sum \rho(\mathbf{x}_i\mathbf{\theta})\right\} \left[\prod \rho(\mathbf{x}_i\mathbf{\theta})^{y_i}\right] \Pi(d\mathbf{\theta}),
\end{displaymath}
where the product and the sum are taken over the index set associated with $(\mathbf{Y},\mathbf{X})$, so will be over $\{1,2,\ldots,n\}$ for $(\mathbf{Y},\mathbf{X})$, and $\{0,1,2,\ldots,n\}$ for $(\mathbf{Y}_0,\mathbf{X}_0)$. Generally, there will be no family of conjugate prior distributions on $\Theta$ with respect to the Poisson regression model, so the function $H$ will not be in a closed analytical form, so that it has to be computed numerically, for instance, using Markov Chain Monte Carlo (MCMC) algorithms. Nevertheless, upon obtaining the posterior predictive distribution of $Y_0$ given in (\ref{posterior predictive in regression}), a prediction region is then obtained by using this PMF $p(y_0|\mathbf{x}_0, (\mathbf{Y},\mathbf{X}))$ in lieu of the Poisson PMF in (\ref{smallest cardinality prediction interval0}), analogously to the development of the prediction region $\check{\Gamma}_5$ in the intercept only model. The prior distribution $\Pi$ will involve hyper-parameters, for example, if $\Theta = \Re_{p+1}$, $\Pi$ could be specified to be a multivariate normal distribution with mean vector $\mathbf{\mu}$ and covariance matrix $\mathbf{\Sigma}$, so $(\mathbf{\mu},\mathbf{\Sigma})$ will be the hyper-parameters. For the Bayesian, these hyper-parameters will be assigned values, unless an improper prior distribution (e.g., Lebesgue measure), which need not involve unknown hyper-parameters, is adopted; whereas, for the empirical Bayesian, these hyper-parameters will be estimated using the data $(\mathbf{Y},\mathbf{X})$. Because of the need to approximate the posterior predictive PMF through numerical methods, these Bayesian and EB approaches to constructing a prediction region for $Y_0$ are clearly computationally-intensive, especially if used in a simulation study to investigate their properties, such as their coverage probabilities and their lengths. Because of the need to specify a non-conjugate prior and its hyper-parameters and the need for intensive computations, these Bayesian and EB procedures are not included in the illustrations, simulations, and applications. It is clear, though, that they are highly viable alternative procedures and should be further explored.

\subsection{Illustrations of Prediction  Regions}

We demonstrate these prediction regions, depicted as intervals in the plots, via the following experiment. We specify a sample size $n$ and an order $p$. We then generate \IID\ realizations $w_i, i=1,2,\ldots,n,n+1,$ from either a $N(\mu,\sigma^2)$ distribution or a standard uniform distribution, and form the covariate vectors $\mathbf{x}_i = (1, w_i, w_i^2, \ldots, w_i^p), i=1,2,\ldots,n, n+1$. For a specified $\mathbf{\theta} = (\theta_0, \theta_1, \ldots, \theta_p)\trp$, the Poisson rates
\begin{displaymath}
\lambda_i = \lambda(\mathbf{x}_i;\mathbf{\theta}) = \rho(\mathbf{x}_i\mathbf{\theta}) = \exp\{\mathbf{x}_i\mathbf{\theta}\}, i=1,2,\ldots,n,n+1.
\end{displaymath}
are computed. The $i$th response $y_i$ is a realization of a random draw from a $POI(\lambda_i)$. The response vector is $\mathbf{y} = (y_1,y_2,\ldots,y_n,y_{n+1})\trp$. The goal is to construct a prediction region for $Y_0 \equiv Y_{n+1}$, given the data $\{(y_i,\mathbf{x}_i), i=1,2,\ldots,n\}$ and $\mathbf{x}_0 \equiv \mathbf{x}_{n+1}$. We construct 95\% prediction intervals $\tilde{\Gamma}_0(U,\mathbf{x}_0,(\mathbf{Y},\mathbf{X});.05)$, $\check{\Gamma}_1(\mathbf{x}_0,(\mathbf{Y},\mathbf{X});.05)$, and $\check{\Gamma}_2(\mathbf{x}_0,(\mathbf{Y},\mathbf{X});.05)$  in (\ref{tilde Gamma0 cov}),  (\ref{check Gamma1 cov}), and (\ref{check Gamma2 cov}), respectively. The procedures were coded into {\tt R} functions and these will be made available publicly in due time. We present the results pictorially via a scatterplot of $\{(w_i,y_i), i=1,2,\ldots,n,n+1\}$. The realized value $y_0 \equiv y_{n+1}$ of $Y_0$ is highlighted and the three prediction intervals for $Y_0$ are also plotted. Included in the plot is the theoretical curve for the $\lambda(\mathbf{x};\mathbf{\theta})$ as a function of $w$ and we also super-impose the fitted curve. Prediction regions for $Y_i$ at $w_i$, $i=1,2,\ldots,n$, are also depicted in the plot. 

\begin{figure}[h]
\caption{A realization with  $p=1, n=30, \theta = (2,3)$, and with $W \sim U[0,1]$. The resulting prediction regions were $\tilde{\Gamma}_0 = [1091,1223], \check{\Gamma}_1 = [1087,1226], \check{\Gamma}_2 = [1088,1227]$. The realized $Y_0$ was $y_0 = 1222$ associated with $w_0 = .8139$.}
\label{Case1}
\includegraphics[width=\textwidth]{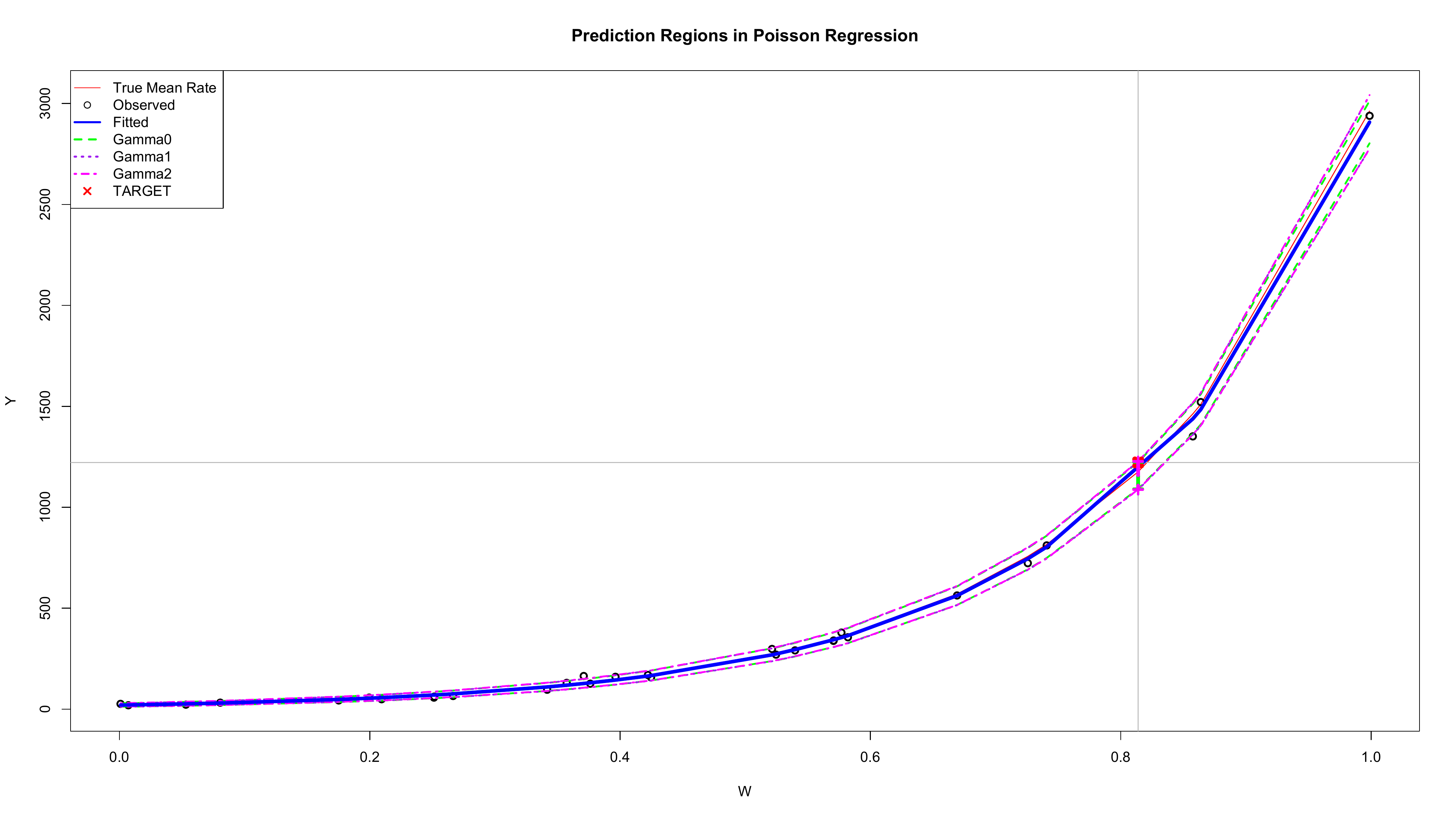}
\end{figure}

\begin{figure}[h]
\caption{A realization with  $p=2, n=100, \theta = (.3,-.2,.05)$, and with $W \sim N(\mu=2, \sigma=2)$.  The resulting prediction regions were $\tilde{\Gamma}_0 =[0,3], \check{\Gamma}_1 = [0,3], \check{\Gamma}_2 = [1,4]$.  The realized $Y_0$ was $y_0 = 1$ associated with $w_0 = 5.0622$.}
\label{Case3}
\includegraphics[width=\textwidth,height=4in]{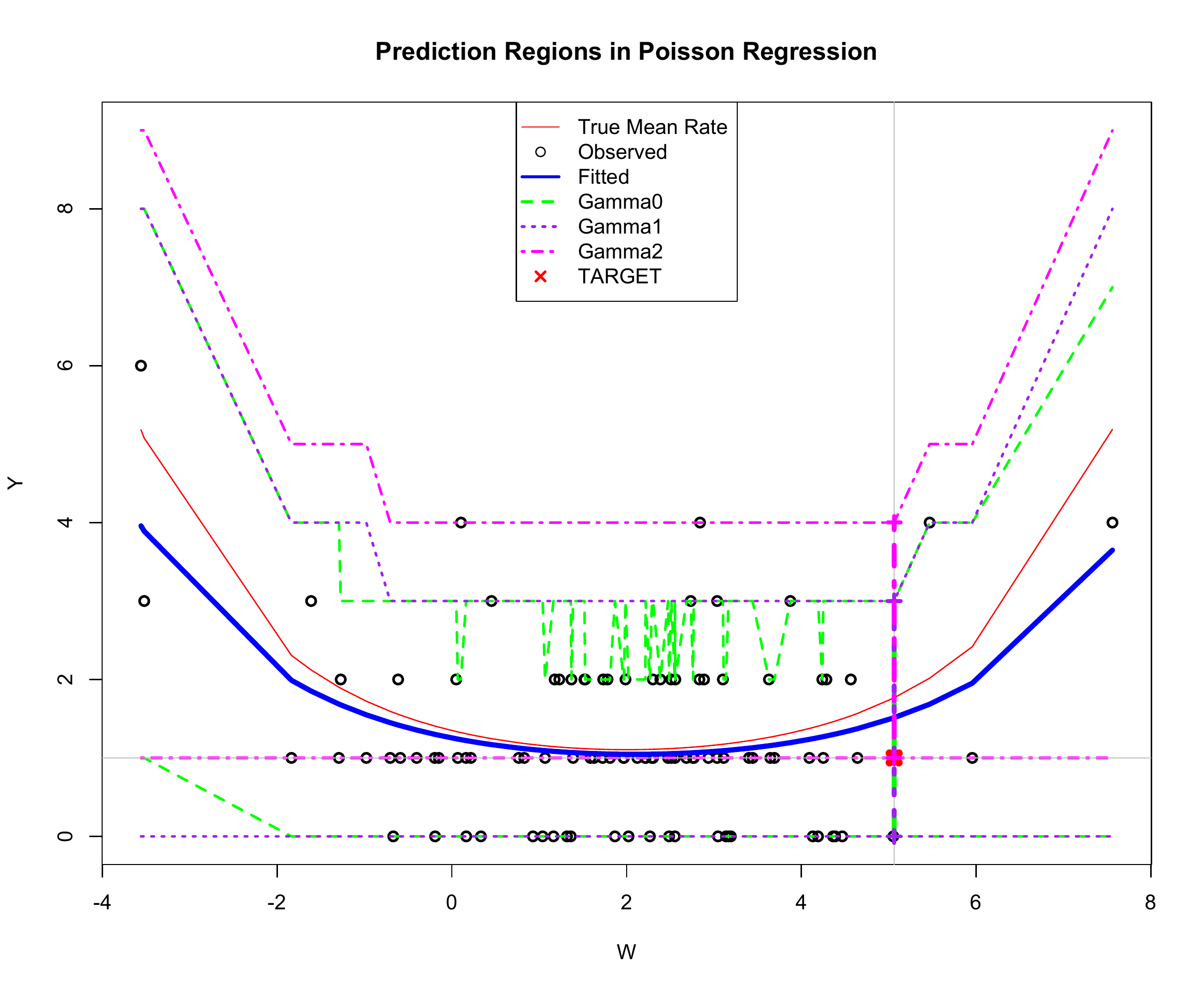}
\end{figure}

\begin{figure}[h]
\caption{A realization with  $p=3, n=200, \theta = (3,.2,-.1,-.05)$,  and with $W \sim N(\mu=1, \sigma=2)$.  The resulting prediction regions were $\tilde{\Gamma}_0 =[10,25], \check{\Gamma}_1 = [9,26], \check{\Gamma}_2 = [10,27]$. The realized $Y_0$ was $y_0 = 23$ associated with $w_0 = -3.2509$.}
\label{Case4}
\includegraphics[width=\textwidth]{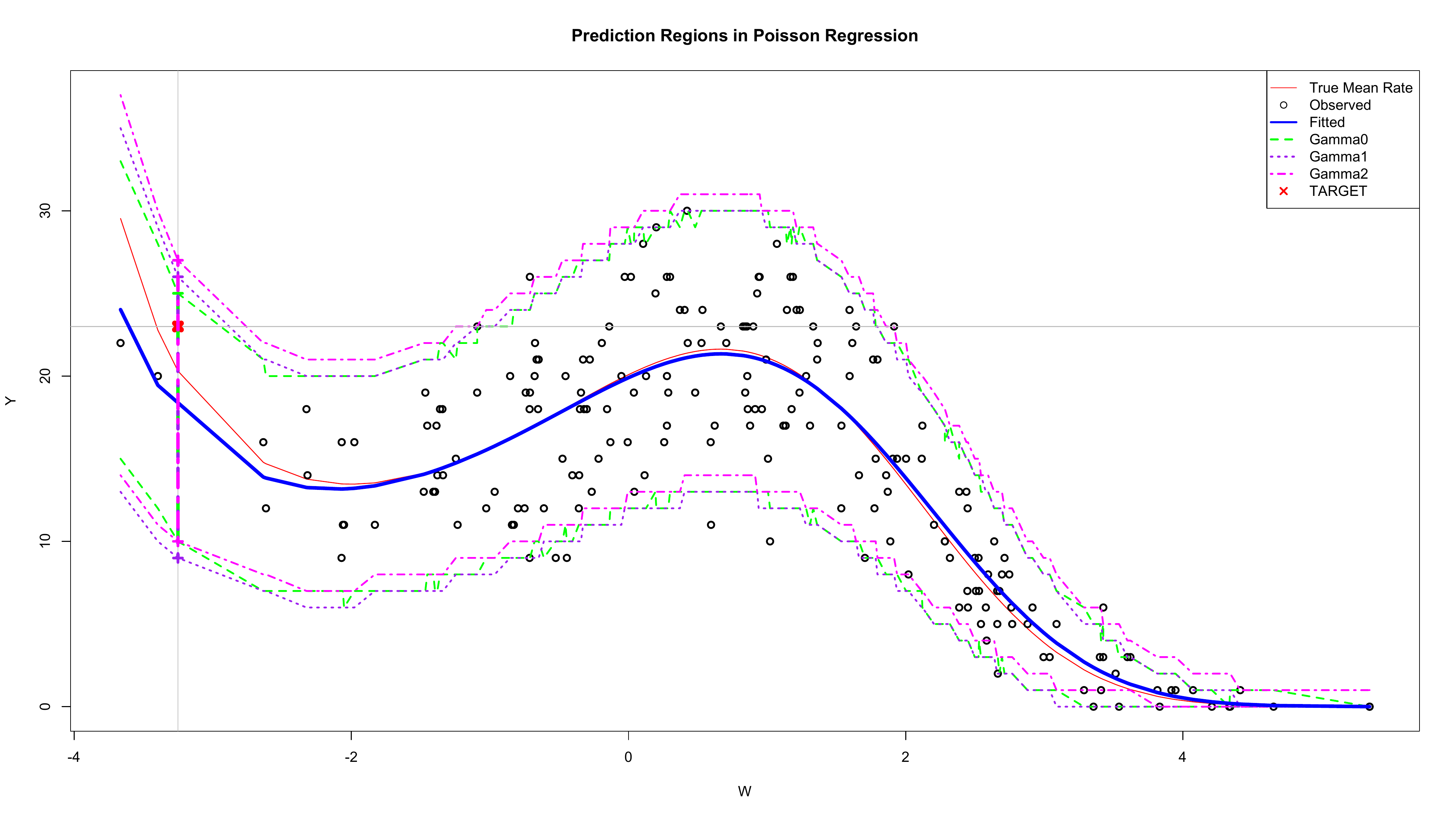}
\end{figure}

\begin{figure}[h]
\caption{A realization with $p=5, n=100, \theta = (3,-1,3,-2,1,-.5),$ and with $W \sim U[0,1]$. The resulting prediction regions were $\tilde{\Gamma}_0 = [21,42], \check{\Gamma}_1 = [20,42], \check{\Gamma}_2 = [22,43]$. The realized $Y_0$ was $y_0 = 29$ associated with $w_0 = .8944$.}
\label{Case0}
\includegraphics[width=\textwidth]{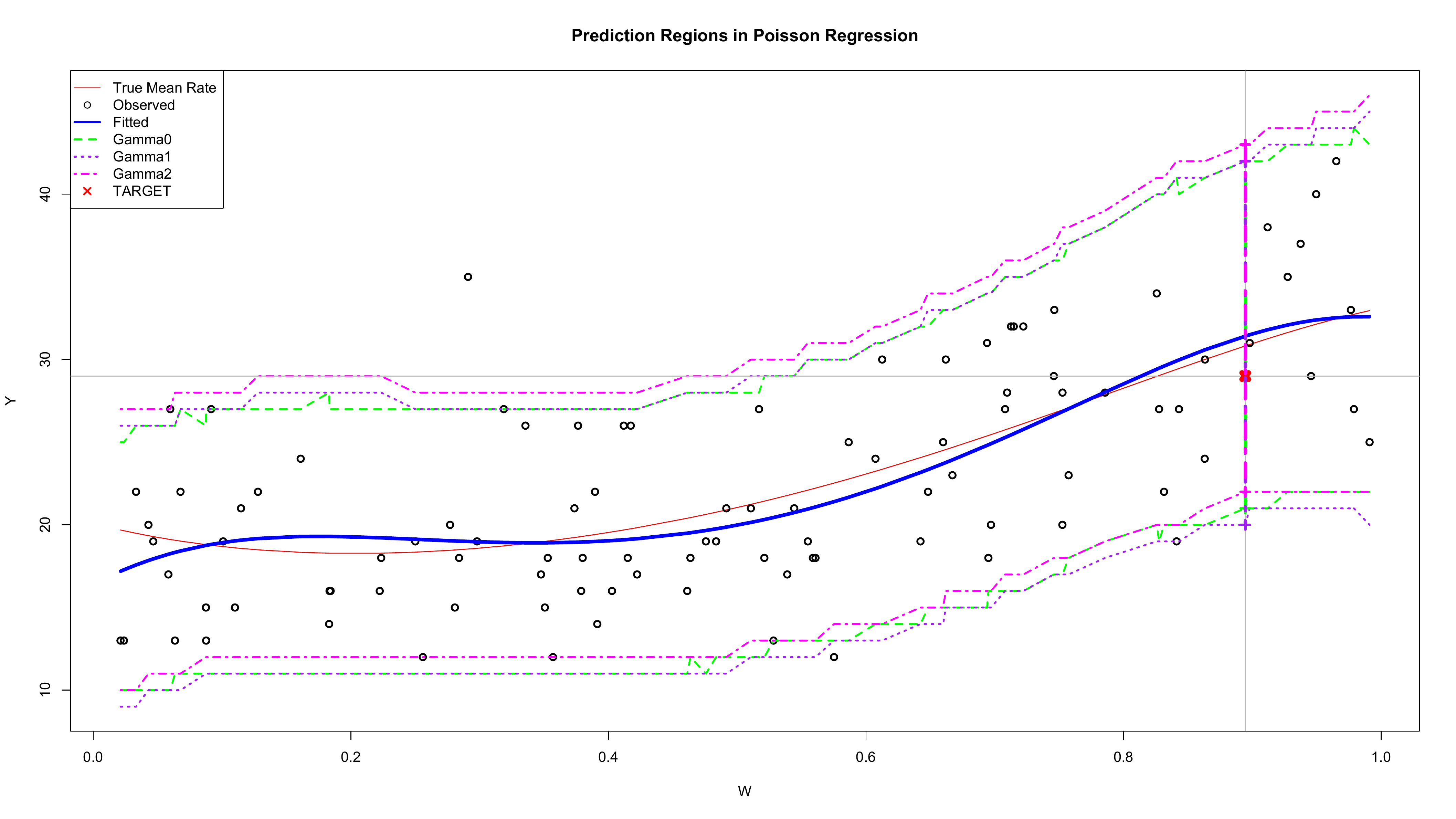}
\end{figure}

Figure \ref{Case1} shows a realization for a model with $n = 30$, $p=1$, with $W_i \sim U[0,1], i=1,2,\ldots,31$, and with $\mathbf{\theta} = (2,3)\trp$, so that $\lambda_i = \exp\{2 + 3w_i\}, i=1,2,\ldots,31$, a sharply and exponentially increasing function of $w_i$. The $Y_{31}$ being predicted was at $w_{31} = .8139$ and the realized value was $y_{31} = 1222$, which ended up being contained in all three realized prediction intervals $\tilde{\Gamma}_0 = [1091,1223]$, $\check{\Gamma}_1 = [1087,1226]$, and $\check{\Gamma}_2 = [1088,1227]$. 
The second realization in Figure \ref{Case3} is from a model with $n=100$, $p=2$, and where $W_i \sim N(\mu=2,\sigma=2), i=1,2,\ldots,101$, and with $\mathbf{\theta} = (.3,-.2,.05)$. The value being predicted is at $w_{101} = 5.0622$ and the realized value of $Y_{101}$ was $y_{101} = 1$. The three prediction regions were $\tilde{\Gamma}_0 = [0,3]$, $\check{\Gamma}_1 = [0,3]$, and $\check{\Gamma}_2 = [1,4]$, which all contained $y_{101}$. Two things to observe from this plot are (1) the prediction region $\check{\Gamma}_2$ was shifted to the right relative to $\check{\Gamma}_1$, and (2) the prediction regions $\tilde{\Gamma}_0$ with respect to the $w$-values are scissor-like or jagged. The latter is a consequence of the randomization approach in the construction of the prediction regions and this non-smooth behavior becomes more apparent since the realized values of the $Y_i$s are small. The third realization, depicted in Figure \ref{Case4}, is from a model with $n = 200$, $p = 3$, with $W_i \sim N(\mu=1,\sigma=2)$, and with $\mathbf{\theta} = (3,.2,-.1,-.05)$. The rate curve $\lambda(w)$ as a function of $w$ goes to zero as $w$ increases, but goes to $\infty$ as $w$ decreases, with a local minimum and maximum close to $w = -2$ and $w =1$, respectively. The target of the prediction regions was $Y_{201}$ which took value $y_{201} = 23$ at $w_{201} = -3.2509$. The realized prediction regions were $\tilde{\Gamma}_0 = [10,25]$, $\check{\Gamma}_1 = [9,26]$, and $\check{\Gamma}_2 = [10,27]$, so all regions contained $y_{201} = 23$. Once again observe that $\check{\Gamma}_2$ tends to be shifted to the right relative to $\check{\Gamma}_1$. Figure \ref{Case0} contains a realization from a model with $n = 100$, $p = 5$, with $W_i \sim U[0,1], i=1,2,\ldots,101$, and with $\mathbf{\theta} = (3,-1,3,-2,1,-.5)$. The rate curve is not highly varying compared to that in the preceding model since $w$ is restricted on $[0,1]$. The target is the value of $Y_{101}$, which turned out to be $y_{101} = 29$ at $w_{101} = .8944$. This realized value of $Y_{101}$ was contained in the realized prediction regions $\tilde{\Gamma}_0 = [21,42]$, $\check{\Gamma}_1 = [20,42]$, and $\check{\Gamma}_2 = [22,43]$.


%

\subsection{Simulation Studies}

\begin{figure}[h]
\caption{An extreme realization from the third model with  $p=3, n=30, \theta = (3,.2,-.1,-.05)$,  and with $W \sim N(\mu=1, \sigma=2)$.  The resulting prediction regions were $\Gamma_0 =[257,323], \Gamma_1 = [0,1100], \Gamma_2 = [0,1665]$. The realized value of $Y_{31}$ was $y_{31} = 47$ associated with $w_{31} = -4.0493$.}
\label{ExtremePlot}
\includegraphics[width=\textwidth]{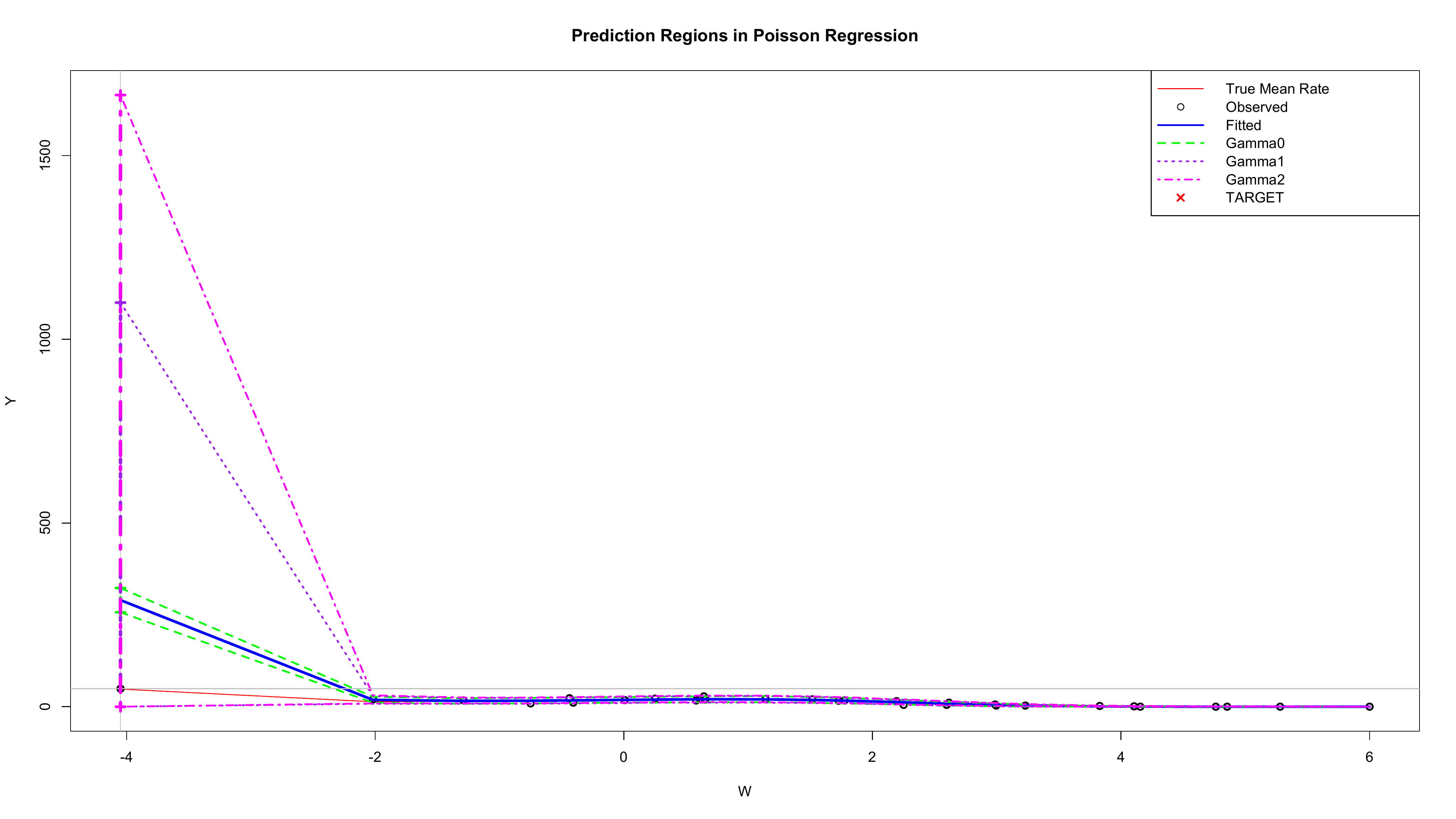}
\end{figure}

In each of these illustrative realizations, the three prediction regions did not vary much from each other in terms of their sizes, except in the first case where $\tilde{\Gamma}_0$ was shorter but barely covered the value being predicted. A question that now arises is how their coverage probabilities and their mean lengths compare with each other? To gain some insights into these comparisons, we performed simulation studies under the four different models described above, with each simulation run having 10000 replications. The sample sizes considered were $n \in \{30, 50, 100, 200\}$. In the Appendix, Tables \ref{BigSimulCase 1}, \ref{BigSimulCase 2},  \ref{BigSimulCase 3} and \ref{BigSimulCase 4} summarize the results of these simulations where we report the coverage probabilities (CP), mean lengths (ML), and standard deviation of lengths (SL). Examining these tables, it appears that $\tilde{\Gamma}_0$ has coverage probabilities that are below the nominal level (between 3\% and 4\% below in Table \ref{BigSimulCase 4} when $n = 30$), with the discrepancy being more pronounced when the sample size is small. As the sample size is increased, these observed coverage probabilities get closer to the nominal level. This deficiency is due to the estimation of the $\mathbf{\theta}$ parameter and, as previously noted, the $\tilde{\Gamma}_0$ does not take into consideration the variability in the resulting estimator of $\lambda(\mathbf{x}_0;\mathbf{\theta})$. On the other hand, $\check{\Gamma}_1$ and $\check{\Gamma}_2$ both achieve coverage probabilities that are quite close to the nominal level, especially $\check{\Gamma}_1$, when $n$ is large. $\tilde{\Gamma}_0$, on the other hand, tends to have a lower mean length compared to the mean lengths of $\check{\Gamma}_1$ and $\check{\Gamma}_2$, with the differences in mean lengths becoming alarmingly large for the model in Table \ref{BigSimulCase 3}. Recall that for this model, the rate curve increases to $\infty$ as $w$ decreases to $-\infty$, and since the $w_i$'s are generated from a normal distribution, on some occasions, $w_{n+1}$ falls outside the range of $\{w_i, i=1,2,\ldots,n\}$. Depending on how different $w_{n+1}$ is from the mean of $w_1, w_2, \ldots, w_n$, this could lead to a large estimate of the standard error of $\hat{\lambda}_{n+1}$, thus leading to very wide prediction regions for $\check{\Gamma}_1$ and $\check{\Gamma}_2$. Since $\tilde{\Gamma}_0$ simply utilized the estimate of $\hat{\lambda}_{n+1}$, but was totally oblivious to its variability, it was not much affected in such a situation. However, because of its rigidness with respect to this added variability, it could dramatically suffer. We demonstrate this situation by plotting an extreme realization in Figure \ref{ExtremePlot} from the third model with $n = 30$ and with the $w_{31}$ outside and far to the left of the values of $w_1, \ldots, w_{30}$. Observe here that the three prediction regions $\tilde{\Gamma}_0 = [257,323]$, $\check{\Gamma}_1 = [0,1100]$, and $\check{\Gamma}_2 = [0,1665]$ are very different in terms of their lengths (66, 1100, and 1665, respectively). But in this case $\tilde{\Gamma}_0$ did {\em not} cover, by a wide margin, the realized value of $Y_{31}$, which was $y_{31} = 49$ associated with $w_{31} = -4.0493$. This particular demonstration warns us of the danger and pitfalls of making a prediction for a response variable that is associated with a covariate vector outside the convex hull of the covariate vectors used in the construction of the prediction regions and when the Poisson rate hyper-surface generated by the map $\mathbf{x} \mapsto \rho(\mathbf{x}\mathbf{\theta})$ is complex. As a word of caution, when performing extrapolation to do predictions, be forewarned of sinkholes littering the forecasting road --- and, if it could be avoided, make no forecasts on long, especially very long, horizons.  But, alas, this is the type of forecasting problem that is actually {\em realistic} and of most interest, such as that of predicting the number of cases or deaths due to \Covid\ in a future date, given the observed data up to a certain date.  

Based on the results of these simulation studies, the prediction region $\check{\Gamma}_1$ appears to be the most preferable among the three prediction regions. In our illustration using the \Covid\ data set in Section \ref{sec-Applications}, we will therefore just present the prediction region provided by $\check{\Gamma}_1$.

\section{Forecasting Application to \Covid\ Pandemic}
\label{sec-Applications}

We now present in this section an illustration of the potential application of the procedures discussed in the preceding sections. One of the interesting questions during this \Covid\ pandemic is the forecasting of the number of cumulative deaths in the US at a given date, for example, at the end of May 31, 2020, given information up to a certain date, say May 15, 2020. Such forecasts are of critical importance since they could partly be the basis of highly consequential and possibly controversial decisions by federal, state, and local governments officials, school administrators, executives of big corporations and small businesses, religious leaders, and many others.  Such decisions could pertain to when to institute stay-in-place directives, when to issue social distancing or social easing guidelines, when to open business establishments, when to open public places such as shopping malls and ocean beaches, when to allow religious gatherings, etc. Data for daily deaths and cumulative deaths, as well as cases, for different countries and states due to the \Covid\ are available from either the US Center for Disease Control and Prevention \url{https://www.cdc.gov/coronavirus/2019-ncov/cases-updates/cases-in-us.html}, Johns Hopkins University \url{https://coronavirus.jhu.edu/data}, or the European CDC \url{https://www.ecdc.europa.eu/en/publications-data/} \cite{ECDCData}. 

Clearly, the sequence of cumulative deaths does not satisfy the independence assumption, so a non-homogeneous Poisson process model \cite{Res92} is not an appropriate model for cumulative deaths when viewed as a continuous-time stochastic process. However, a non-homogeneous Poisson process could plausibly model the occurrences of deaths in continuous-time, from which it follows that the sequence of daily deaths will be independently Poisson distributed with possibly different rates depending on the number of days from the time origin and the specific day of the week, as well as other features such as, for example, the quality of the health care facilities, which is hard to quantify and not available in the European CDC data set. Our novel idea therefore is to utilize Poisson regression to predict the number of daily deaths according to the methods developed earlier, and then to aggregate these daily forecasts to obtain forecasts of the cumulative deaths.

We will use the data set for the US provided by the European CDC (\cite{ECDCData}) plotted in Figure \ref{deaths and cumulative deaths data USA} 
which are the observed numbers of daily deaths attributed to \Covid\ starting on March 1, 2020, the day after the first reported death due to \Covid, until May 15, 2020. Note that, technically, this will be the deaths data at the end of May 14, 2020. Using this data set on May 15th, and given the cumulative number of deaths until then, the goal is to forecast the cumulative number of deaths in the US due to \Covid\ by the end of May 31, 2020, that is, June 1, 2020. We limit our illustration to simply utilizing the variable {\tt DayNum}, which is the number of days starting from December 31, 2019, {\tt Day} which the day of the week, and {\tt Deaths}, the variable representing the daily number of deaths. We surmise that the deaths data set is the most reliable among the data sets that were compiled, compared, for instance, to the data set pertaining to the number of cases or infected people. However, the deaths data set need not also be totally reliable and could be subject to misclassification error and competing causes of deaths., For example, a patient who contracted \Covid\ who dies primarily because of pneumonia may be classified as having died of \Covid, but could also be classified as having died, not of \Covid, but of pneumonia. See also, for instance, the WSJ article \cite{WSJ416}, \cite{Economist416}, and the BBC news article \url{https://www.bbc.com/news/world-53073046} \cite{BBCNews0618}, the last two discussing the notion of ``excess deaths,'' which are deaths that may have been due to the pandemic, but which are not included in the reported \Covid\ deaths data set. Certainly, we could have used other information such as the number of reported cases; by performing separate  forecasts in each of the 50 states and the District of Columbia, then aggregating; or even by utilizing counties or metropolitan cities as strata, and then combining forecasts from these strata to obtain an overall forecast for the whole US. However, for illustrative purposes, we decided to keep things simple.

From March 1 ({\tt DayNum} = 62) to May 15 ({\tt DayNum} = 137), consisting of 76 days, we have the daily number of deaths, hence also the cumulative number of deaths. The other variable used in the modeling is {\tt Day} (e.g., Sunday, Monday, etc.) associated with each value of {\tt DayNum}, which is a categorical or factor variable. We fitted a Poisson regression model using the {\tt glm} function in {\tt R} with a log-link, with response being $Y = {\tt Deaths}$ and covariate vector $\mathbf{X} = (1, W, W^2, W^3, W^4, W^5, {\tt Day})$, where $W = {\tt DayNum}$, and {\tt Day} is considered as a factor variable hence is converted into six, instead of seven since we already have an intercept term, dummy variables in the design matrix. We chose this 5th-order model with respect to {\tt DayNum} since the Akaike Information Criterion (AIC) values, computed under the Poisson regression model, appear to stabilize starting at this model and adhering to the Law of Parsimony (Occam's Razor). The AIC values associated with the 8th- and 9th-order models were actually smaller than for the 5th-order model; however, these models possess highly unstable predicted values.  Table \ref{AIC values for fitted models} summarizes the AIC values for the different models, both without and with {\tt Day} in the model. It also contains the estimates of $\xi$, the over-dispersion parameter in a model that will be introduced shortly, and as we will then see, larger values of $\xi$ are indicative of the Poisson regression model becoming a more adequate model.

\begin{table}
\caption{AIC values for the fitted model under different order ($p$) for {\tt DayNum}. The second column does not include {\tt Day} in the model, whereas the third column includes {\tt Day} in the model as a categorical variable. The fourth column is the estimate of $\xi$ without {\tt Day}, while the fifth column is the estimate of $\xi$ with {\tt Day}.}
\label{AIC values for fitted models}
\centering
\begin{tabular}{rrrrrr}
  \hline
 & $p$ & $AIC_{ND}$ & $AIC_D$ & $\hat{\xi}_{ND}$ & $\hat{\xi}_D$ \\ 
  \hline
1 & 1.00 & 43755.86 & 40750.61 & 2.66 & 2.99 \\ 
  2 & 2.00 & 12289.82 & 8504.21 & 7.26 & 10.96 \\ 
  3 & 3.00 & 8551.98 & 5117.39 & 9.33 & 16.25 \\ 
  4 & 4.00 & 8486.66 & 5098.87 & 9.40 & 16.27 \\ 
  5 & 5.00 & 8372.16 & 4889.52 & 9.47 & 16.89 \\ 
  6 & 6.00 & 8374.09 & 4875.94 & 9.46 & 17.14 \\ 
  7 & 7.00 & 8374.27 & 4867.00 & 9.47 & 17.20 \\ 
  8 & 8.00 & 8216.72 & 4797.71 & 9.73 & 17.80 \\ 
  9 & 9.00 & 7917.52 & 4600.89 & 10.08 & 18.79 \\ 
   \hline
\end{tabular}
\end{table}

\begin{figure}[h]
\caption{Histogram and Time Plot of the Residuals from the Fitted Fifth-Order Poisson Regression Model of {\tt Deaths} with respect to {\tt DayNum}.}
\label{ResidualPlots}
\includegraphics[width=\textwidth,height=2in]{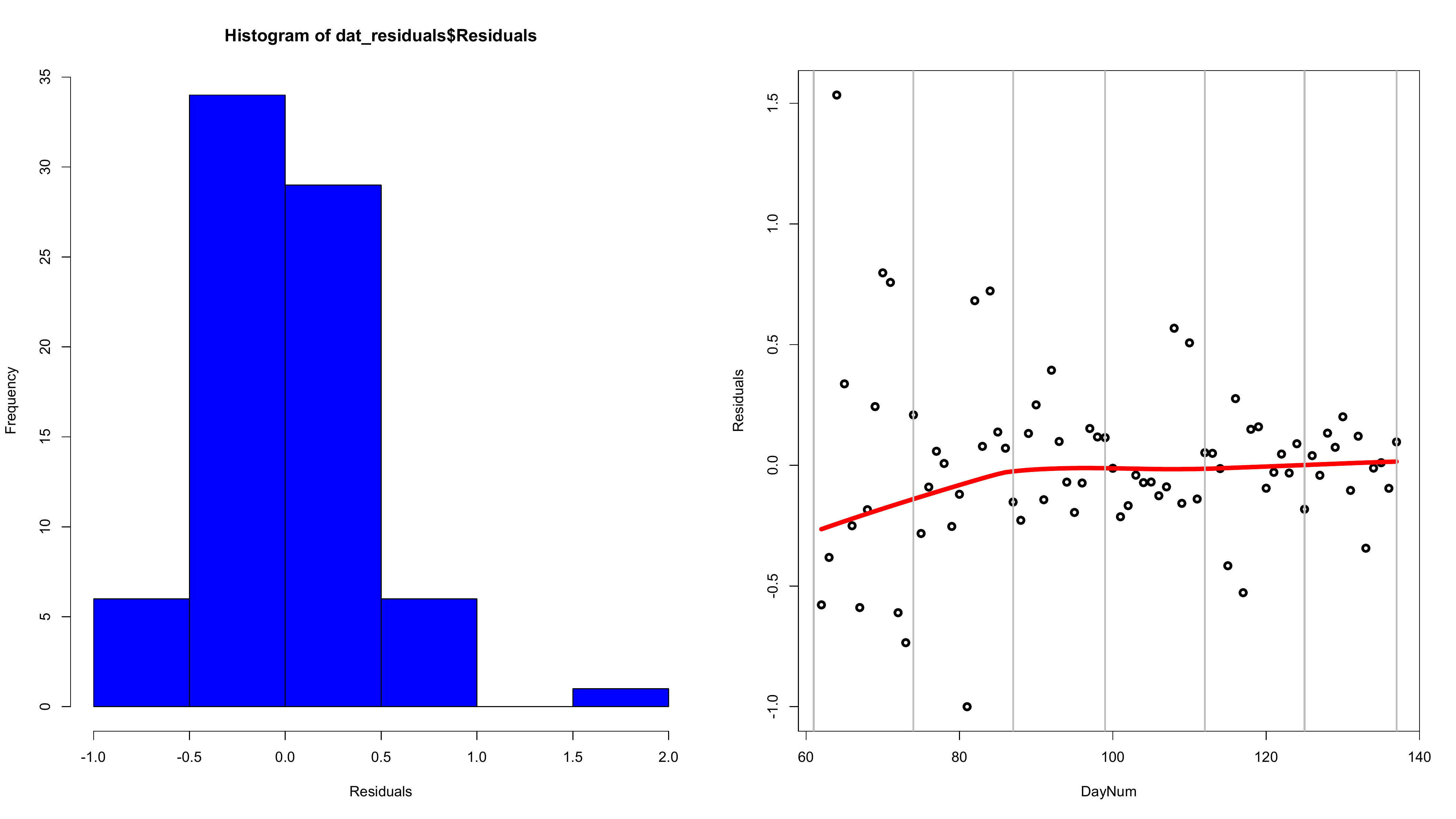}
\end{figure}

The histogram and time plot of the residuals with respect to {\tt DayNum} are provided in Figure \ref{ResidualPlots}. Recall that the $i$th residual in Poisson regression is defined as
\begin{displaymath}
R_i = \frac{Y_i - \hat{\lambda}_i}{\sqrt{\hat{\lambda}}_i}, i=1,2,\ldots,n,
\end{displaymath}
where $\hat{\lambda}_i$ is the fitted value associated with $\mathbf{x}_i$. As such, if the Poisson regression model is adequate, one should see a histogram similar to that associated with a centered (at zero) Poisson distribution with unit rate, but this will just be an approximation since the rates are estimated, hence that affects the distribution of the residual. Similarly, the time plot of the residuals should be randomly distributed on the zero horizontal line. The histogram and time plot in Figure \ref{ResidualPlots}, with the time plot also showing a {\em lowess} fit, do not appear to contradict expected behaviors under the Poisson regression model assumptions of Poisson{\em ness}, independence, and an adequate rate function. In the time plot, observe the outliers at {\tt DayNum} values of 64 and 81, with mild outliers at {\tt DayNum} of 70, 71, 82, 84, and 108. The outlying values at {\tt DayNum} 64 and 81 could be a consequence of the uncertainty on who to count as \Covid\ deaths since they occurred in the early days of recording \Covid\ deaths, while on {\tt Daynum} 81, which was March 20th, there were zero reported \Covid\ deaths, highly suspect since two days before and two days after this day there were 23, 42, 110, and 80 \Covid\ reported deaths. The outlier on {\tt DayNum} 108, corresponding to April 16th, was due to a one-time adjustment of 3778 that was made because of a change in criteria on what are considered as \Covid\ deaths \cite{BBC415,WSJ416}. The two days prior to April 16th, there were 1541 and 2408 deaths reported, while the three days after April 16th, there were 2299, 3770, and 1856 deaths reported. Thus, the reported number on April 16th of 4928, the highest number of daily deaths reported, is an outlier explained by the adjustment made. However, we still included these perceived outliers in the fitting of the fifth-order, with respect to {\tt DayNum}, Poisson regression model. Later, when we consider forecasting for July 15th and August 1st, and since another significant adjustment was made on June 26th, we will re-allocate each of the adjustments proportionately to the observed deaths on the days on or before the adjustment day.

Based on the fitted model's residuals, we further assessed the independence assumption of the daily deaths. We do this by creating a contingency table for {\tt DayNum} with six intervals and with {\tt Residuals} being either negative or positive and then performing a test for independence. The observed contingency table is presented in Table \ref{table-test for independence}. A chi-square test for independence based on this table yielded $\chi_c^2 = 4.4685$ on 5 degrees-of-freedom, with associated $p$-value of 0.4841, hence the null hypothesis of independence cannot be rejected.  Observe, however, that the fit of the model in the early days is not satisfactory, and between {\tt DayNum} 99 to 112, there was a preponderance of negative residuals, possibly owing to the influence of the adjusted reported daily deaths on {\tt DayNum} 108.

%
%
\begin{table}
\caption{Contingency Table of {\tt DayNum} and {\tt Residuals} from the Fifth-Order Poisson Regression Model.} 
\label{table-test for independence}
\centering
\begin{tabular}{rrr}
  \hline
{\tt DayNum} Interval  & Residual $\le 0$ & Residual $> 0$ \\ 
  \hline
61  $<$ DayNum $<$=  74 & 7.00 & 6.00 \\ 
  74  $<$ DayNum $<$=  87 & 6.00 & 7.00 \\ 
  87  $<$ DayNum $<$=  99 & 5.00 & 7.00 \\ 
  99  $<$ DayNum $<$=  112 & 10.00 & 3.00 \\ 
  112  $<$ DayNum $<$=  125 & 7.00 & 6.00 \\ 
  125  $<$ DayNum $<$=  137 & 5.00 & 7.00 \\ 
   \hline
\end{tabular}
\end{table}

It may appear surprising and counter intuitive to include a {\tt Day} effect in the model, since one could argue that when a patient is dying, he/she does not really have any control or choice of which day he/she should die, so that deaths due to \Covid\ should be uniformly distributed over the days of a week. However, the data pertains to {\em reported} daily deaths, and so the number of daily deaths could be affected by reporting delays due possibly to limited health or hospital personnel during certain days of the week. The fitted model based on data up to May 15th did demonstrate that there is a {\tt Day} effect. Based on the estimates of the regression coefficients, Sunday has the lowest {\tt Day} effect, followed by Monday, then Saturday,  then Friday, then almost equally by Tuesday and Thursday, and finally Wednesday, which has the largest effect. 

Recall that our main objective is to obtain a prediction region for the cumulative number of deaths at a specified date, in our case June 1, 2020 (${\tt DayNum} = 154$). This means we were predicting the cumulative deaths at the end of May 31, 2020. From ${\tt DayNum} = 138$ to ${\tt Day Num} = 154$, there were a total of 17 days. If we denote by $Y_j, j=62,63,\ldots,154,$ the random variable denoting the daily number of deaths for ${\tt DayNum} = j$, the random variable denoting the cumulative number of deaths until ${\tt DayNum} = k$ is $S_k = \sum_{j=62}^k Y_k$. Thus, we are seeking a prediction region for $S_{154}$, given that $S_{137} = 85906$.  Under the fitted Poisson regression model, $Y_j$, given 
\begin{eqnarray*}
\mathbf{x}_j & = & \left(1, w_j, w_j^2, w_j^3, w_j^4, w_j^5, d_{1j},d_{2j},d_{3j},d_{4j},d_{5j},d_{6j}\right)
\end{eqnarray*}
with $w_j = {\tt DayNum}_j$ and $d_{kj}, k=1,2,\ldots,6,$ the dummy variables representing whether {\tt Day} is a Tuesday, a Wednesday, a Thursday, a Friday, a Saturday, or a Sunday, respectively, has Poisson distribution with rate 
$\lambda(\mathbf{x}_j;\mathbf{\theta}) = \exp\left\{\mathbf{x}_j \mathbf{\theta}\right\}.$
We mention that in our {\tt R} code for fitting this model, for computational stability, we first centered and standardized the non-constant columns of the $\mathbf{x}_j$'s.
Let $\alpha^* \in (0,1)$, and the $Y_j$s being independent. From the preceding section we know how to construct a $100(1-\alpha^*)\%$ prediction interval $[a_j,b_j]$ for $Y_j$, where, as mentioned earlier, we will simply utilize the prediction region $\check{\Gamma}_1$.  By the independence, we will have
\begin{eqnarray*}
\lefteqn{(1 - \alpha^*)^{17}  \le  \Pr\left\{\bigcap_{j=138}^{154} \left[Y_j \in [a_j,b_j]\right] | S_{137} \right\} } \\
& \le & \Pr\left\{S_{154} - S_{137} = \sum_{j=138}^{154} Y_j \in \left[a_\bullet \equiv \sum_{j=138}^{154} a_j, b_\bullet \equiv \sum_{j=138}^{154} b_j\right] | S_{137}\right\}.
\end{eqnarray*}
Thus, if we wanted $S_{137} + [a_\bullet,b_\bullet]$ to be a $100(1-\alpha)\%$ prediction interval for $S_{154}$, given $S_{137}$, we could choose $\alpha^* = 1 - (1-\alpha)^{1/17}$. This procedure will guarantee a conservative $100(1-\alpha)\%$ prediction interval for $S_{154}$, given $S_{137}$. This is the approach we followed in constructing a (conservative) $95\%$ prediction interval for $S_{154}$, the cumulative number of deaths due to \Covid\ in the US {\em by the end of May 31, 2020}. 

\subsection{Over-Dispersed Poisson Regression Model}

\begin{figure}[h]
\caption{Scatterplot of daily deaths data until May 15, 2020, together with the 95\% prediction intervals, under the Poisson regression model.}
\label{Scatterplot until May 15 with pred ints}
\includegraphics[width=\textwidth]{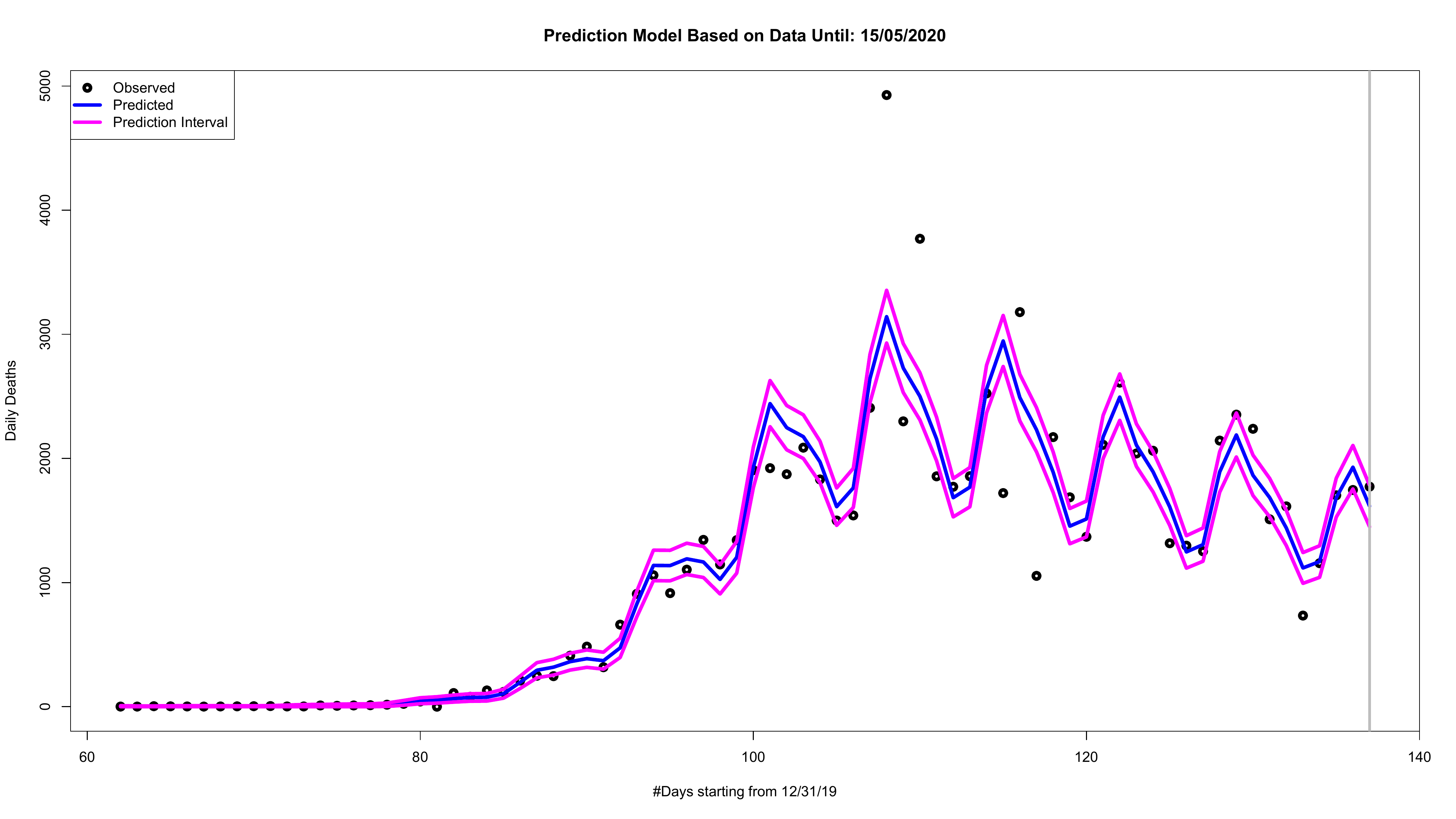}
\end{figure}

We implemented the above procedure and also constructed the prediction intervals at each of the observed {\tt DayNum} which are depicted in Figure \ref{Scatterplot until May 15 with pred ints}. 
%
%
%
Examining this figure note that there are more observed {\bf Deaths} outside the prediction curves than what is expected nominally. This indicates that either there is more variability inherent in the stochastic mechanism generating the observed number of daily deaths relative to a purely Poisson regression model, or the fifth-order Poisson rate model is still inadequate, or both.  We propose an approach that introduces over-dispersion with the Poisson regression model serving as a hidden model. We mention that our 
Occam's Razor-type solution is motivated by frailty modeling in Survival Analysis (see, for instance, \cite{ABGK93}).  Our model assumes the existence of an unobserved positive latent variable $Z_j$ of mean 1 at $w_j = {\tt DayNum}_j$, and the reported number of deaths $Y_j$ is the integer part of $Z_j Y_j^*$, with $\{Y_j^*\}$ arising from a Poisson regression model and with $Z_j$ and $Y_j^*$ independent. Recall that frailty models in Survival Analysis are used specifically to model correlations among observations; whereas, in our model it serves as an unobserved random contamination component in the observed number of daily deaths.   In our implementation, we shall take $Z_j$ to have a gamma distribution with mean one and variance $1/\xi$ (see \cite{CasBer90}). This $\xi$ is then an additional parameter in the regression model aside from the parameter vector $\mathbf{\theta}$. Such a model leads to an over-dispersed Poisson regression model, with the purely Poisson regression model embedded in this model and obtainable as a limiting case when $\xi \rightarrow \infty$. Inference for such a model requires further study, with possible use of an EM-type algorithm, though this could be difficult to implement since the distributions of the $Y_i$s are not in closed forms. However, we may implement a $Z$-estimation approach (see, for instance, \cite{vanderVaart98}). We first note that
\begin{eqnarray*}
E(Y_i) & \approx & E[E(Y_i|Z_i)] = E[Z_i \lambda_i] = \lambda_i \\
Var(Y_i) & \approx & E[Var(Y_i|Z_i)] + Var[E(Y_i|Z_i)] \\ & = & E[Z_i^2\lambda_i] + Var[Z_i \lambda_i] = 
\lambda_i(1 + 1/\xi) + \lambda_i^2 (1/\xi) = 
\lambda_i\left[1 + \left(\frac{1+\lambda_i}{\xi}\right)\right].
\end{eqnarray*}
The approximate higher-order moments of the $Y_i$s are also obtainable. Based on these moments, we could form the set of estimating equations, where we recall that $\rho(w) = \exp(w)$:
\begin{eqnarray*}
& \frac{1}{n} \sum_{i=1}^n \mathbf{x}_i\trp {[Y_i - \rho(\mathbf{x}_i\mathbf{\theta})]}  =  0; & \\
& \frac{1}{n} \sum_{i=1}^n \left\{[Y_i - \rho(\mathbf{x}_i\mathbf{\theta})]^2 - \rho(\mathbf{x}_i\mathbf{\theta})[1 + (1 + \rho(\mathbf{x}_i\mathbf{\theta}))/\xi]\right\}  =  0. &
\end{eqnarray*}
By $Z$-Estimation Theory (\cite{vanderVaart98}) and under regularity conditions, it will follow that, for some $(p+1) \times (p+1)$ matrix 
$$\mathbf{\Xi} = \left[\begin{array}{cc} \mathbf{\Xi}_{11} & \mathbf{\Xi}_{12} \\ \mathbf{\Xi}_{21} & \Xi_{22}\end{array}\right]$$
we have
\begin{displaymath}
\left(\begin{array}{c}
\tilde{\mathbf{\theta}} \\ \tilde{\xi}
\end{array}
\right)  \sim AN\left[\left(\begin{array}{c} \mathbf{\theta} \\ \xi \end{array}\right),
\frac{1}{n} \mathbf{\Xi} \right].
\end{displaymath}
In fact, let us introduce the $(p+1) \times 1$ vector functions:
\begin{eqnarray*}
\mathbf{U}((y,\mathbf{x});\mathbf{\theta},\xi) & = & \left[ \begin{array}{c} \mathbf{U}_1((y,\mathbf{x});\mathbf{\theta},\xi) \\ U_2((y,\mathbf{x});\mathbf{\theta},\xi) \end{array} \right] = 
\left[
\begin{array}{c}
\mathbf{x}\trp {[y - \rho(\mathbf{x}\mathbf{\theta})]} \\
\left[y - \rho(\mathbf{x}\mathbf{\theta})\right]^2 
- \rho(\mathbf{x}\mathbf{\theta})[1 + (1 + \rho(\mathbf{x}\mathbf{\theta}))/\xi]
\end{array}
\right].
\end{eqnarray*}
Denote by $\mathbf{H}$ the $(p+1) \times (p+1)$ matrix function consisting of the derivatives of $\mathbf{U}$ with respect to $(\mathbf{\theta}, \xi)$. The components of this matrix function are:
\begin{eqnarray*}
\mathbf{H}_{11}((y,\mathbf{x});\mathbf{\theta},\xi) & = & -\frac{1}{n} \sum_{i=1}^n \mathbf{x}_i^{\otimes 2} \rho(\mathbf{x}_i\mathbf{\theta}) \quad\quad \mbox{and} \quad\quad
\mathbf{H}_{12}((y,\mathbf{x});\mathbf{\theta},\xi)  =  \mathbf{0}; \\
\mathbf{H}_{21}((y,\mathbf{x});\mathbf{\theta},\xi) & = & -\frac{1}{n}\sum_{i=1}^n \mathbf{x}_i\trp \rho(\mathbf{x}_i\mathbf{\theta}) \left\{2(y_i-\rho(\mathbf{x}_i\mathbf{\theta})) + 
[1+(1+\rho(\mathbf{x}_i\mathbf{\theta}))/\xi] + \rho(\mathbf{x}_i\mathbf{\theta})/\xi\right\}; \\
H_{22}((y,\mathbf{x});\mathbf{\theta},\xi) & = & \frac{1}{n} \frac{1}{\xi^2} \sum_{i=1}^n \rho(\mathbf{x}_i\mathbf{\theta}) [1 + \rho(\mathbf{x}_i\mathbf{\theta})].
\end{eqnarray*}
We could then obtain the estimates via the Newton-Raphson (NR) method with iteration step
\begin{displaymath}
\left[\begin{array}{c} \mathbf{\theta} \\ \xi\end{array}\right] \leftarrow \left[\begin{array}{c} \mathbf{\theta} \\ \xi\end{array}\right] - \left[\frac{1}{n} \sum_{i=1}^n \mathbf{H}((y_i,\mathbf{x}_i);\mathbf{\theta},\xi)\right]^{-1} 
\left[ \frac{1}{n} \sum_{i=1}^n \mathbf{U}((y_i,\mathbf{x}_i);\mathbf{\theta},\xi) \right].
\end{displaymath}
It turns out that a simpler way to obtain the estimates of $\mathbf{\theta}$ and $\xi$ is to first obtain the estimate $\tilde{\mathbf{\theta}}$ of $\mathbf{\theta}$ from the first estimating equation. This could be done by using the {\tt glm} object function in {\tt R} \cite{R} with the Poisson family and logarithm link. The estimate $\tilde{\xi}$ of $\xi$ is then obtained from the second estimating equation using a one-variable NR iteration with $\mathbf{\theta}$ replaced by $\tilde{\mathbf{\theta}}$.
Define the $(p+1) \times (p+1)$ matrices $\mathbf{\Sigma}$ and $\mathbf{\Omega}$ according to
\begin{eqnarray*}
& \mathbf{\Sigma}  =  \mbox{plim} \frac{1}{n} \sum_{i=1}^n \mathbf{U}((Y_i,\mathbf{x}_i);\mathbf{\theta},\xi)^{\otimes 2} \quad \mbox{and} \quad
\mathbf{\Omega}  =  \mbox{plim} \frac{1}{n}  \sum_{i=1}^n H((Y_i,\mathbf{x}_i);\mathbf{\theta},\xi), &
\end{eqnarray*}
with $\mbox{plim}$ denoting ``in-probability limit'' as $n \rightarrow \infty$. 
Then, the asymptotic covariance matrix of $(\tilde{\mathbf{\theta}}\trp, \tilde{\xi})\trp$ is
\begin{displaymath}
\mathbf{\Xi} = \left[\begin{array}{cc} \mathbf{\Xi}_{11} & \mathbf{\Xi}_{12} \\ \mathbf{\Xi}_{21} & \Xi_{22}\end{array}\right] = \mathbf{\Omega}^{-1} \mathbf{\Sigma} \left[\mathbf{\Omega}^{-1}\right]\trp,
\end{displaymath}
which could be consistently estimated by $\hat{\mathbf{\Xi}} = \hat{\mathbf{\Omega}}^{-1} \hat{\mathbf{\Sigma}}  [\hat{\mathbf{\Omega}}^{-1}]\trp$, with
\begin{eqnarray*}
& \hat{\mathbf{\Sigma}}  =  \frac{1}{n} \sum_{i=1}^n \mathbf{U}((Y_i,\mathbf{x}_i);\tilde{\mathbf{\theta}},\tilde{\xi})^{\otimes 2} \quad \mbox{and} \quad
\hat{\mathbf{\Omega}}  =  \frac{1}{n}  \sum_{i=1}^n H((Y_i,\mathbf{x}_i);\tilde{\mathbf{\theta}},\tilde{\xi}). &
\end{eqnarray*}
We note that another way of obtaining estimates of $\mathbf{\Sigma}$ and $\mathbf{\Omega}$ is to obtain their theoretical expressions using higher-order moments of the $Y_j$s. These expressions depend on $\mathbf{\theta}$ and $\xi$, so estimates could then be obtained by replacing $\mathbf{\theta}$ and $\xi$ by $\hat{\mathbf{\theta}}$ and $\hat{\xi}$. We also point out that $\Xi_{11}$ is generally not equal to $\Sigma_{11}^{-1}$ when ${\xi}$ is finite. In fact, it is imperative that $\Xi_{11}$ should be used instead of $\Sigma_{11}^{-1}$ since it takes into account the {\em impact} of the estimation of $\xi$.

Applying the Delta-Method \cite{vanderVaart98}, we then have the pivotal quantity result, as $n \rightarrow \infty$, given by
\begin{displaymath}
\frac{Y_0 - \hat{\lambda}_0}{\sqrt{\hat{\lambda}_0[1 + (1+\hat{\lambda}_0)/\tilde{\xi}] + \hat{\lambda}_0^2 \hat{\psi}_0\trp \hat{\mathbf{\Xi}}_{11} \hat{\psi}_0/n}} \stackrel{\bullet}{\sim} N(0,1),
\end{displaymath}
where quantities with\ `$\hat{\ }$'\ are estimates obtained by plugging in $\tilde{\mathbf{\theta}}$ and $\tilde{\xi}$ for $\mathbf{\theta}$ and $\xi$ in their respective expressions.
From this pivotal quantity, it follows that an approximate $100(1-\alpha)\%$ prediction interval for $Y_0$ is given by
\begin{equation}
\label{prediction interval in over-dispersed model}
Y_0 \in \check{\Gamma}_6 \equiv \left[ \hat{\lambda}_0  \pm z_{\alpha/2} \sqrt{\frac{\hat{\lambda}_0(1+\hat{\lambda}_0)}{\hat{\xi}} + {\hat{\lambda}_0} + \frac{1}{n} \hat{\lambda}_0^2 \hat{\psi}_0\trp \hat{\mathbf{\Xi}}_{11} \hat{\psi}_0}\right] \bigcap \mathbb{Z}_{0,+}.
\end{equation}
%
%
Because of the term $1/\hat{\xi}$, when $\xi$ is small, this prediction interval will be wider than the prediction intervals under the purely Poisson regression model. As $\xi \rightarrow \infty$, which makes the model approach the Poisson regression model, then this prediction interval will approach $\check{\Gamma}_1$.  

\begin{figure}[h]
\caption{Scatterplot of deaths data until May 15, 2020 together with the prediction intervals under the over-dispersed 5th-order Poisson regression model.}
\label{Scatterplot until May 15 with over-dispersed pred ints}
\includegraphics[width=\textwidth]{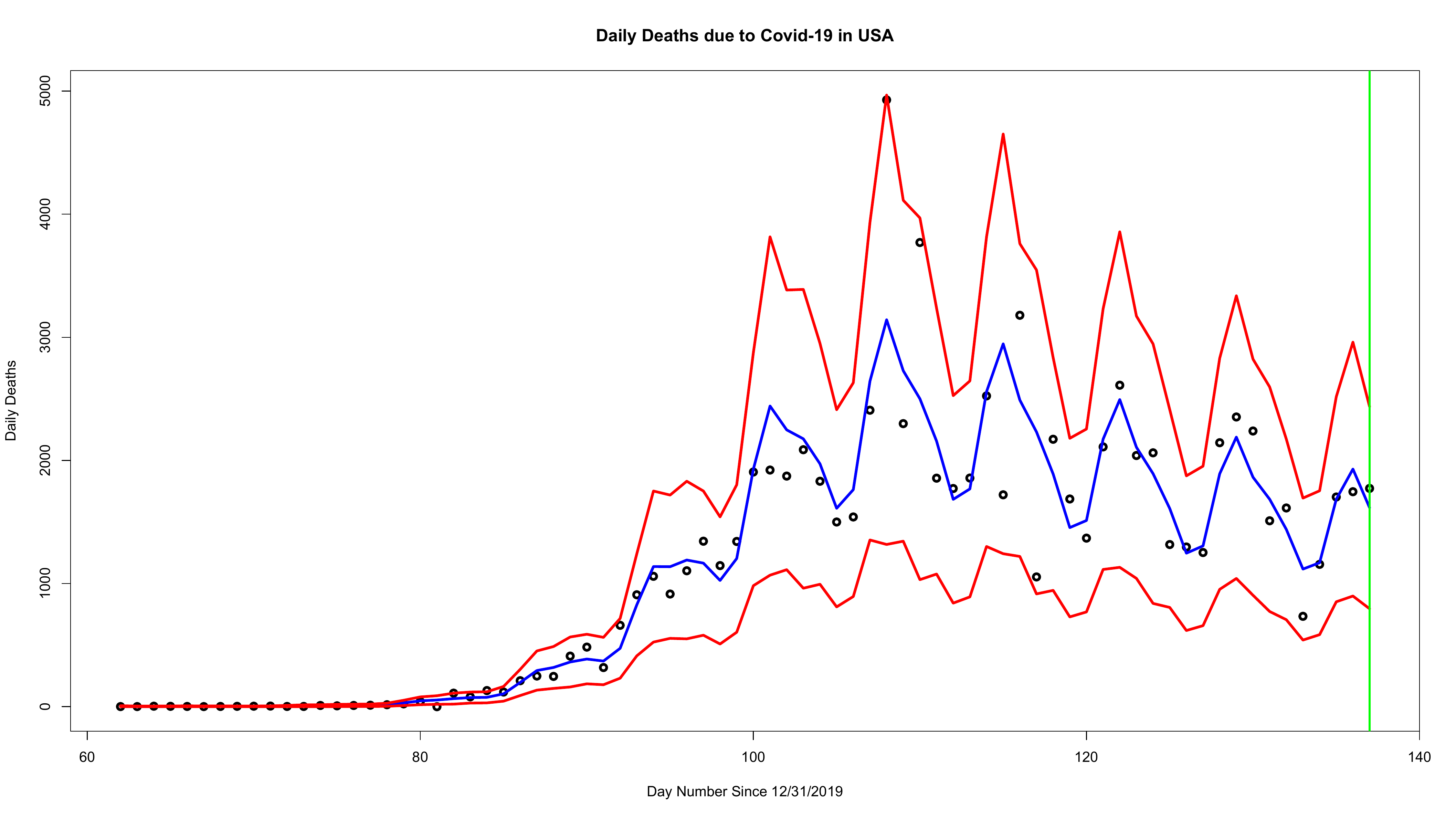}
\end{figure}

\begin{figure}[h]
\caption{Prediction intervals for daily and cumulative deaths under the over-dispersed 5th-order Poisson regression model using data from ${\tt DayNum} = 62$ (March 1, 2020) until ${\tt DayNum} = 137$ (May 15, 2020). Red points in the plot were the subsequent observed number of deaths after ${\tt Daynum} = 137$. The last value of {\tt DayNum} is 154 coinciding with the end of May 31, 2020.}
\label{fig-complete plot daily}
\begin{tabular}{cc}
\includegraphics[width=3in]{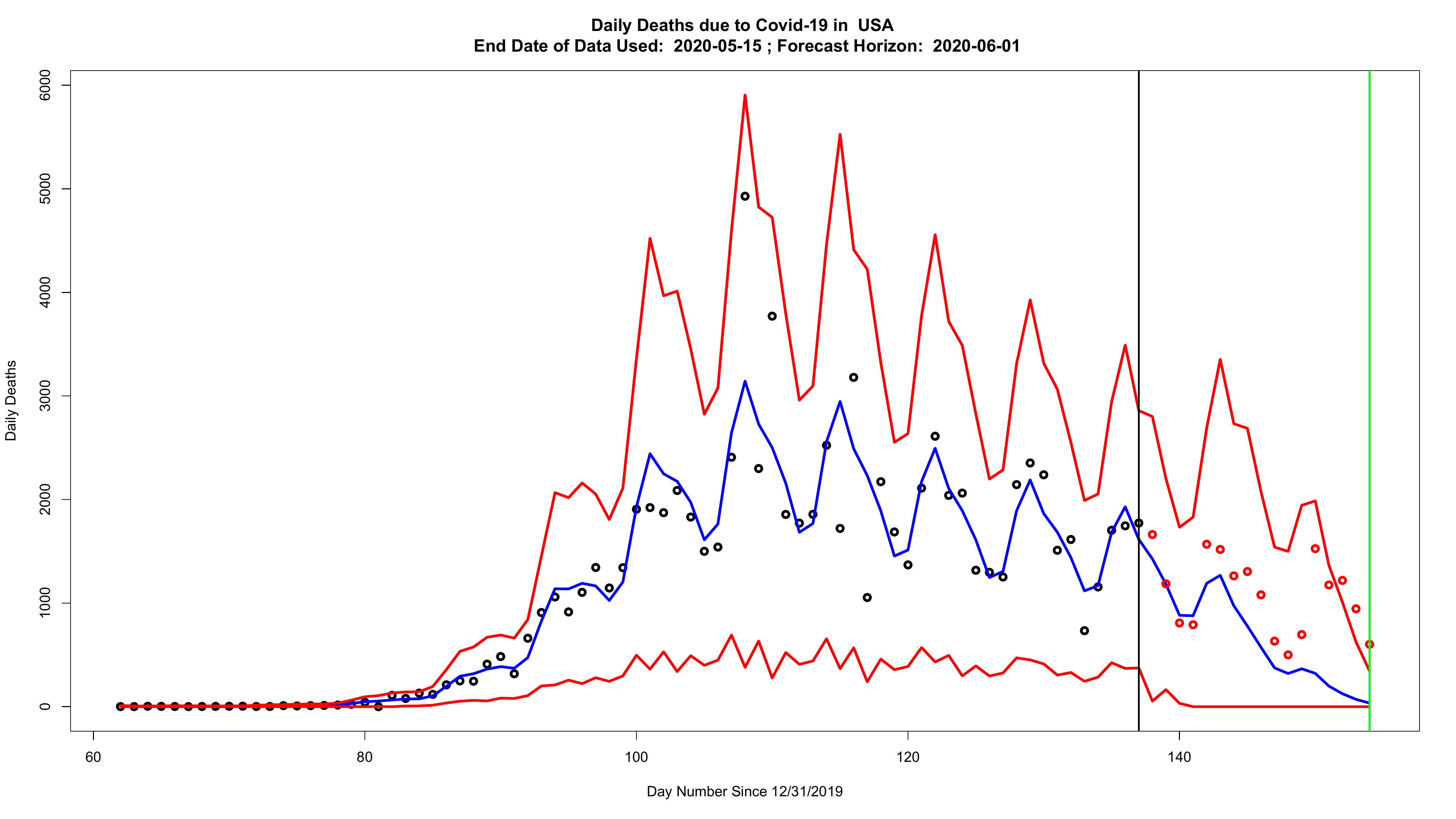} &
%
\includegraphics[width=3in]{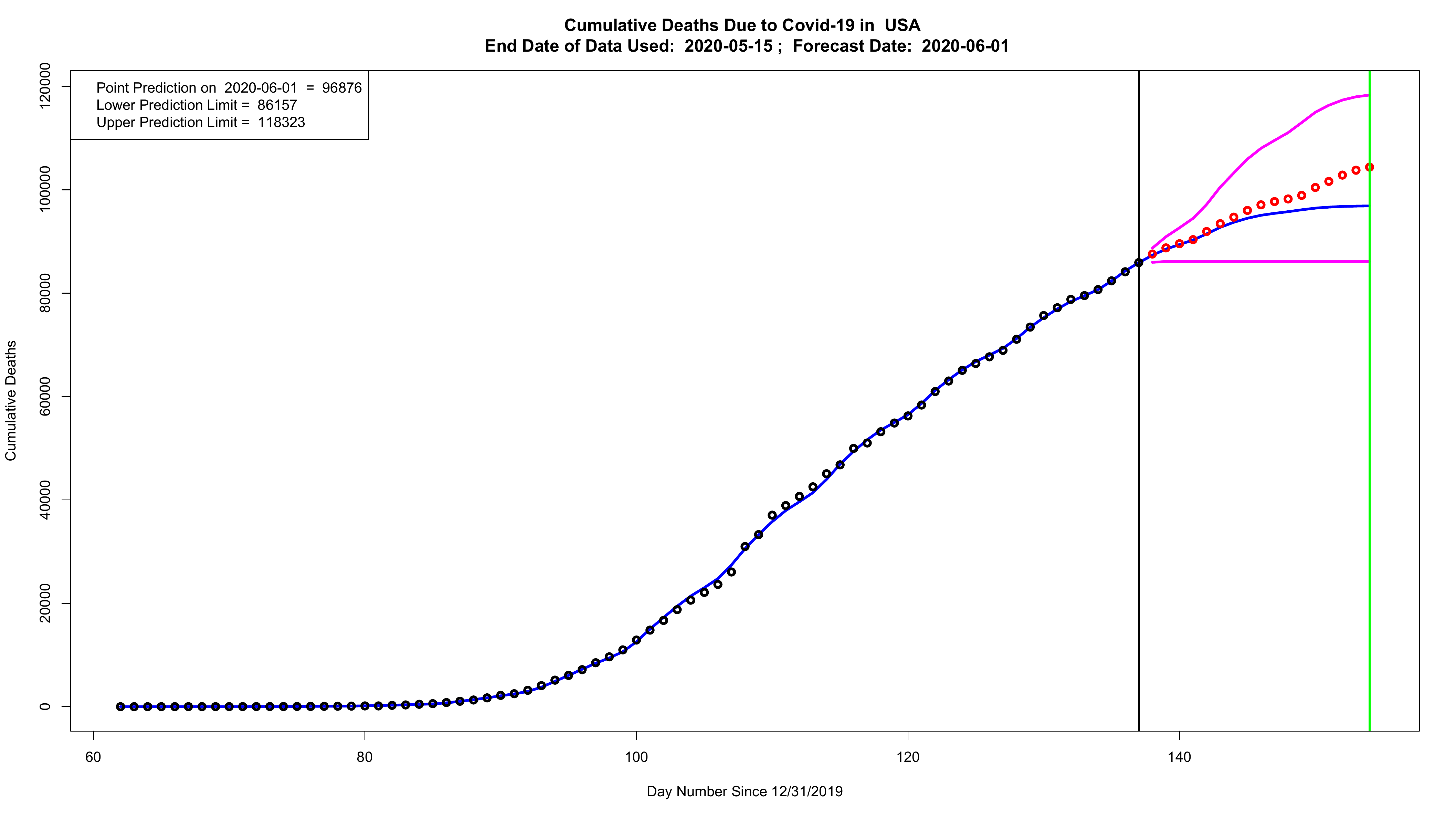}
\end{tabular}
\end{figure}

Implementing this procedure based on this over-dispersed Poisson regression model, we first examine the prediction intervals at each of the observed {\tt DayNum} values, which are shown in Figure \ref{Scatterplot until May 15 with over-dispersed pred ints}. We now see that these approximate 95\% prediction intervals cover most of the observed daily deaths. This indicates that the over-dispersed Poisson regression model provides a better fit to the observed daily deaths data than the purely Poisson regression model whose approximate 95\% prediction intervals are shown in Figure \ref{Scatterplot until May 15 with pred ints}. The estimate of $\xi$ turned out to be $\hat{\xi} = 16.89016$.
The left-panel of Figure \ref{fig-complete plot daily} is the scatterplot of daily deaths, but now including the actual observed values after {\tt DayNum} 137 and until 154 (these are the red dots) and the prediction intervals for the daily deaths past 137. Observe that the prediction intervals for {\tt DayNum} between 62 and 137 are wider than those in Figure \ref{Scatterplot until May 15 with over-dispersed pred ints} and the reason for this is the prediction coefficient used is now adjusted for the goal of constructing a prediction interval for $S_{154}$. In this case, $\alpha^* =  0.003012$. The right-panel of Figure \ref{fig-complete plot daily} displays the prediction interval for $S_{154}$, given $S_{137}$; in fact, this also displays the prediction intervals for $S_{k}, k=138,139,\ldots,153,$ given $S_{137}$. The red dots are the actual observed values past {\tt DayNum} equal to 137. The predicted cumulative deaths on ${\tt DayNum} = 154$ was $\hat{S}_{154} = 96876$ and the conservative prediction interval for $S_{154}$ was $[86157, 118323]$. In both of these plots, notice that the 5th-order prediction model did not perform well past ${\tt DayNum} =  150$, though the prediction interval for $S_{154}$ did cover what was {\em actually} observed, which was 104383. From Figure \ref{fig-complete plot daily}, an elevated number of daily deaths occurred starting at ${\tt DayNum} = 150$ (May 28). The grim milestone of 100000 cumulative deaths due to \Covid\ in the US was also surpassed on this day.
 
\subsection{Sensitivity Analysis}

\begin{figure}[h]
\caption{Prediction intervals for daily and cumulative deaths under the over-dispersed 5th-order Poisson regression model using data from ${\tt DayNum} = 62$ (March 1, 2020) until ${\tt DayNum} = 145$ (May 23, 2020). Red points in the plot were the subsequent observed number of deaths and cumulative deaths after ${\tt Daynum} = 145$. The last value of {\tt DayNum} is 154 coinciding with the end of May 31, 2020.}
\label{fig-complete plot daily 5/23}
\begin{tabular}{cc}
\includegraphics[width=3in]{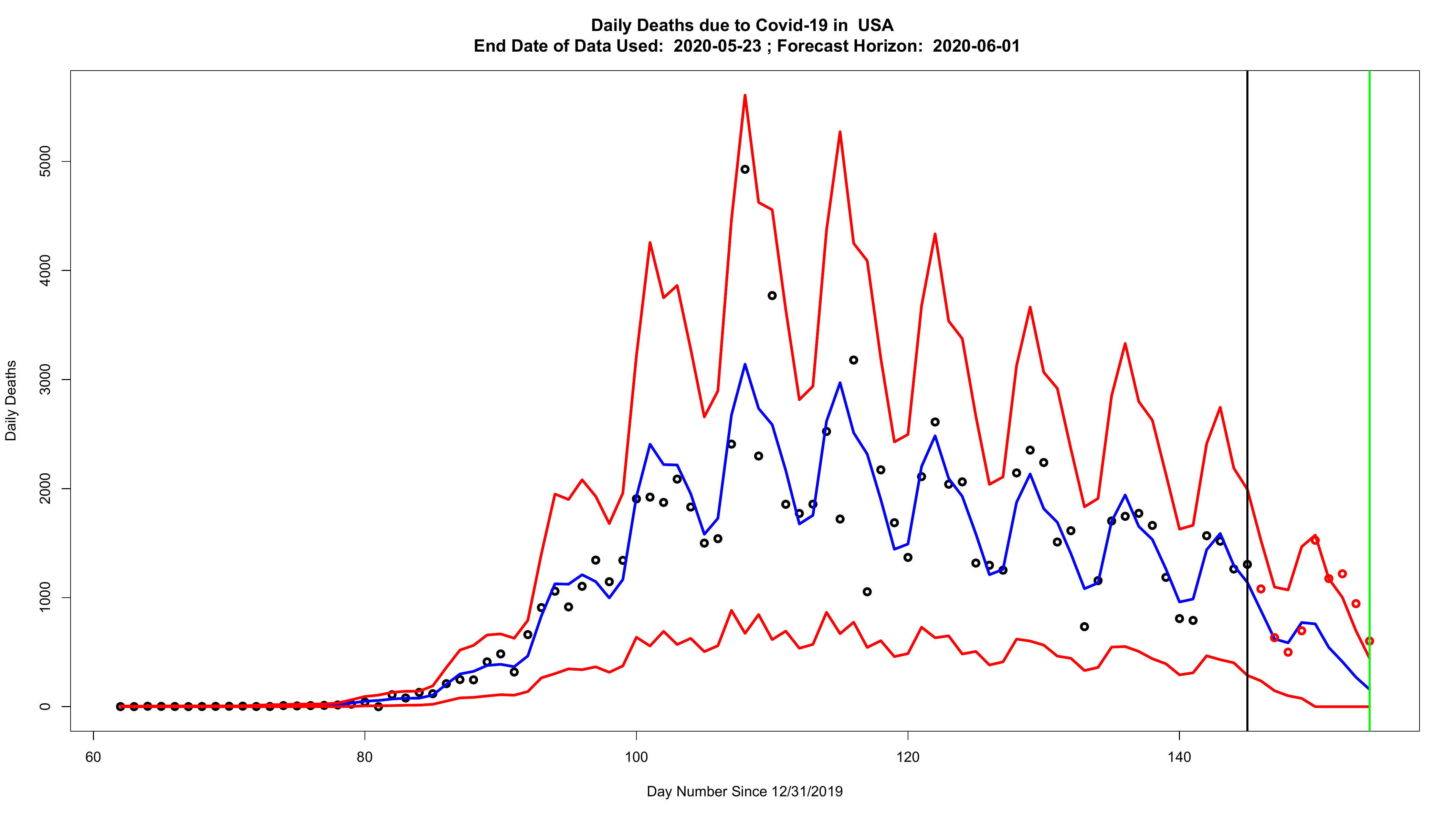} &
%
\includegraphics[width=3in]{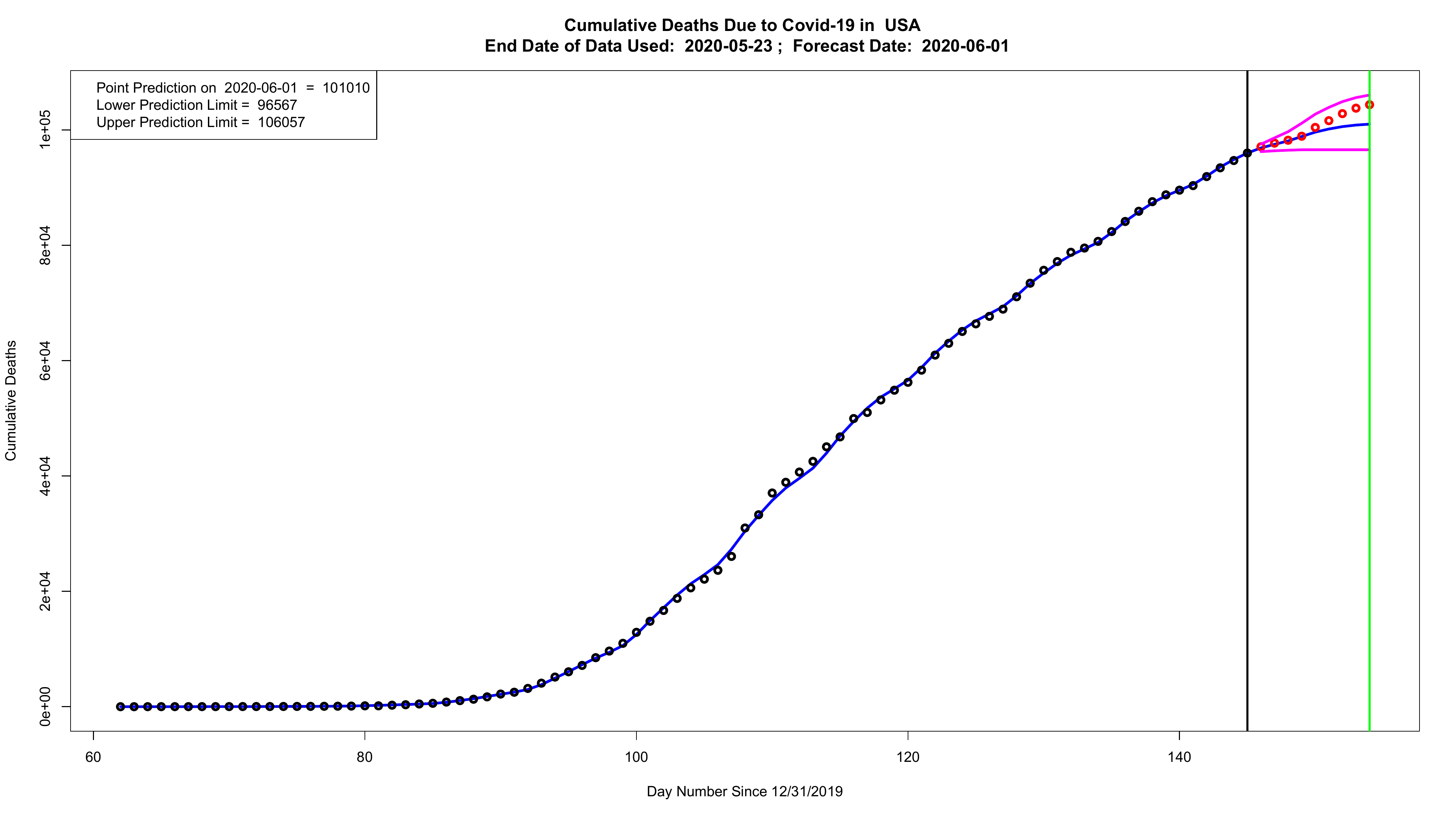}
\end{tabular}
\end{figure}

\begin{figure}[h]
\caption{Prediction curve and intervals for daily deaths and observed cumulative deaths together with its associated fitted prediction curve under the over-dispersed 5th-order Poisson regression model using data from ${\tt DayNum} = 62$ (March 1, 2020) until ${\tt DayNum} = 154$ (June 1, 2020)..}
\label{fig-complete plot daily 6/01}
\begin{tabular}{cc}
\includegraphics[width=3in]{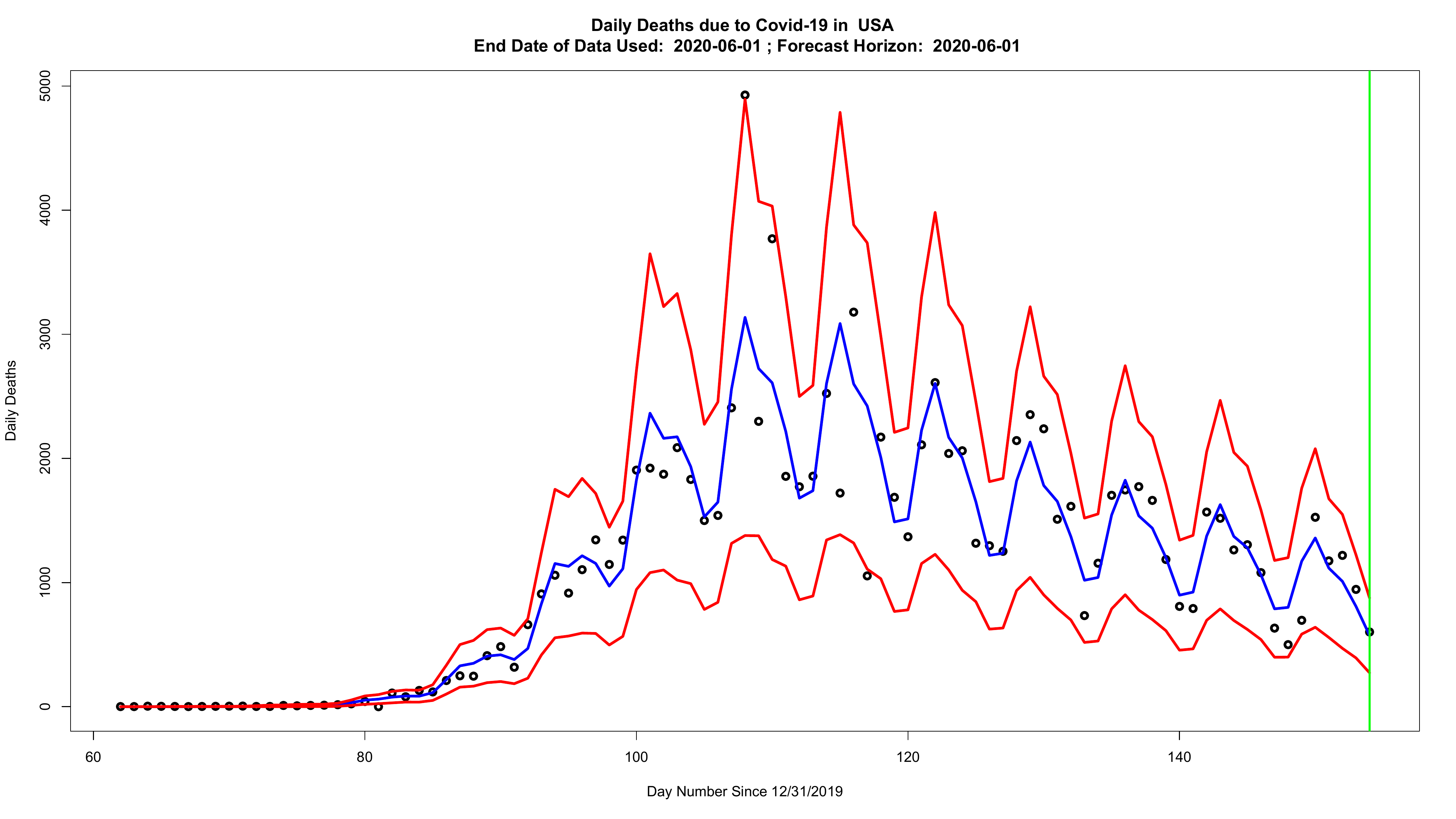} &
%
\includegraphics[width=3in]{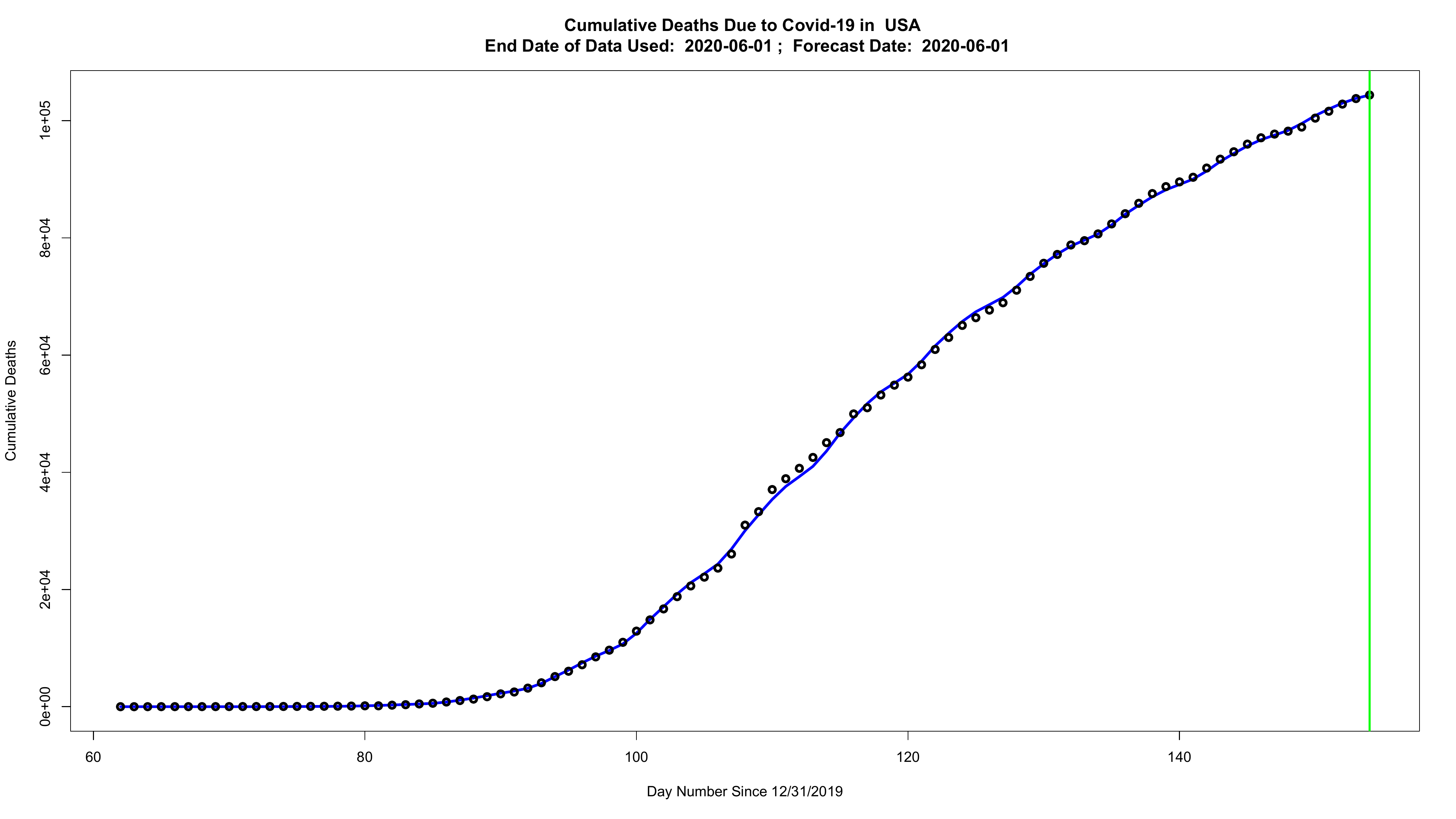}
\end{tabular}
\end{figure}

\begin{figure}[h]
\caption{Predictions and prediction intervals for the cumulative deaths by June 1, 2020 when the data used in the model fitting is up until the different days between May 15, 2020 and June 1, 2020. The gray horizontal line represents the actual observed cumulative number of deaths on June 1, 2020, which was 104383.}
\label{plot-sensitivity}
\includegraphics[width=\textwidth,height=2in]{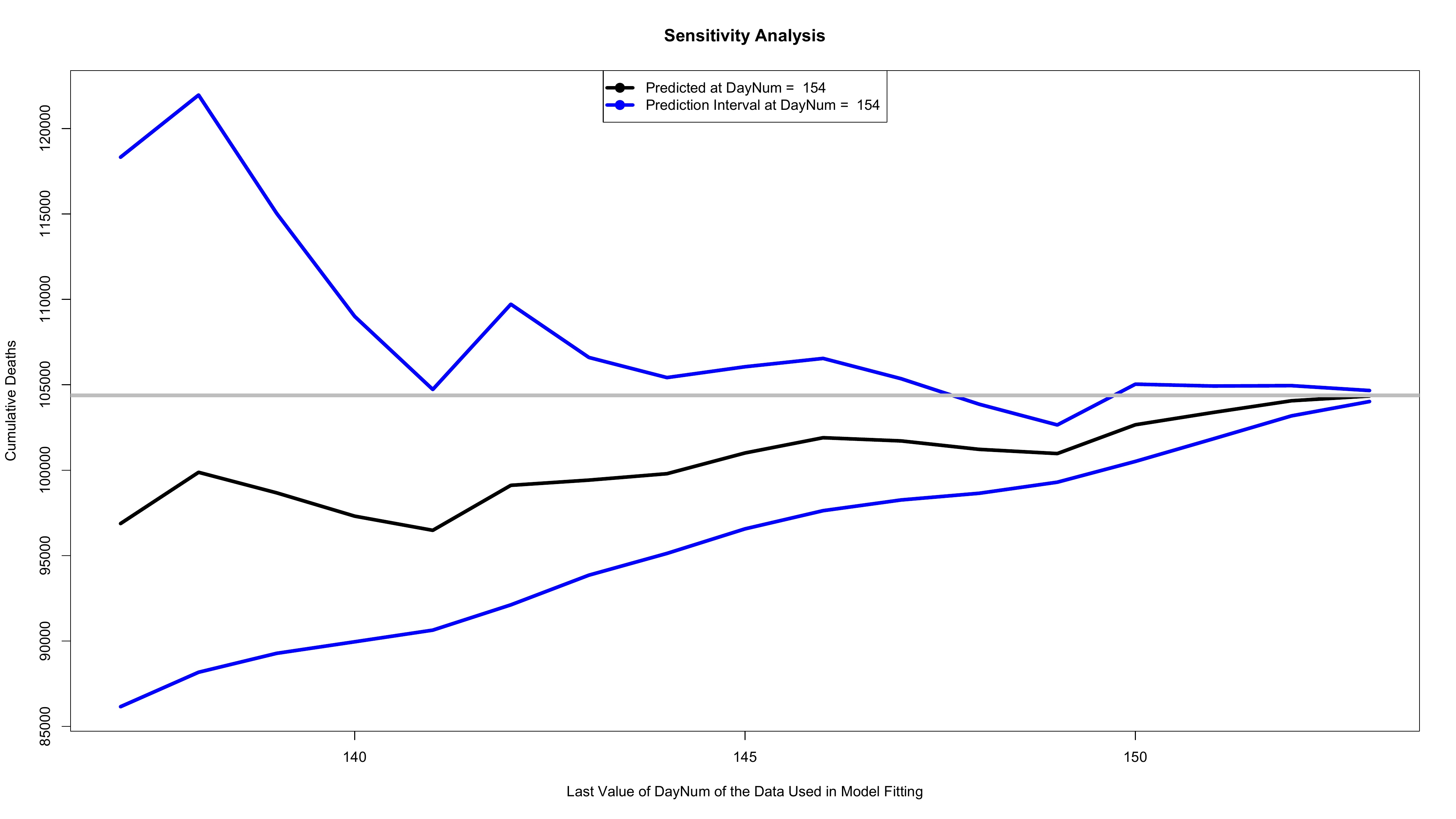}
\end{figure}

To partly assess the sensitivity of the procedure, if we had used the data until ${\tt DayNum} = 145$ (May 23rd), the predicted value for $S_{154}$ is $\hat{S}_{154} = 101010$ and the conservative 95\% prediction interval, given $S_{145}$, is $[96567, 106057]$. The associated plots in this case are provided in Figure \ref{fig-complete plot daily 5/23}.  On the other hand, if we had used the data until June 1, 2020 ({\tt DayNum} = 154), the fitted value is 104383, which coincided with the {\em observed} cumulative number of deaths. That the fitted cumulative number of deaths and the observed cumulative number of deaths on the last day were equal is actually a consequence of the estimating equation, hence in hindsight is not a surprising result. The associated plots in this case are provided in Figure \ref{fig-complete plot daily 6/01}. Observe in the right-panel of Figure \ref{fig-complete plot daily 6/01} that the model for the cumulative deaths based on this 5th-order model is quite excellent, lending support to our novel approach of modeling the daily deaths data, instead of the cumulative deaths data, for the purpose of making predictions for the cumulative deaths. We summarize the results of this sensitivity analysis in Table \ref{summ-sensitivity} and Figure \ref{plot-sensitivity}, where we report the predicted values and the prediction intervals for $S_{154}$ under scenarios where data used in the model fitting is up to the different days from {\tt DayNum}  equal to 137 up to 153. Based on this analysis, the fifth-order model appears to possess stability since the predictions and the prediction intervals for $S_{154}$ remain somewhat consistent as the amount of data being used in the model fitting varies.  As to be expected, note that as the forecasting horizon shortens, then the prediction interval also narrows. This is the case, {\em even} though we still considered the outlier on April 16th, which was the adjusted {\tt Deaths} data point, as a legitimate observation.

{\em However}, any model, especially higher-order models, will have a breakdown point in the sense of yielding seemingly unreasonable predictions, perhaps due to a long forecasting horizon, insufficient amount of data,  or wildly changing data points drastically altering estimates which highly impact forecasts. For this 5th-order model, it appears to break down when the data used is on or before {\tt DayNum} 132. One possible cause appears to be sharp increases and decreases in the observed daily deaths. For instance, on ${\tt DayNum}=120$ the reported daily deaths was 1369, but on the next two days they were 2110 and 2611, and these led to a huge jump in the predicted value for $S_{154}$. Also, from {\tt DayNum} ranging from 125 to 133, the reported daily deaths were 1317, 1297, 1252, {\bf 2144}, {\bf 2353}, {\bf 2239}, 1510, 1624, and 734, and the predictions were highly unstable and only started to stabilize after ${\tt DayNum} = 132$. This is a clear warning on the danger of fitting higher-order models or extrapolating with a long forecasting horizon.  As the adage goes, attributed to Niels Bohr  \cite{Ulam}, with similar versions attributed to Mark Twain, Yogi Berra,  and others: 
{\em  It is difficult to make predictions, especially about the future.}
%
%

\begin{table}[h]
\caption{Examining the fitted and prediction intervals for the cumulative number of deaths  $S_{154}$ at {\tt DayNum} = 154 (June 1, 2020) under different scenarios for the amount of data used in the model-fitting.}
\label{summ-sensitivity}
\centering
\begin{tabular}{rrrrr}
  \hline
 & Until\_DayNum & Predicted & PI: Lower & PI: Upper \\ 
  \hline
1 & 137.00 (5/15) & 96876.00 & 86157.00 & 118323.00 \\ 
  2 & 138.00 & 99878.00 & 88174.00 & 121963.00 \\ 
  3 & 139.00 & 98676.00 & 89281.00 & 115037.00 \\ 
  4 & 140.00 & 97311.00 & 89957.00 & 109003.00 \\ 
  5 & 141.00 & 96482.00 & 90639.00 & 104727.00 \\ 
  6 & 142.00 & 99116.00 & 92119.00 & 109717.00 \\ 
  7 & 143.00 & 99421.00 & 93857.00 & 106601.00 \\ 
  8 & 144.00 & 99796.00 & 95130.00 & 105423.00 \\ 
  9 & 145.00 & 101010.00 & 96567.00 & 106057.00 \\ 
  10 & 146.00 & 101903.00 & 97632.00 & 106545.00 \\ 
  11 & 147.00 & 101715.00 & 98260.00 & 105356.00 \\ 
  12 & 148.00 & 101221.00 & 98651.00 & 103863.00 \\ 
  13 & 149.00 & 100975.00 & 99299.00 & 102651.00 \\ 
  14 & 150.00 & 102661.00 & 100515.00 & 105037.00 \\ 
  15 & 151.00 & 103384.00 & 101840.00 & 104928.00 \\ 
  16 & 152.00 & 104066.00 & 103182.00 & 104951.00 \\ 
  17 & 153.00 (5/31) & 104344.00 & 104022.00 & 104665.00 \\ 
   \hline
\end{tabular}
\end{table}

\subsection{Forecasting Fifteen Days Forward}

With dire thoughts of the inherent dangers of forecasting over a long horizon, and fully cognizant of the many eventualities (e.g., re-opening of economy; nationwide protests and riots due to police brutality; changing hotspots; adjustments on counts; etc.), which we are not taking into account in order to be purely data-driven, but which could drastically alter trajectories of daily and cumulative deaths, on the basis of the observed data up to July 2, 2020, in which the cumulative deaths 128062, we seek to forecast the cumulative deaths fifteen days forward, which will be July 16, 2020.  We should mention that on June 26th, there was an  adjustment of 1854 which occurred ``following a state review of death certificates and prior outbreaks'' in the State of New Jersey \cite{NJ625}. Together with the 3778 adjustment made by the State of New York on April 16th \cite{BBC415}, this is the second documented non-trivial adjustment made on the daily deaths counts. We provide two point predictions and prediction regions for the target date: -- the first one based on considering the observed {\tt Deaths} values on April 16th and June 26th as legitimate values, and the second one based on re-allocating the adjustment values on those days proportional to the observed {\tt Deaths} on the days on or before the day of adjustment. Without additional information, such a proportional re-allocation of the adjustment values appears to be most sensible, though this approach is not immune to criticism.

\begin{figure}
\caption{Based on the available data until July 2, 2020, point predictions and prediction intervals of the deaths and cumulative deaths by July 16, 2020 (${\tt DayNum} = 199$) based on an analyses where adjustment values were not re-allocated (top plots) and re-allocated (bottom plots).}
\label{July 16 Predictions}
\centering
\begin{tabular}{cc}
\includegraphics[width=3in,height=3in]{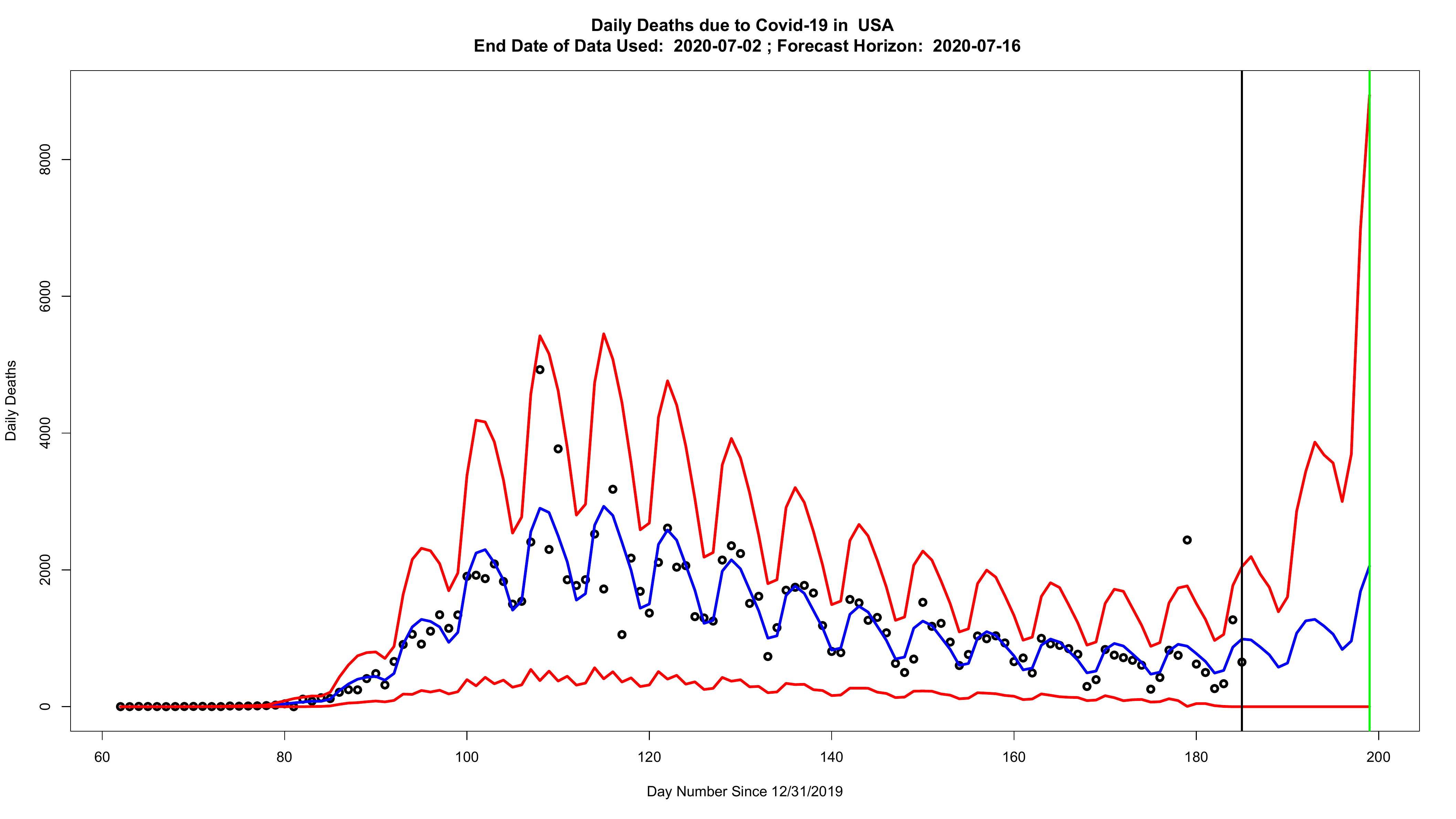}  &
\includegraphics[width=3in,height=3in]{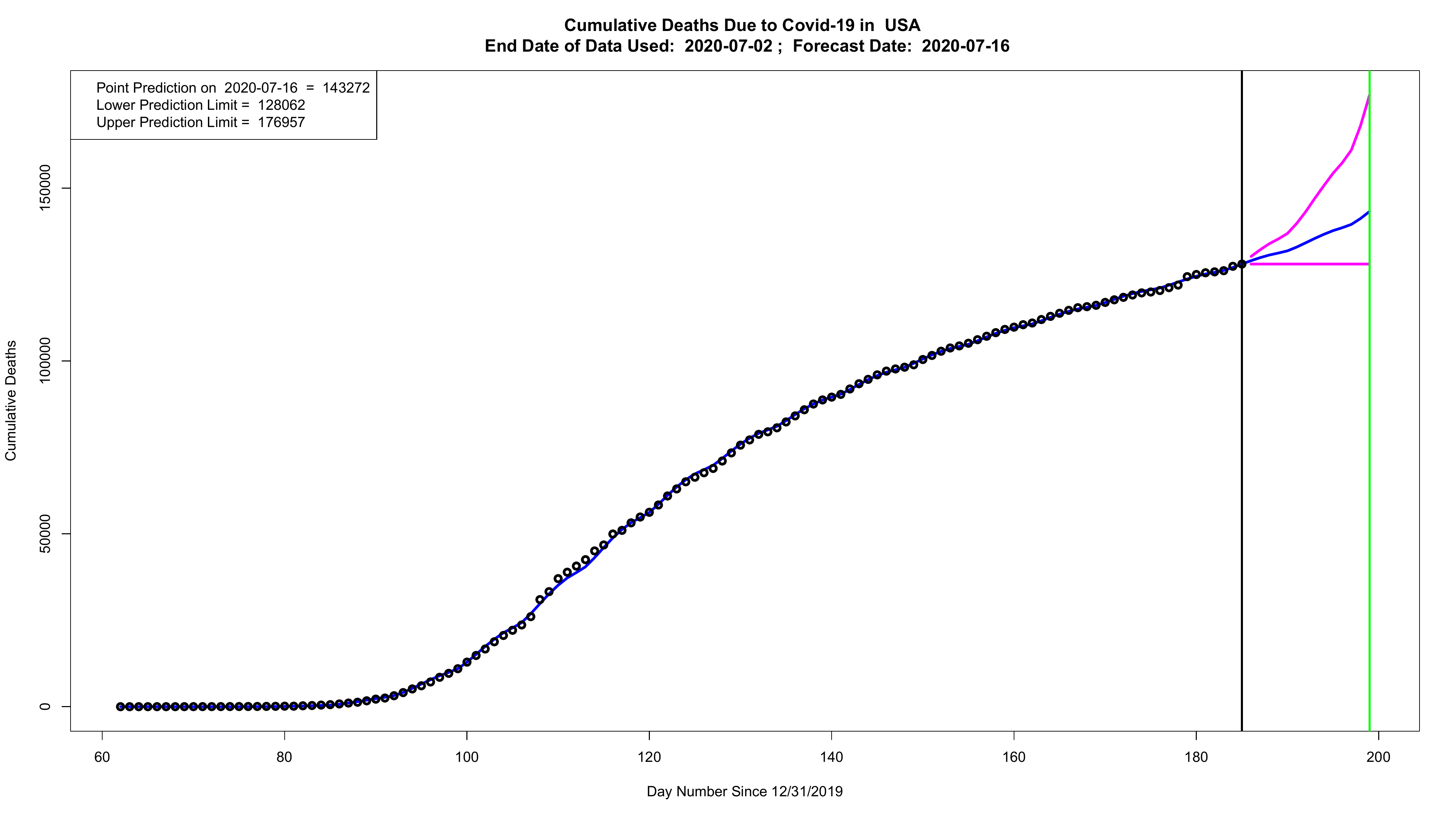} \\
\includegraphics[width=3in,height=3in]{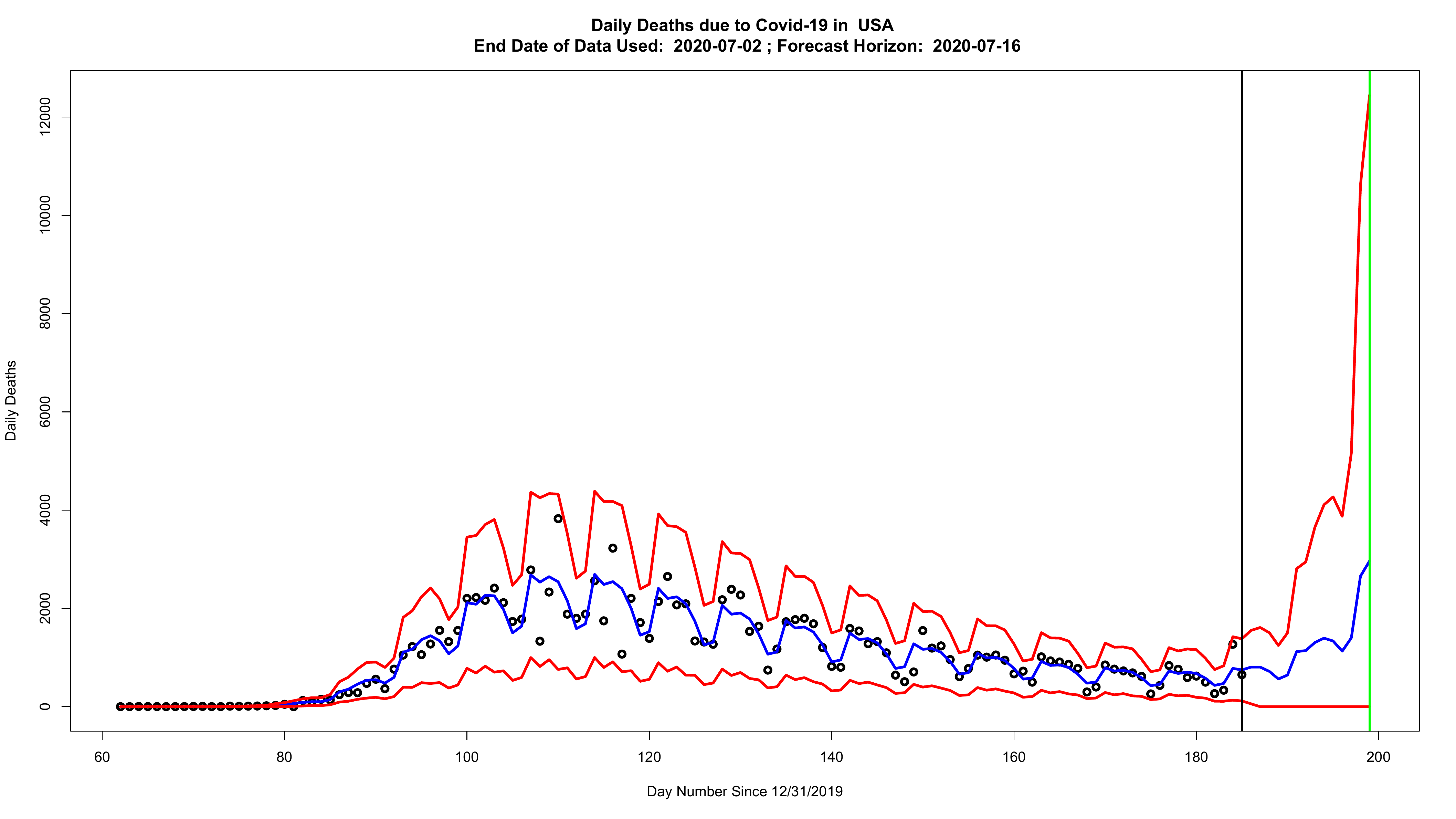}  &
\includegraphics[width=3in,height=3in]{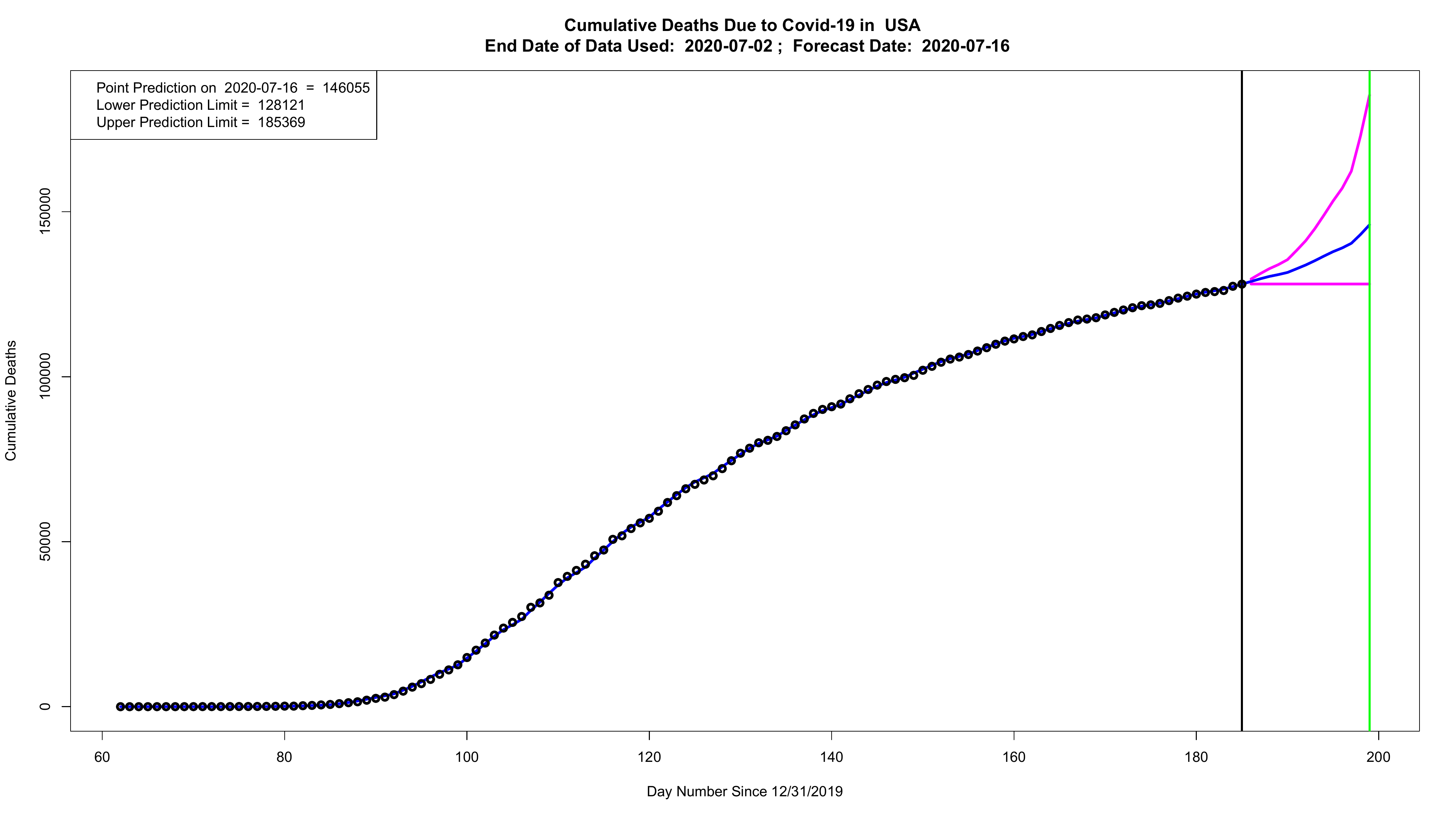} 
\end{tabular}
\end{figure}

Figure \ref{July 16 Predictions} presents the point prediction and the prediction interval for July 16, 2020 based on these two analyses on with a 5th-order model. When the adjustment values are not re-allocated, the predictions and prediction intervals are depicted in the two top plots, whereas when they are re-allocated, they are in the two bottom plots. With no re-allocation, the point prediction is 143272 together with an associated  prediction interval  of $[128062, 176957]$ for the cumulative deaths. With re-allocation, the point prediction is 146055 with an associated prediction interval of $[128121, 185369]$. Observe that without re-allocation, the observed {\tt Deaths} of 2437 on June 26th fell outside of the prediction interval on that date, while the observed {\tt Deaths} of 4928 on April 16th barely fell inside the prediction interval on that date. Notice the wider prediction intervals for {\tt Deaths} when no re-allocations were performed compared to those with re-allocations for the observed {\tt DayNum} values. Observe also the widening prediction intervals as we go farther away from July 2nd, indicating high uncertainty on what may happen moving forward. Interestingly, the prediction interval for the cumulative deaths on July16th when re-allocations were performed is wider than that without re-allocations, and the point prediction is also tad higher. Ominously, notice that the prediction curve for {\tt Deaths} appears to be acquiring an increasing trend past July 2nd, in contrast to the decreasing trend from {\tt DayNum} 120 (April 28th) to 183 (June 30th). It remains to be seen if this is the effect of the lessening of social distancing guidelines, re-opening of business establishments and beaches, people gathering because of the current social unrest, or changing hotspots in the country. Of course, in forecasting settings with new data points accruing frequently -- daily in this \Covid\ pandemic -- forecasts should be updated as each new data point accrues.  We intend to provide a publicly-accessible software applet to enable interested users to update forecasts with the latest updated data.

\section{Concluding Remarks}
\label{sec-Concluding Remarks}

Motivated by the \Covid\ pandemic, we examined the problem of constructing prediction regions for a Poisson distributed random variable, both under the no-covariate (that is, intercept only) and with-covariate settings. We compared the performances of the different prediction regions through simulation studies. In the regression setting, we also introduced an over-dispersed Poisson regression model upon observing over-dispersion in the \Covid\ reported {\tt Deaths} data relative to a purely Poisson regression  model. With the ultimate goal of predicting cumulative deaths due to \Covid\ at a future date, we first studied how to construct prediction intervals for the daily deaths data, and then utilized these prediction intervals to construct the prediction interval for the cumulative deaths. The final fitted models involved a 5th-order model in the variable {\tt DayNum}, and also included the factor variable {\tt Day}. Based on data until July 2, 2020, prediction and prediction interval for the July 16, 2020 cumulative deaths were obtained. The methodologies developed have the potential to be used in the monitoring of daily and cumulative deaths during epidemics or pandemics through the construction of prediction regions, which could then be used by decision-makers regarding implementation of social distancing/easing guidelines and deciding on the closure/opening of business, educational, government, and other establishments. However, further studies are needed to compare our methodologies to other methods that have been proposed during this pandemic.

The prediction and prediction region procedures we developed also possess limitations. First, in contrast to the Susceptible-Exposed-Infected-Recovered (SEIR) compartment model cf., \cite{Allen17}, based on a continuous-time Markov Chain, our model does not posit an upper bound to the number of people that could die, which clearly is not the case.  Second, especially since it involves higher-order terms in {\tt DayNum}, they are highly sensitive to outliers, such as when huge adjustments are made as in the cases for the States of New York and New Jersey \cite{WSJ416,NJ625}. It would be desirable to develop procedures that are robust to such non-trivial adjustments, or to procedures that impose constraints on the rate of increase or decrease of the prediction curve via regularization to curb the influential impact of such adjustments, though this will entail developing new theory for the construction of prediction regions. Due to these first two limitations, the proposed methods are not suitable for use in long-horizon forecasting, hence our decision to simply forecast 15 days forward. Third, the procedures are not adaptive in its choice of the prediction model. Possible improvements may occur by choosing the prediction model in a data-dependent manner, but then model choice uncertainty needs to be accounted for in constructing prediction regions.  Fourth, there could be an advantage in utilizing other bases functions to transform the variable {\tt DayNum}, such as by using Laguerre polynomials, Legendre polynomials, trigonometric functions, or even splines or wavelets. These limitations of the proposed methods generate several potential research avenues for further studies.

\bibliographystyle{plain}
\bibliography{PoissonRegression}


\appendix

\section{Appendices: Supplementary Materials}

\subsection{About the European CDC}
\label{supp-ECDC}

Copyright Statement from the European CDC Website can be found in \url{https://www.ecdc.europa.eu/en/copyright}.
%
%
%
%
%
%
Information about the European CDC \Covid\ Data Set as stated in ECDC Website: {\em The downloadable data file is updated daily and contains the latest available public data on \Covid. Each row/entry contains the number of new cases reported per day and per country. You may use the data in line with ECDC's copyright policy.}


\subsection{Tables of Simulation Results}
\label{sec-supplement}



\begin{table}
\caption{Simulated coverage probabilities, mean of the lengths, and standard deviation of the lengths of the prediction intervals $\tilde{\Gamma}_0$ and $\check{\Gamma}_j, j=1,2,3,4,5$, for different $\lambda$'s and $n$'s. For each combination of $(n,\lambda)$, 10000 replications were performed. For $\lambda=1$.}
\label{table-simulation 1 results}
\centering
\begin{tabular}{rrrrrrrr}
  \hline
 & n & Gam0CP & Gam1CP & Gam2CP & Gam3CP & Gam4CP & Gam5CP \\ 
  \hline
1 & 5.00 & 92.01 & 93.47 & 84.29 & 89.33 & 89.79 & 93.62 \\ 
  2 & 10.00 & 93.18 & 94.06 & 83.29 & 92.10 & 92.45 & 94.59 \\ 
  3 & 15.00 & 94.08 & 94.81 & 83.03 & 93.10 & 93.15 & 94.48 \\ 
  4 & 20.00 & 94.05 & 95.12 & 82.95 & 93.75 & 93.77 & 95.06 \\ 
  5 & 30.00 & 94.72 & 94.89 & 79.72 & 94.36 & 94.36 & 95.00 \\ 
  6 & 50.00 & 94.46 & 94.81 & 77.30 & 94.35 & 94.37 & 94.75 \\ 
  7 & 70.00 & 94.91 & 95.33 & 77.26 & 94.92 & 95.06 & 95.26 \\ 
  8 & 100.00 & 95.01 & 94.83 & 75.31 & 95.03 & 94.87 & 95.07 \\ 
   \hline
 & n & Gam0ML & Gam1ML & Gam2ML & Gam3ML & Gam4ML & Gam5ML \\ 
  \hline
1 & 5.00 & 2.41 & 2.61 & 3.32 & 2.10 & 2.17 & 2.70 \\ 
  2 & 10.00 & 2.42 & 2.56 & 3.18 & 2.29 & 2.30 & 2.58 \\ 
  3 & 15.00 & 2.44 & 2.55 & 3.12 & 2.34 & 2.35 & 2.55 \\ 
  4 & 20.00 & 2.44 & 2.54 & 3.10 & 2.37 & 2.38 & 2.53 \\ 
  5 & 30.00 & 2.44 & 2.45 & 2.93 & 2.40 & 2.40 & 2.49 \\ 
  6 & 50.00 & 2.44 & 2.46 & 2.83 & 2.43 & 2.43 & 2.49 \\ 
  7 & 70.00 & 2.46 & 2.48 & 2.86 & 2.44 & 2.44 & 2.49 \\ 
  8 & 100.00 & 2.48 & 2.44 & 2.72 & 2.46 & 2.46 & 2.48 \\ 
   \hline
 & n & Gam0SL & Gam1SL & Gam2SL & Gam3SL & Gam4SL & Gam5SL \\ 
  \hline
1 & 5.00 & 0.92 & 0.97 & 0.75 & 0.85 & 0.88 & 0.99 \\ 
  2 & 10.00 & 0.71 & 0.75 & 0.46 & 0.68 & 0.69 & 0.71 \\ 
  3 & 15.00 & 0.62 & 0.58 & 0.37 & 0.61 & 0.61 & 0.63 \\ 
  4 & 20.00 & 0.58 & 0.54 & 0.33 & 0.58 & 0.58 & 0.57 \\ 
  5 & 30.00 & 0.54 & 0.51 & 0.27 & 0.54 & 0.54 & 0.54 \\ 
  6 & 50.00 & 0.51 & 0.50 & 0.38 & 0.51 & 0.51 & 0.51 \\ 
  7 & 70.00 & 0.50 & 0.50 & 0.35 & 0.51 & 0.50 & 0.51 \\ 
  8 & 100.00 & 0.50 & 0.50 & 0.45 & 0.50 & 0.50 & 0.50 \\ 
   \hline
\end{tabular}
\end{table}
\begin{table}
\caption{Table \ref{table-simulation 1 results} continued. For $\lambda=5$.}
\centering
\begin{tabular}{rrrrrrrr}
  \hline
 & n & Gam0CP & Gam1CP & Gam2CP & Gam3CP & Gam4CP & Gam5CP \\ 
  \hline
1 & 5.00 & 92.47 & 94.99 & 94.01 & 88.29 & 89.03 & 94.62 \\ 
  2 & 10.00 & 92.93 & 94.77 & 93.53 & 91.33 & 91.61 & 94.24 \\ 
  3 & 15.00 & 93.93 & 95.40 & 93.91 & 92.96 & 93.08 & 94.93 \\ 
  4 & 20.00 & 94.45 & 95.68 & 94.02 & 93.80 & 93.78 & 95.29 \\ 
  5 & 30.00 & 94.23 & 95.56 & 94.18 & 93.81 & 93.90 & 94.71 \\ 
  6 & 50.00 & 94.63 & 95.52 & 94.33 & 94.33 & 94.26 & 94.82 \\ 
  7 & 70.00 & 94.68 & 95.53 & 94.52 & 94.39 & 94.35 & 94.73 \\ 
  8 & 100.00 & 94.48 & 95.50 & 94.38 & 94.44 & 94.66 & 94.75 \\ 
   \hline
 & n & Gam0ML & Gam1ML & Gam2ML & Gam3ML & Gam4ML & Gam5ML \\ 
  \hline
1 & 5.00 & 7.58 & 8.57 & 8.53 & 6.56 & 6.75 & 8.32 \\ 
  2 & 10.00 & 7.59 & 8.19 & 8.16 & 7.15 & 7.21 & 7.99 \\ 
  3 & 15.00 & 7.60 & 8.04 & 8.03 & 7.32 & 7.35 & 7.88 \\ 
  4 & 20.00 & 7.61 & 7.95 & 7.97 & 7.39 & 7.42 & 7.81 \\ 
  5 & 30.00 & 7.62 & 7.93 & 7.93 & 7.48 & 7.50 & 7.76 \\ 
  6 & 50.00 & 7.64 & 7.91 & 7.91 & 7.56 & 7.56 & 7.71 \\ 
  7 & 70.00 & 7.64 & 7.92 & 7.89 & 7.58 & 7.58 & 7.69 \\ 
  8 & 100.00 & 7.65 & 7.92 & 7.90 & 7.61 & 7.61 & 7.69 \\ 
   \hline
 & n & Gam0SL & Gam1SL & Gam2SL & Gam3SL & Gam4SL & Gam5SL \\ 
  \hline
1 & 5.00 & 0.98 & 1.02 & 0.98 & 0.89 & 0.88 & 1.10 \\ 
  2 & 10.00 & 0.76 & 0.78 & 0.77 & 0.72 & 0.74 & 0.80 \\ 
  3 & 15.00 & 0.68 & 0.73 & 0.74 & 0.66 & 0.66 & 0.71 \\ 
  4 & 20.00 & 0.62 & 0.68 & 0.69 & 0.61 & 0.61 & 0.64 \\ 
  5 & 30.00 & 0.57 & 0.61 & 0.62 & 0.56 & 0.57 & 0.58 \\ 
  6 & 50.00 & 0.52 & 0.51 & 0.51 & 0.52 & 0.52 & 0.52 \\ 
  7 & 70.00 & 0.50 & 0.44 & 0.46 & 0.50 & 0.50 & 0.50 \\ 
  8 & 100.00 & 0.49 & 0.37 & 0.38 & 0.49 & 0.49 & 0.48 \\ 
   \hline
\end{tabular}
\end{table}
\begin{table}
\caption{Table \ref{table-simulation 1 results} continued. For $\lambda=15$.}
\centering
\begin{tabular}{rrrrrrrr}
  \hline
 & n & Gam0CP & Gam1CP & Gam2CP & Gam3CP & Gam4CP & Gam5CP \\ 
  \hline
1 & 5.00 & 92.37 & 94.83 & 94.66 & 88.18 & 88.94 & 94.52 \\ 
  2 & 10.00 & 93.65 & 94.81 & 94.59 & 91.88 & 92.18 & 94.68 \\ 
  3 & 15.00 & 93.96 & 94.88 & 94.38 & 92.97 & 93.09 & 94.68 \\ 
  4 & 20.00 & 94.25 & 94.96 & 94.56 & 93.63 & 93.75 & 94.98 \\ 
  5 & 30.00 & 94.81 & 95.29 & 94.67 & 94.28 & 94.38 & 95.08 \\ 
  6 & 50.00 & 94.83 & 95.22 & 94.80 & 94.58 & 94.57 & 95.11 \\ 
  7 & 70.00 & 94.82 & 94.98 & 94.47 & 94.71 & 94.72 & 94.97 \\ 
  8 & 100.00 & 94.88 & 95.18 & 94.54 & 94.82 & 94.68 & 95.03 \\ 
   \hline
 & n & Gam0ML & Gam1ML & Gam2ML & Gam3ML & Gam4ML & Gam5ML \\ 
  \hline
1 & 5.00 & 14.07 & 15.68 & 15.58 & 12.17 & 12.52 & 15.42 \\ 
  2 & 10.00 & 14.07 & 14.86 & 14.91 & 13.21 & 13.32 & 14.80 \\ 
  3 & 15.00 & 14.09 & 14.66 & 14.69 & 13.54 & 13.60 & 14.57 \\ 
  4 & 20.00 & 14.10 & 14.56 & 14.54 & 13.69 & 13.73 & 14.46 \\ 
  5 & 30.00 & 14.10 & 14.44 & 14.43 & 13.84 & 13.85 & 14.35 \\ 
  6 & 50.00 & 14.10 & 14.32 & 14.31 & 13.95 & 13.95 & 14.25 \\ 
  7 & 70.00 & 14.09 & 14.26 & 14.26 & 13.99 & 13.98 & 14.20 \\ 
  8 & 100.00 & 14.10 & 14.22 & 14.21 & 14.01 & 14.02 & 14.17 \\ 
   \hline
 & n & Gam0SL & Gam1SL & Gam2SL & Gam3SL & Gam4SL & Gam5SL \\ 
  \hline
1 & 5.00 & 0.97 & 1.04 & 1.02 & 0.87 & 0.89 & 1.16 \\ 
  2 & 10.00 & 0.75 & 0.75 & 0.78 & 0.71 & 0.72 & 0.78 \\ 
  3 & 15.00 & 0.66 & 0.68 & 0.68 & 0.65 & 0.64 & 0.68 \\ 
  4 & 20.00 & 0.60 & 0.61 & 0.61 & 0.59 & 0.59 & 0.61 \\ 
  5 & 30.00 & 0.55 & 0.54 & 0.54 & 0.54 & 0.54 & 0.55 \\ 
  6 & 50.00 & 0.47 & 0.48 & 0.48 & 0.48 & 0.48 & 0.49 \\ 
  7 & 70.00 & 0.44 & 0.44 & 0.44 & 0.44 & 0.45 & 0.45 \\ 
  8 & 100.00 & 0.42 & 0.41 & 0.41 & 0.41 & 0.42 & 0.43 \\ 
   \hline
\end{tabular}
\end{table}
\begin{table}
\caption{Table \ref{table-simulation 1 results} continued. For $\lambda=30$.}
\centering
\begin{tabular}{rrrrrrrr}
  \hline
 & n & Gam0CP & Gam1CP & Gam2CP & Gam3CP & Gam4CP & Gam5CP \\ 
  \hline
1 & 5.00 & 92.80 & 95.08 & 94.73 & 88.18 & 89.10 & 94.98 \\ 
  2 & 10.00 & 93.66 & 94.94 & 94.51 & 92.08 & 92.26 & 94.95 \\ 
  3 & 15.00 & 94.51 & 95.34 & 95.02 & 93.55 & 93.69 & 95.21 \\ 
  4 & 20.00 & 94.36 & 95.15 & 94.63 & 93.86 & 93.86 & 95.00 \\ 
  5 & 30.00 & 94.67 & 95.28 & 94.97 & 94.42 & 94.29 & 95.03 \\ 
  6 & 50.00 & 94.75 & 95.11 & 94.76 & 94.53 & 94.47 & 94.92 \\ 
  7 & 70.00 & 95.00 & 95.29 & 94.98 & 94.83 & 94.79 & 95.25 \\ 
  8 & 100.00 & 95.00 & 95.15 & 94.98 & 94.84 & 94.84 & 95.08 \\ 
   \hline
 & n & Gam0ML & Gam1ML & Gam2ML & Gam3ML & Gam4ML & Gam5ML \\ 
  \hline
1 & 5.00 & 20.42 & 22.49 & 22.50 & 17.68 & 18.18 & 22.34 \\ 
  2 & 10.00 & 20.42 & 21.51 & 21.52 & 19.17 & 19.33 & 21.45 \\ 
  3 & 15.00 & 20.41 & 21.18 & 21.19 & 19.62 & 19.70 & 21.11 \\ 
  4 & 20.00 & 20.40 & 21.01 & 21.00 & 19.82 & 19.88 & 20.92 \\ 
  5 & 30.00 & 20.42 & 20.83 & 20.83 & 20.03 & 20.06 & 20.77 \\ 
  6 & 50.00 & 20.42 & 20.69 & 20.69 & 20.19 & 20.20 & 20.63 \\ 
  7 & 70.00 & 20.41 & 20.62 & 20.62 & 20.25 & 20.26 & 20.57 \\ 
  8 & 100.00 & 20.42 & 20.58 & 20.58 & 20.31 & 20.31 & 20.52 \\ 
   \hline
 & n & Gam0SL & Gam1SL & Gam2SL & Gam3SL & Gam4SL & Gam5SL \\ 
  \hline
1 & 5.00 & 0.97 & 1.05 & 1.05 & 0.87 & 0.89 & 1.31 \\ 
  2 & 10.00 & 0.75 & 0.77 & 0.76 & 0.71 & 0.72 & 0.77 \\ 
  3 & 15.00 & 0.66 & 0.68 & 0.68 & 0.64 & 0.64 & 0.67 \\ 
  4 & 20.00 & 0.61 & 0.61 & 0.59 & 0.59 & 0.60 & 0.61 \\ 
  5 & 30.00 & 0.55 & 0.55 & 0.55 & 0.53 & 0.53 & 0.55 \\ 
  6 & 50.00 & 0.52 & 0.50 & 0.51 & 0.48 & 0.48 & 0.51 \\ 
  7 & 70.00 & 0.50 & 0.49 & 0.49 & 0.47 & 0.48 & 0.50 \\ 
  8 & 100.00 & 0.50 & 0.49 & 0.49 & 0.48 & 0.48 & 0.50 \\ 
   \hline
\end{tabular}
\end{table}
\begin{table}
\caption{Table \ref{table-simulation 1 results} continued. For $\lambda=50$.}
\centering
\begin{tabular}{rrrrrrrr}
  \hline
 & n & Gam0CP & Gam1CP & Gam2CP & Gam3CP & Gam4CP & Gam5CP \\ 
  \hline
1 & 5.00 & 92.46 & 94.79 & 94.97 & 88.00 & 88.98 & 94.53 \\ 
  2 & 10.00 & 93.54 & 94.69 & 94.69 & 91.71 & 91.91 & 94.70 \\ 
  3 & 15.00 & 93.99 & 94.87 & 94.89 & 92.99 & 93.09 & 94.87 \\ 
  4 & 20.00 & 94.45 & 95.12 & 95.08 & 93.69 & 93.74 & 95.12 \\ 
  5 & 30.00 & 94.70 & 95.22 & 94.96 & 94.16 & 94.27 & 95.16 \\ 
  6 & 50.00 & 94.53 & 94.93 & 94.79 & 94.30 & 94.32 & 94.82 \\ 
  7 & 70.00 & 95.13 & 95.32 & 95.09 & 95.02 & 95.06 & 95.31 \\ 
  8 & 100.00 & 94.94 & 95.14 & 95.07 & 94.83 & 94.78 & 95.05 \\ 
   \hline
 & n & Gam0ML & Gam1ML & Gam2ML & Gam3ML & Gam4ML & Gam5ML \\ 
  \hline
1 & 5.00 & 26.65 & 29.37 & 29.37 & 23.12 & 23.76 & 29.13 \\ 
  2 & 10.00 & 26.65 & 28.04 & 28.04 & 25.04 & 25.24 & 27.99 \\ 
  3 & 15.00 & 26.67 & 27.61 & 27.63 & 25.63 & 25.74 & 27.56 \\ 
  4 & 20.00 & 26.68 & 27.41 & 27.42 & 25.92 & 25.98 & 27.35 \\ 
  5 & 30.00 & 26.67 & 27.18 & 27.18 & 26.18 & 26.21 & 27.13 \\ 
  6 & 50.00 & 26.68 & 26.99 & 27.00 & 26.39 & 26.40 & 26.95 \\ 
  7 & 70.00 & 26.67 & 26.91 & 26.91 & 26.47 & 26.48 & 26.86 \\ 
  8 & 100.00 & 26.67 & 26.85 & 26.86 & 26.54 & 26.54 & 26.81 \\ 
   \hline
 & n & Gam0SL & Gam1SL & Gam2SL & Gam3SL & Gam4SL & Gam5SL \\ 
  \hline
1 & 5.00 & 0.96 & 1.02 & 1.02 & 0.87 & 0.88 & 1.50 \\ 
  2 & 10.00 & 0.75 & 0.77 & 0.78 & 0.71 & 0.72 & 0.78 \\ 
  3 & 15.00 & 0.65 & 0.66 & 0.67 & 0.64 & 0.64 & 0.67 \\ 
  4 & 20.00 & 0.60 & 0.62 & 0.62 & 0.59 & 0.59 & 0.61 \\ 
  5 & 30.00 & 0.55 & 0.55 & 0.55 & 0.54 & 0.54 & 0.55 \\ 
  6 & 50.00 & 0.50 & 0.45 & 0.46 & 0.51 & 0.51 & 0.47 \\ 
  7 & 70.00 & 0.48 & 0.44 & 0.44 & 0.50 & 0.50 & 0.44 \\ 
  8 & 100.00 & 0.47 & 0.43 & 0.44 & 0.50 & 0.50 & 0.44 \\ 
   \hline
\end{tabular}
\end{table}
\begin{table}[ht]
\caption{Table \ref{table-simulation 1 results} continued. For $\lambda=100$.}
\centering
\begin{tabular}{rrrrrrrr}
  \hline
 & n & Gam0CP & Gam1CP & Gam2CP & Gam3CP & Gam4CP & Gam5CP \\ 
  \hline
1 & 5.00 & 92.22 & 94.90 & 94.77 & 87.69 & 88.75 & 94.60 \\ 
  2 & 10.00 & 93.35 & 94.49 & 94.66 & 91.88 & 92.11 & 94.52 \\ 
  3 & 15.00 & 94.14 & 94.89 & 94.86 & 93.13 & 93.20 & 94.82 \\ 
  4 & 20.00 & 94.53 & 95.29 & 95.17 & 93.73 & 93.84 & 95.24 \\ 
  5 & 30.00 & 94.79 & 95.17 & 95.08 & 94.22 & 94.33 & 95.11 \\ 
  6 & 50.00 & 94.62 & 94.89 & 94.88 & 94.34 & 94.28 & 94.86 \\ 
  7 & 70.00 & 94.91 & 95.08 & 95.03 & 94.75 & 94.75 & 95.04 \\ 
  8 & 100.00 & 94.69 & 94.87 & 94.84 & 94.58 & 94.60 & 94.82 \\ 
   \hline
 & n & Gam0ML & Gam1ML & Gam2ML & Gam3ML & Gam4ML & Gam5ML \\ 
  \hline
1 & 5.00 & 38.17 & 41.98 & 41.93 & 33.15 & 34.05 & 41.68 \\ 
  2 & 10.00 & 38.16 & 40.11 & 40.10 & 35.85 & 36.16 & 40.06 \\ 
  3 & 15.00 & 38.17 & 39.49 & 39.49 & 36.70 & 36.85 & 39.45 \\ 
  4 & 20.00 & 38.16 & 39.17 & 39.17 & 37.09 & 37.17 & 39.13 \\ 
  5 & 30.00 & 38.17 & 38.85 & 38.85 & 37.47 & 37.52 & 38.82 \\ 
  6 & 50.00 & 38.16 & 38.59 & 38.59 & 37.76 & 37.78 & 38.56 \\ 
  7 & 70.00 & 38.17 & 38.49 & 38.48 & 37.87 & 37.89 & 38.45 \\ 
  8 & 100.00 & 38.17 & 38.39 & 38.38 & 37.96 & 37.97 & 38.37 \\ 
   \hline
 & n & Gam0SL & Gam1SL & Gam2SL & Gam3SL & Gam4SL & Gam5SL \\ 
  \hline
1 & 5.00 & 0.97 & 1.04 & 1.05 & 0.87 & 0.88 & 1.91 \\ 
  2 & 10.00 & 0.75 & 0.75 & 0.78 & 0.70 & 0.71 & 0.77 \\ 
  3 & 15.00 & 0.66 & 0.67 & 0.67 & 0.64 & 0.64 & 0.67 \\ 
  4 & 20.00 & 0.60 & 0.61 & 0.61 & 0.59 & 0.60 & 0.61 \\ 
  5 & 30.00 & 0.54 & 0.55 & 0.56 & 0.55 & 0.55 & 0.54 \\ 
  6 & 50.00 & 0.48 & 0.51 & 0.51 & 0.49 & 0.49 & 0.52 \\ 
  7 & 70.00 & 0.45 & 0.50 & 0.50 & 0.44 & 0.44 & 0.51 \\ 
  8 & 100.00 & 0.43 & 0.49 & 0.49 & 0.40 & 0.40 & 0.49 \\ 
   \hline
\end{tabular}
\end{table}
\begin{table}
\caption{Table \ref{table-simulation 1 results} continued. For $\lambda=200$.}
\centering
\begin{tabular}{rrrrrrrr}
  \hline
 & n & Gam0CP & Gam1CP & Gam2CP & Gam3CP & Gam4CP & Gam5CP \\ 
  \hline
1 & 5.00 & 92.80 & 95.03 & 95.01 & 88.65 & 89.44 & 94.77 \\ 
  2 & 10.00 & 93.66 & 94.75 & 94.62 & 92.00 & 92.29 & 94.69 \\ 
  3 & 15.00 & 93.95 & 94.69 & 94.75 & 92.80 & 92.95 & 94.70 \\ 
  4 & 20.00 & 94.11 & 94.81 & 94.90 & 93.47 & 93.51 & 94.85 \\ 
  5 & 30.00 & 94.96 & 95.35 & 95.41 & 94.61 & 94.66 & 95.29 \\ 
  6 & 50.00 & 94.84 & 95.09 & 95.11 & 94.60 & 94.63 & 95.08 \\ 
  7 & 70.00 & 95.19 & 95.36 & 95.11 & 94.99 & 94.99 & 95.30 \\ 
  8 & 100.00 & 94.65 & 94.78 & 94.85 & 94.50 & 94.53 & 94.77 \\ 
   \hline
 & n & Gam0ML & Gam1ML & Gam2ML & Gam3ML & Gam4ML & Gam5ML \\ 
  \hline
1 & 5.00 & 54.41 & 59.71 & 59.70 & 47.29 & 48.56 & 59.37 \\ 
  2 & 10.00 & 54.41 & 57.13 & 57.13 & 51.15 & 51.57 & 57.09 \\ 
  3 & 15.00 & 54.41 & 56.23 & 56.27 & 52.33 & 52.53 & 56.21 \\ 
  4 & 20.00 & 54.42 & 55.81 & 55.81 & 52.89 & 53.01 & 55.77 \\ 
  5 & 30.00 & 54.43 & 55.35 & 55.35 & 53.43 & 53.49 & 55.32 \\ 
  6 & 50.00 & 54.41 & 54.97 & 54.98 & 53.83 & 53.86 & 54.95 \\ 
  7 & 70.00 & 54.42 & 54.83 & 54.83 & 54.01 & 54.03 & 54.81 \\ 
  8 & 100.00 & 54.42 & 54.72 & 54.72 & 54.13 & 54.14 & 54.69 \\ 
   \hline
 & n & Gam0SL & Gam1SL & Gam2SL & Gam3SL & Gam4SL & Gam5SL \\ 
  \hline
1 & 5.00 & 0.97 & 1.06 & 1.03 & 0.86 & 0.88 & 2.51 \\ 
  2 & 10.00 & 0.74 & 0.77 & 0.76 & 0.71 & 0.71 & 0.84 \\ 
  3 & 15.00 & 0.65 & 0.65 & 0.66 & 0.64 & 0.64 & 0.66 \\ 
  4 & 20.00 & 0.61 & 0.61 & 0.61 & 0.58 & 0.60 & 0.61 \\ 
  5 & 30.00 & 0.55 & 0.55 & 0.55 & 0.55 & 0.55 & 0.55 \\ 
  6 & 50.00 & 0.51 & 0.47 & 0.47 & 0.48 & 0.48 & 0.47 \\ 
  7 & 70.00 & 0.50 & 0.46 & 0.45 & 0.42 & 0.44 & 0.46 \\ 
  8 & 100.00 & 0.50 & 0.46 & 0.46 & 0.41 & 0.42 & 0.47 \\ 
   \hline
\end{tabular}
\end{table}

\begin{table}
\caption{Simulation Case \#1 :  Parameters:  $p= 1$, $\theta= (3, 5)$,  $W \sim U[0,1]$.} 
\label{BigSimulCase  1}
\centering
\begin{tabular}{rrrrr}
  \hline
 & n & Gam0CP & Gam1CP & Gam2CP \\ 
  \hline
1 & 30.00 & 94.36 & 95.26 & 94.98 \\ 
  2 & 50.00 & 94.36 & 94.99 & 94.78 \\ 
  3 & 100.00 & 94.95 & 95.24 & 95.15 \\ 
  4 & 200.00 & 94.82 & 94.92 & 94.84 \\ 
   \hline
  \hline
 & n & Gam0ML & Gam1ML & Gam2ML \\ 
  \hline
1 & 30.00 & 77.92 & 82.29 & 82.28 \\ 
  2 & 50.00 & 77.77 & 80.29 & 80.29 \\ 
  3 & 100.00 & 76.94 & 78.15 & 78.15 \\ 
  4 & 200.00 & 78.74 & 79.37 & 79.37 \\ 
   \hline
 & n & Gam0SL & Gam1SL & Gam2SL \\ 
  \hline
1 & 30.00 & 54.04 & 60.53 & 60.52 \\ 
  2 & 50.00 & 54.40 & 58.01 & 58.01 \\ 
  3 & 100.00 & 53.84 & 55.50 & 55.50 \\ 
  4 & 200.00 & 54.39 & 55.22 & 55.22 \\ 
   \hline
\end{tabular}
\end{table}

\begin{table}
\caption{Simulation Case \#2:  Parameters:  $p= 2$, $\theta= (3, -0.2, 0.05)$,  $W \sim N(\mu=2,\sigma=2)$.} 
\label{BigSimulCase  2}
\centering
\begin{tabular}{rrrrr}
  \hline
 & n & Gam0CP & Gam1CP & Gam2CP \\ 
  \hline
1 & 30.00 & 93.10 & 95.27 & 95.01 \\ 
  2 & 50.00 & 94.13 & 94.99 & 94.79 \\ 
  3 & 100.00 & 94.10 & 94.68 & 94.41 \\ 
  4 & 200.00 & 94.52 & 94.76 & 94.59 \\ 
   \hline
 & n & Gam0ML & Gam1ML & Gam2ML \\ 
  \hline
1 & 30.00 & 16.78 & 19.18 & 19.21 \\ 
  2 & 50.00 & 16.69 & 17.86 & 17.87 \\ 
  3 & 100.00 & 16.73 & 17.30 & 17.30 \\ 
  4 & 200.00 & 16.72 & 17.01 & 17.01 \\ 
   \hline
 & n & Gam0SL & Gam1SL & Gam2SL \\ 
  \hline
1 & 30.00 & 3.66 & 21.32 & 23.37 \\ 
  2 & 50.00 & 3.30 & 11.27 & 11.62 \\ 
  3 & 100.00 & 3.32 & 6.08 & 6.08 \\ 
  4 & 200.00 & 3.12 & 4.05 & 4.05 \\ 
   \hline
\end{tabular}
\end{table}

\begin{table}
\caption{Simulation Case \#3:  Parameters:  $p= 3$, $\theta= (3, 0.2, -0.1, -0.05)$,  $W \sim N(\mu=1,\sigma=2)$.} 
\label{BigSimulCase  3}
\centering
\begin{tabular}{rrrrr}
  \hline
 & n & Gam0CP & Gam1CP & Gam2CP \\ 
  \hline
1 & 30.00 & 92.33 & 94.74 & 94.44 \\ 
  2 & 50.00 & 92.67 & 94.30 & 93.48 \\ 
  3 & 100.00 & 94.13 & 95.08 & 94.22 \\ 
  4 & 200.00 & 94.91 & 95.51 & 94.36 \\ 
   \hline
 & n & Gam0ML & Gam1ML & Gam2ML \\ 
  \hline
1 & 30.00 & 13.62 & 2011.21 & 5679.31 \\ 
  2 & 50.00 & 13.28 & 150.32 & 237.33 \\ 
  3 & 100.00 & 13.44 & 617.14 & 1010.62 \\ 
  4 & 200.00 & 13.07 & 47.16 & 47.30 \\ 
   \hline
 & n & Gam0SL & Gam1SL & Gam2SL \\ 
  \hline
1 & 30.00 & 57.83 & 197118.78 & 561658.43 \\ 
  2 & 50.00 & 22.62 & 7945.90 & 12852.71 \\ 
  3 & 100.00 & 46.33 & 58624.75 & 96369.77 \\ 
  4 & 200.00 & 17.05 & 3327.74 & 3327.77 \\ 
   \hline
\end{tabular}
\end{table}

\begin{table}
\caption{Simulation Case \#4:  Parameters:  $p= 5$, $\theta= (3, -1, 3, -2, 1, -0.5)$, $W \sim U[0,1]$.} 
\label{BigSimulCase  4}
\centering
\begin{tabular}{rrrrr}
  \hline
 & n & Gam0CP & Gam1CP & Gam2CP \\ 
  \hline
1 & 30.00 & 91.32 & 94.96 & 94.81 \\ 
  2 & 50.00 & 92.96 & 94.99 & 94.75 \\ 
  3 & 100.00 & 94.36 & 95.13 & 94.99 \\ 
  4 & 200.00 & 94.71 & 95.08 & 94.84 \\ 
   \hline
 & n & Gam0ML & Gam1ML & Gam2ML \\ 
  \hline
1 & 30.00 & 17.68 & 21.43 & 22.93 \\ 
  2 & 50.00 & 17.65 & 19.11 & 19.12 \\ 
  3 & 100.00 & 17.65 & 18.35 & 18.35 \\ 
  4 & 200.00 & 17.63 & 17.99 & 17.99 \\ 
   \hline
 & n & Gam0SL & Gam1SL & Gam2SL \\ 
  \hline
1 & 30.00 & 2.50 & 35.46 & 124.53 \\ 
  2 & 50.00 & 2.12 & 3.70 & 3.77 \\ 
  3 & 100.00 & 2.02 & 2.39 & 2.38 \\ 
  4 & 200.00 & 1.97 & 2.08 & 2.08 \\ 
   \hline
\end{tabular}
\end{table}

\end{document}